\documentclass[11pt]{article}
\usepackage{graphicx} \usepackage[margin=1in]{geometry}
\usepackage{drl}
\usepackage{setspace}
\pgfplotsset{compat=1.18}
\usepackage{tcolorbox}
\usepackage{times}

\usepackage{algorithmic}
\usepackage{authblk}

\usepackage[normalem]{ulem}

\title{
Causal Invariance Learning via Efficient Nonconvex Optimization
}
\author[1,*]{Zhenyu Wang}
\author[1,*]{Yifan Hu}
\author[2,†]{Peter B\"{u}hlmann}
\author[3,†]{Zijian Guo}
\affil[1]{Department of Statistics, Rutgers University, USA}
\affil[2]{Seminar for Statistics, ETH Z\"urich, Switzerland}
\affil[3]{Center for Data Science, Zhejiang University, China}
\affil[*]{Equal contribution \quad † Co-corresponding authors}

\date{\today}

\begin{document}
\maketitle

\begin{abstract}
Identifying the causal relationship among variables from observational data is an important yet challenging task. This work focuses on identifying the direct causes of an outcome and estimating their magnitude, i.e., learning the causal outcome model. Data from multiple environments provide valuable opportunities to uncover causality by exploiting the invariance principle that the causal outcome model holds across heterogeneous environments.  Based on the invariance principle, we propose the Negative Weighted Distributionally Robust Optimization (NegDRO) framework to learn an invariant prediction model. NegDRO minimizes the worst-case combination of risks across multiple environments and enforces invariance by allowing potential negative weights. Under the additive interventions regime, we establish three major contributions: (i) On the statistical side, we provide sufficient and nearly necessary identification conditions under which the invariant prediction model coincides with the causal outcome model; (ii) On the optimization side, despite the nonconvexity of NegDRO, we establish its benign optimization landscape, where all stationary points lie close to the true causal outcome model; (iii) On the computational side, we develop a gradient-based algorithm that provably converges to the causal outcome model, with non-asymptotic convergence rates in both sample size and gradient-descent iterations.
In particular, our method avoids exhaustive combinatorial searches over exponentially many subsets of covariates found in 
the literature, ensuring scalability even when the dimension of the covariates is
large. To our knowledge, this is the first causal invariance learning method that finds the approximate global optimality for a nonconvex optimization problem efficiently. 
\end{abstract}

\section{Introduction}
\label{sec: intro}
\begin{figure}[ht!]
    \centering
    \includegraphics[width=0.6\linewidth]{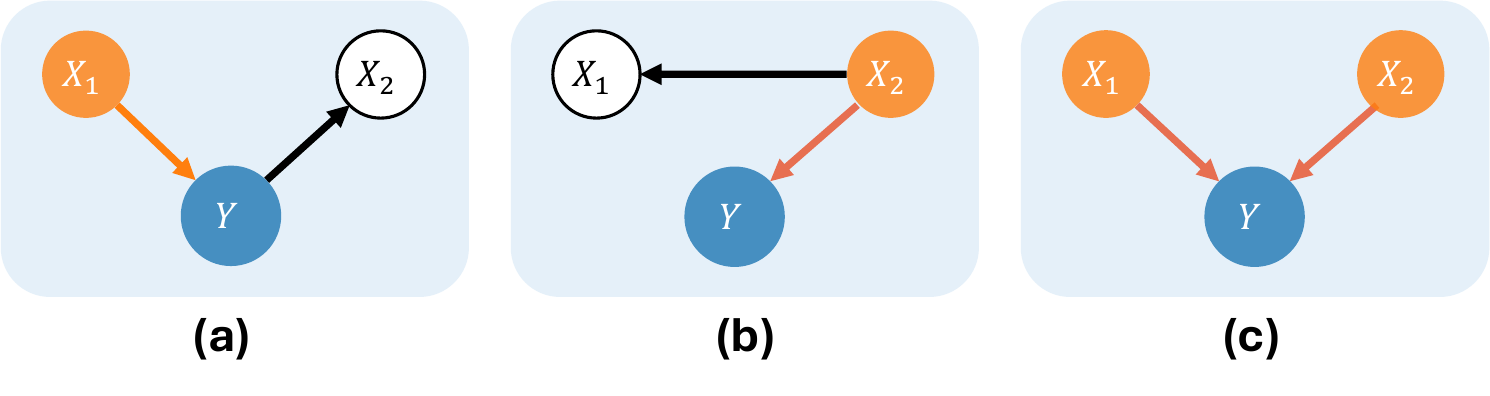}
    \caption{Illustration of plausible causal structures, where orange nodes denote the direct causes for $Y$. }
    \label{fig: eg-1}
\end{figure}

Establishing causal relationships between an outcome and multiple covariates is a fundamental objective in various fields, including marketing and economics \citep{duflo2007using, ludwig2013long, varian2016causal}, epidemiology \citep{baicker2013oregon}, and computer science \citep{scholkopf2021toward, scholkopf2022causality}. In many cases, the causal structure is unknown: some covariates may affect the outcome, others may be influenced by the outcome, some may share unobserved confounders, and others may have no causal connection at all. In such settings, methods that focus on minimizing prediction error often capture spurious correlations rather than true causal effects. %
To facilitate the discussion, we introduce a concrete example that will serve as a running illustration throughout the paper.
\begin{Example}
    Consider a company aiming to identify which advertising channels causally drive sales $Y$. Suppose it tracks two key engagement metrics: $X_1$, the clicks from search ads on Google, and $X_2$, the engagements from social media ads. Several plausible causal structures could underlie these variables, as illustrated in Figure \ref{fig: eg-1}: (a) $X_1 \to Y \to X_2$: search ad clicks increase sales, and higher sales bring more customers to generate social engagement; (b) $X_2 \to Y, X_2\to X_1$: social media engagement drives sales and also leads customers to perform additional search activity; (c) $X_1\to Y\leftarrow X_2$: both search ad clicks and social media engagements directly boost sales.
In the figure, each arrow represents the causal relationship from one variable to another where the orange arrow represents true cause of $Y$.
\label{eg: concrete}
\end{Example}
For this example, a model trained solely to minimize prediction risk, across all three causal structures, would include both $X_1$ and $X_2$ to capture the associations with $Y$. However, such a model cannot distinguish among the above causal structures and thus fails to reveal which ad channel causes sales. As a result, it could lead to a waste of resources if one invests in non-causal metrics.  This limitation underscores the need for methodologies that go beyond predictive accuracy to uncover the underlying causes of the outcome. Throughout this work, we refer to the task of identifying the causes of the outcome and quantifying their magnitude as \textbf{causal identification}.

Solving causal identification is not easy, as it often requires randomized trials and controlled experiments. The emergence of heterogeneous multi-environment data offers opportunities for causal identification using observational data through identifying invariance among differences, which motivates causal invariance learning \citep{peters2016causal, ghassami2017learning, pfister2021stabilizing, fan2023environment,taeb2024learning}. 
The central idea is that while the joint data distribution varies across environments, the causal relationship from the true causal variables to the outcome remains invariant. Such type of multi-source dataset is widely available in healthcare, marketing, and operations.

Returning to Example \ref{eg: concrete}, suppose we collect ads and sales data from different environments, such as across distinct time slots or regional markets. Although the observed data may exhibit distributional shifts due to differences in measurement conditions or market characteristics, the true causal drivers of sales remain unchanged.

This perspective distinguishes causal invariance learning from much of the existing causality literature. A large body of causal inference literature assumes a known causal structure (i.e., causal graph of different variables) and focuses on estimating causal effects under assumptions such as ignorability or leveraging instrumental variables \citep{wooldridge2009introductory, imbens2015causal, wager2018estimation, wager2024causal}. Meanwhile, DAG-based approaches aim to recover the full causal structure among variables using conditional independence testing, but mainly up to a Markov equivalence class \citep{verma1990equivalence, spirtes2001causation,chickering2002optimal,
hauser2012characterization, zheng2018dags}. %
In contrast, causal invariance learning aims to identify an invariant prediction model that captures the underlying causal relationship from $X$ to $Y$, leveraging distributional heterogeneity across environments.

\subsection{Problem Formulations and Gaps} %
\label{subsec: gaps}

Consider a collection of environments denoted as $\Ec=\{1,2,..., |\Ec|\}$. For each $e\in \Ec$, we observe independent and identically distributed (i.i.d.) samples from the distribution of $(\X{e},\Y{e})$, where $\X{e}\in \Rb^p$ and $\Y{e}\in \Rb$ represent the covariates and the outcome for the $e$-th environment, respectively. 
Among the $p$ covariates, an unknown subset $S^*$ consists of the direct causes of the outcome, while the remaining covariates may be associated with the outcome but are not its direct causes, e.g., variables influenced by the outcome. Notably, this framework allows for the existence of hidden confounding variables; that is, unobserved factors that may influence both the observed covariates and the outcome.

The central assumption of causal invariance learning is that the causal outcome model remains invariant across environments,  while the joint distributions of $(\X{e}, \Y{e})$ vary with $e\in \Ec$.  To identify the causal outcome model from the observed multi-environment data, the standard strategy is to define an invariant prediction model, establish causal identification conditions under which this invariant prediction model coincides with the causal outcome model, and then develop algorithms to learn the invariant prediction model. %
However, there are two major challenges for the causal invariance learning: %
\begin{enumerate}
\item[\textit{(Q1)}] \textbf{Causal Identification:} \textit{When does the defined invariant prediction model recover the causal outcome model?}
\item[\textit{(Q2)}] 
\textbf{Computational Efficiency:} \textit{How can we learn the invariant prediction model in a computationally efficient manner?}
\end{enumerate}

For \textit{(Q1)}, existing methods either rely on abstract conditions or impose overly restrictive conditions, both of which offer limited practical guidance for collecting data or designing heterogeneous environments.
For instance, \citet{fan2023environment} and \citet{yin2024optimization} require the collected environments to exhibit sufficient heterogeneity so that any subset of variables involving non-causal ones violates their proposed invariance property. 
This assumption can be difficult to translate into concrete experimental procedures, especially when researchers or practitioners must decide which variables to intervene on to uncover the underlying causal structure\citep{he2008active, hauser2015jointly}.
On the other hand, \citet{rojas2018invariant} and \citet{arjovsky2019invariant} propose more explicit conditions on the number of environments, but these are highly demanding: the former requires infinitely many environments covering all possible covariate interventions, while the latter requires at least \( |\Ec| \geq p \) environments. However, for large $p$, creating such a diverse collection of environments is prohibitively costly or even infeasible, limiting the applicability of these methods. This gap motivates us to establish concrete conditions under which the invariant prediction model recovers the causal outcome model that remains valid, even with a limited number of environments, providing practical guidance on the data collection. We denote such conditions as causal identification conditions. %

For \textit{(Q2)}, since the true causal subset of covariates is unknown, existing methods \citep{peters2016causal, ghassami2017learning} rely on exhaustive searches over all possible (exponentially many) subsets of covariates to test for invariance.
While \citet{fan2023environment} proposed finding invariant prediction models using a regularized least squares optimization, due to the discontinuity and nonconvexity of the regularizer, solving the problem still necessitates an exhaustive enumeration of subsets. All these approaches, despite theoretical guarantees that the invariant prediction models recover the causal outcome model, admit computational costs that scale exponentially with the dimension $p$; see more discussions in Section \ref{subsec: negdro formulation}. 
This motivates the development of alternative approaches that can compute invariant prediction models in time polynomial in $p$, which allows us to handle much larger problems.

\subsection{Results and Contributions}

 Leveraging the fact that the causal outcome model achieves the same prediction risks across all environments under linear models,  We propose a novel risk invariance principle that a prediction model is risk invariant if it achieves the same prediction model risks in all environments. Building upon this risk invariance principle, we introduce a continuous minimax formulation for causal invariance learning, called \textbf{Neg}ative Weighted \textbf{D}istributionally \textbf{R}obust \textbf{O}ptimization ({NegDRO}). NegDRO minimizes the worst-case combination of risks across multiple environments, where the weights are allowed to be negative. Such negative weights are critical in enforcing risk invariance across environments and building risk-invariant prediction models; yet, they also break convexity, as they lead to a difference-of-convex objective function. 
Unlike existing approaches that rely on exhaustive subset enumeration which usually comes from an integer programming, NegDRO aims to solve the continuous nonconvex optimization problem to global optimality.

Under the additive intervention regime, a widely used framework for modeling heterogeneous multiple environment data \citep{rothenhausler2019causal, rothenhausler2021anchor, shen2023causality} that characterizes the data generation process via structural equation models, we address \textit{(Q1)} by establishing concrete conditions for causal identification, i.e., when does the invariant prediction model, defined as the optimal solution of NegDRO, recover the causal outcome model.  Specifically, we derive a sufficient and nearly necessary condition under which the invariant prediction model is unique across environments. Since the causal outcome model itself is an invariant prediction model, this uniqueness guarantees that the learned invariant prediction model exactly recovers the causal outcome model. Unlike the abstract heterogeneity requirements in prior works \citep{fan2023environment, yin2024optimization}, our condition is stated directly in terms of covariate distributions across environments.
Moreover, compared to approaches that demand prohibitively many environments \citep{rojas2018invariant, arjovsky2019invariant}, our condition can achieve causal identification with as few as two environments; see Condition \ref{cond: strict positive} for more details.

Building on the proposed identification condition, we establish the benign nonconvex landscape of NegDRO, by demonstrating that any (generalized) stationary point of NegDRO lies close to the causal outcome model. Leveraging this insight, we design a gradient descent maximization algorithm that efficiently computes stationary points of NegDRO, and thereby closely approximates the causal outcome model. Unlike exhaustive search approaches that incur a computational burden scaling exponentially, our algorithm easily scales to scenarios involving a large number of variables, addressing \textit{(Q2)}. Furthermore, we provide theoretical guarantees for the convergence of our algorithm in terms of both the sample size and the number of iterations needed to find a prediction model that is $\epsilon$ close to the causal outcome model. 

Under a more general limited intervention regime where  the invariant prediction model may not be unique, we further establish a weaker condition under which NegDRO still achieves causal identification. The key insight is that NegDRO not only enforces risk invariance; it also selects the invariant prediction model that achieves the smallest prediction risk. This weaker condition becomes particularly useful under limited intervention regimes, where some covariates are never intervened upon across environments. Such scenarios often arise in practice due to cost constraints or ethical considerations; for example, in clinical trials where certain variables cannot be manipulated \citep{banerjee2009experimental}. This result underscores the importance of incorporating an optimization objective instead of focusing purely on invariance testing~\citep{peters2016causal}, which can be seen as a constraint feasibility problem.

Lastly, we validate NegDRO through empirical studies, demonstrating its superior performance in terms of both estimation accuracy and computational efficiency within the additive intervention regime, compared to existing causal invariance learning methods.

To summarize, the main contributions of this paper are as follows:
\begin{enumerate}
    \item We propose a novel continuous nonconvex NegDRO formulation for causal invariance learning.
    \item We establish the sufficient and nearly necessary causal identification condition under which the invariant prediction model is unique in the additive intervention regime, ensuring alignment between the invariant prediction model and the causal outcome model.
    \item Despite nonconvexity, we show that any (generalized) stationary point of NegDRO is nearly globally optimal, and propose a computationally efficient algorithm with convergence rates in both sample size and iteration complexity to recover the causal outcome model.
    \item Under a more general limited intervention regime, we establish  a weaker causal identification condition for NegDRO and provide insights on how optimization outperforms invariance testing.
\end{enumerate}

\subsection{Related Literature}
\label{sec: further liter}

This section reviews relevant research and highlights how the proposed NegDRO relates to and differs from existing works.

\paragraph{Multi-source Learning.}
The literature has extensively studied multi-source data, including multi-task learning \citep{duan2023adaptive, xu2025multitask},
Group DRO \citep{sagawa2019distributionally, wang2023distributionally, zhang2024optimal}, and Maximin Effects \citep{meinshausen2015maximin, guo2024statistical}.
These methods are designed to construct predictive models that perform well under distributional shifts, instead of identifying the causal outcome model. In contrast, NegDRO minimizes the worst-case combination of risks by allowing \emph{negative} combination weights, which enforces risk invariance, a key property for identifying the causal outcome model. The introduction of negative weights, however, leads to a nonconvex optimization problem that does not arise in the aforementioned methods. As one of our key contributions, we develop a computationally efficient algorithm to solve this nonconvex optimization problem and thus achieve causal identification.

\paragraph{Causal-oriented Representation Learning.} Inspired by causal invariance learning, approaches such as “Invariant Risk Minimization” (IRM) \citep{arjovsky2019invariant} and its variants \citep{chuang2020estimating, lu2021nonlinear, liu2021heterogeneous} focus on finding data representations where optimal prediction models remain invariant across all environments. Such techniques have been applied extensively in machine learning. ``Minimax Risk Extrapolation'' (MM-REx) \citep{krueger2021out} leverages the risk invariance principle, aligning conceptually with our NegDRO but emphasizing minimizing the variance of environmental risks. Despite some reported and partially debated empirical successes, their
theoretical understanding remains limited \citep{rosenfeld2020risks, kamath2021does}. In contrast, our work focuses on linear models, providing theoretical analysis including identification conditions, as well as computationally efficient algorithms with rigorous convergence guarantees. Our results serve as a stepping stone for advancing causal invariance learning in more complex, nonlinear scenarios.

\paragraph{Global Optimality of Nonconvex Optimization.} 
Achieving global optimality in nonconvex optimization is challenging.
Recent advancements have leveraged structured landscape properties to establish global convergence for gradient-based methods. Key conditions facilitating such results include hidden convexity \citep{ben1996hidden}, the Polyak-Lojasiewicz (PL) condition \citep{polyak1963gradient,lojasiewicz1963topological}, the Kurdyka-Lojasiewicz (KL) condition \citep{kurdyka1998gradients}, and other gradient dominance properties \citep{karimi2016linear}. We refer interested readers to \citet{sun2021nonconvex} for a more comprehensive list. Such conditions have been established in various applications such as revenue management~\citep{chen2023network,chen2024efficient,miao2025network} and reinforcement learning~\citep{bhandari2024global, zhang2020variational,fatkhullin2023stochastic2,chen2024landscape}. For
the causal invariance learning problem considered in this work, exploiting the unique structure of the causal outcome model and the associated identification conditions, we develop a novel analysis demonstrating that any (generalized) stationary point of the NegDRO problem is close to the true causal outcome model, establishing a benign landscape. Note that the derived landscape characterization does not admit any of the above-mentioned existing landscape conditions. Indeed, the landscape characterization resembles the error bound condition~\citep{karimi2016linear} but it is built towards the true causal outcome model instead of the optimal solution.

\paragraph{Minimax Optimization.} 
The proposed NegDRO identifies the causal outcome model by solving a nonconvex concave minimax optimization problem. In contrast to standard convex concave minimax problems, where a vast amount of literature has demonstrated global optimality through gradient-based algorithms \citep{korpelevich1976extragradient, chen1997convergence, nemirovski2004prox, auslender2009projected, nedic2009subgradient, lin2020near, 
rahimian2022frameworks}, nonconvex concave minimax optimization typically only guarantees convergence to local minima, leaving the global optimum unexplored \citep{heusel2017gans, rafique2022weakly, lin2024two, zhang2024generalization}. We emphasize that the nonconvexity in the proposed NegDRO problem arises from the possibly negative weights instead of the generic nonconvex function as studied in the previous works. Leveraging the proposed identification condition, we obtain global optimizers of the nonconvex-concave NegDRO problem. This fundamental distinction sets our work apart from prior studies, where achieving global optimality remains an unsolved challenge.

\subsection{Preliminaries and Notations}
We summarize several important concepts that facilitate the discussion of the current paper, and introduce the notations that are used throughout this paper. 
\begin{Definition}[Structural Equation Models] SEMs are used to characterize the causal relationship among variables \citep{pearl2009causality, peters2017elements}.
    We consider the SEM on variables $Z = (Z_1,Z_2,...,Z_{p+1})$ with 
    $$Z_j = f_j({\rm Pa}(Z_j), \varepsilon_j) \quad \text{for}\quad 1\leq j\leq p+1,$$ where the set ${\rm Pa}(Z_j)\subseteq \{Z_1,...,Z_{p+1}\}$ denotes the set of direct causes of $Z_j$, or \emph{parents}, of the variables, and $\varepsilon_j$ represents the random error or disturbance due to omitted factors. 
    In our following discussions, the first component $Z_1$ is specified as the outcome $Y$, while other components $Z_{2:p+1}$ correspond to the covariates $X\in \Rb^p$. When all functions $f_j$'s are linear, these SEMs are referred to as linear SEMs.
    \label{def: SEM}
\end{Definition}
\begin{Definition}[Stationary Point]
    A stationary point of a differentiable function $f(\cdot)$ over $\Rb^p$ is a point $x^*$ with $\nabla f(x^*)=0.$
    \label{def: stationary point}
\end{Definition}

We define $[m] = \{1,2,...,m\}$ for the positive integer $m$. 
For real numbers $a$ and $b$, define $a\wedge b = \min\{a,b\}$ and $a\vee b = \max\{a,b\}$. For positive sequences $a(n)$ and $b(n)$, we use $a(n) \lesssim b(n)$, $a(n) = \mathcal{O}(b(n))$ or $b(n) = \Omega(a(n))$ to represent that there exists some universal constant $C > 0$ such that $a(n) \leq C \cdot b(n)$ for all $n \geq 1$, 
and denote $a(n) \asymp b(n)$ if $a(n) \lesssim b(n)$ and $b(n) \lesssim a(n)$. We use notations $a(n) \ll b(n)$ or $a(n) = o(b(n))$ if $\limsup_{n\to\infty}(a(n)/b(n)) = 0$. For a set $S$, we use $|S|$ to denote its cardinality.
For a vector $x\in \Rb^p$ and a set $S\subseteq [p]$, $x_{S}$ represents the $|S|$-dimensional sub-vector of $x$ consisting of $x_j$'s for all $j\in S$. For $q\geq 0,$ let $\|x\|_q = (\sum_{i=1}^p |x_i|^q)^{1/q}$ be its $\ell_q$ norm.
For a matrix $A = [A_{i,j}]_{i\in [n], j\in [m]}$, we denote $A_{S_1,S_2} = [A_{i,j}]_{i\in S_1, j\in S_2}$ as the sub-matrix of $A$. We let $\|A\|_2$ and $\|A\|_F$ be the spectral norm and Frobenius norm of matrix $A$, respectively.  For a symmetric matrix $A$, we use $\lambda_{\rm min}(A)$ to denote its smallest eigenvalue.
We use $\Ibf_p$ to denote the $p$-dimensional identity matrix. We use $c$ and $C$ to denote generic positive constants that may vary from place to place.

\subsection{Organization}
The structure of the paper is as follows. 
Section \ref{sec: causal inv} introduces the causal invariance learning framework and presents the formulation of the proposed NegDRO method.
In Section \ref{sec: additive intervention}, we focus on the additive intervention regime and establish concrete conditions under which the invariant prediction model is unique and coincides with the causal outcome model.
Building on this, Section \ref{sec: alg} characterizes the benign nonconvex landscape of NegDRO and develops a computationally efficient algorithm for solving it.
Section \ref{sec: minimization} then relaxes the identification condition and demonstrates that NegDRO continues to achieve causal identification even when multiple invariant prediction models exist.
Finally, Section \ref{sec: numerical} reports numerical experiments that validate our theoretical results and highlight the advantages of the proposed method.

\section{Multi-environment Causal Invariance Learning and NegDRO}
\label{sec: causal inv}

Recall that we consider data collected from multiple environments, denoted by $\Ec=\{1,\cdots, |\Ec|\}$, with $|\Ec|$ denoting the number of environments.
For the $e$-th environment, the i.i.d. data $\{\x{e}_i, \y{e}_i\}_{i=1}^{n_e}$ are drawn from the joint distribution of $(\X{e}, \Y{e})$, where $\X{e}\in\Rb^p$ and $\Y{e}\in \Rb$ denote the covariates and the outcome, respectively. We assume that the causal relationship from the covariates $\X{e}$ to the outcome $\Y{e}$ is invariant across all environments. In the current paper, we focus on the linear causal relationship parameterized by $\beta^*$ and use $S^*=\{j\in [p]: \beta^*_j\neq 0\}$ to denote the support of $\beta^*$, that is,  %
\begin{equation}
    \Y{e} = (\beta^*_{S^*})^\intercal \X{e}_{S^*} + \varepsilon_Y^{e},\quad \textrm{for all $e\in \Ec$},
    \label{eq: invariant causal model}
\end{equation}
where $\beta^*_{S^*}$ denotes the sub-vector of $\beta^*$ with indexes belonging to $S^*$, $\X{e}_{S^*}$ denotes the direct causes of $\Y{e}$, and the noise term $\varepsilon_Y^{e}$ encodes other (unobserved) factors that affect the outcome. It is clear that $\beta^*_{(S^*)^\complement}=0$. Our goal is to learn such a causal relationship $\beta^*$, referred to as the \textbf{causal outcome model}, using samples of $\{(\X{e}, \Y{e})\}_{e\in\Ec}$. 

A central challenge in identifying $\beta^*$ is that we only observe a full set of covariates $\X{e}$ but do not know the exact causes $\X{e}_{S^*}$ of the outcome $\Y{e}$. Other covariates $\X{e}_j$ for $j\notin S^*$ can be correlated with $\Y{e}$; for instance, because they are consequences of $\Y{e}$, making the standard regression of $\Y{e}$ on $\X{e}$ fail to recover $S^*$ and $\beta^*$. To address this challenge, we leverage the structure in \eqref{eq: invariant causal model}: while the joint distribution of $(\X{e},\Y{e})$ may vary across environments $e \in \Ec$, the causal outcome model $\beta^*$ remains invariant. Therefore, we aim to find an invariant relationship from $\X{e}$ to $\Y{e}$ in the face of changing environments and, under suitable conditions, establish that it coincides with the causal outcome model $\beta^*$.

\begin{figure}[ht!]
    \centering
    \includegraphics[width=0.85\linewidth]{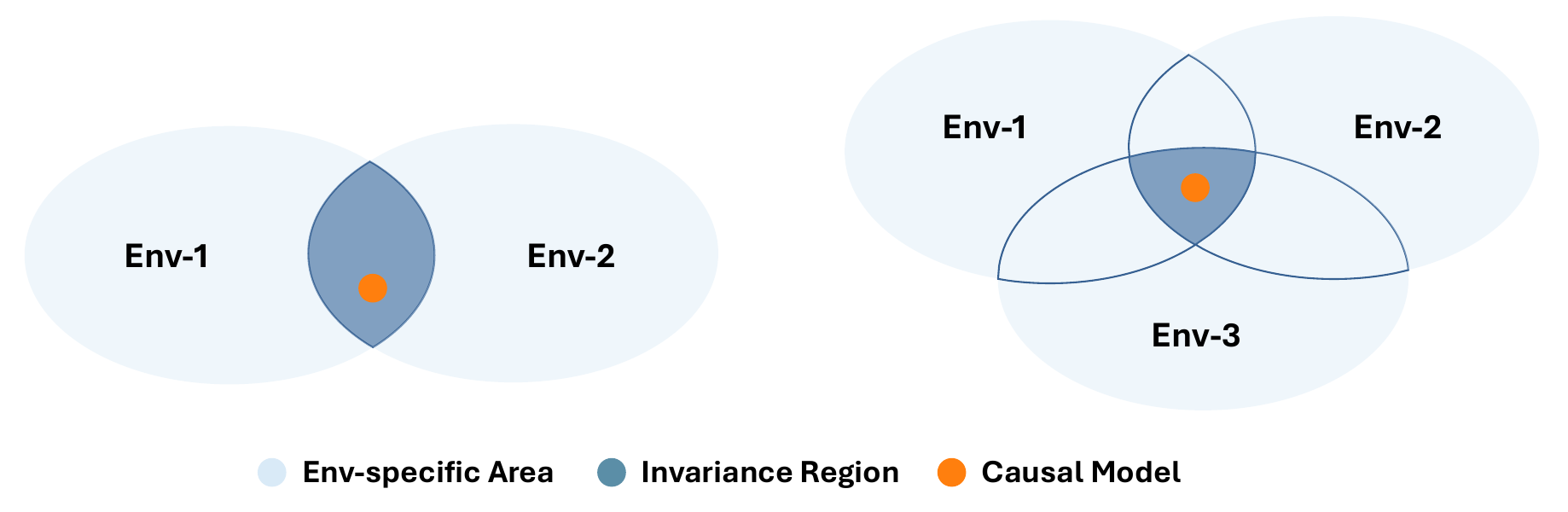}
    \caption{Illustration of Invariance and Causality.}
    \label{fig:illu-envs_for_causal}
\end{figure}

Figure \ref{fig:illu-envs_for_causal} illustrates how heterogeneous environments enable the identification of $\beta^*$ by leveraging invariance among differences. The left panel depicts the case of two environments ($|\Ec| = 2$), where each ellipse represents the set of prediction models that could potentially generate the observed data in its own environment.  The overlap forms the invariance region, consisting of prediction models that could potentially generate both environments simultaneously. The causal outcome model $\beta^*$ (highlighted in orange) lies in this invariance region, as specified in \eqref{eq: invariant causal model}. The right panel illustrates the case when a third environment is involved and the invariance region becomes smaller.  Intuitively, with sufficiently many heterogeneous environments or two sufficiently different environments, the {invariance region} progressively narrows down to precisely the {causal outcome model} under certain conditions, which we will specify in  Section \ref{subsec: ident conditions}. Note that with only a single environment and thus no heterogeneity, it is generally impossible to identify $\beta^*$ without prior knowledge of the causal structure  between covariates and outcome.

To materialize the intuition of identifying $\beta^*$, 
Section \ref{subsec: invariant causal model} formalizes the {invariance region} of Figure \ref{fig:illu-envs_for_causal}, and clarifies how our formulation differs from existing approaches in the literature. Section \ref{subsec: illu} then uses an illustrative example to show how prior methods recover $\beta^*$, which suffer from a computational cost that grows exponentially with dimension due to their combinatorial nature.  To overcome these computational challenges, we introduce a nonconvex continuous NegDRO optimization problem in Section \ref{subsec: negdro formulation}, aiming to enable the efficient identification of $\beta^*$.

\subsection{Risk-invariance Principle}
\label{subsec: invariant causal model}

To define the invariance region, which  by construction must include the causal outcome model $\beta^*$ (see Figure \ref{fig:illu-envs_for_causal}), we first need to identify the properties of the causal relationship \eqref{eq: invariant causal model} that remain invariant across different environments. Such properties are referred to as the Invariance Principle. Specifying a proper principle directly determines the invariance region of the prediction models.

The pioneering causal invariance learning work \citep{peters2016causal} adopts a strong requirement of invariance principle,
requiring that the noise $\eps{e}_Y$ in \eqref{eq: invariant causal model} is independent of $X^e_{S^*}$ and has an identical distribution for all $e\in \Ec$.
In contrast, we impose a weaker condition on the invariance principle: instead of requiring identical noise distributions, we assume only that the noise level remains constant across environments, i.e.,
\begin{equation}
    \textrm{The model \eqref{eq: invariant causal model} holds for all $e\in \Ec$ with $\Eb[\eps{e}_Y]^2=\sigma_Y^2$ for some constant $\sigma_Y^2>0$}.
    \label{eq: causal outcome model identical noise level}
\end{equation}
We denote \eqref{eq: causal outcome model identical noise level} as the \textbf{risk-invariance principle}, since it ensures that the causal outcome model $\beta^*$ yields the same prediction risk across environments, i.e., $\Eb[\Y{e}-(\beta^*)^\intercal\X{e}]^2 = \sigma_Y^2$ for all $e\in \Ec$. Motivated by this, we define the invariant prediction model set, corresponding to the invariance region in Figure \ref{fig:illu-envs_for_causal}, as:
\begin{equation}
    \Bc_{\rm inv} := \left\{b\in \Rb^p: \;\; \Eb[\Y{e} - b^\intercal\X{e}]^2 = \Eb[\Y{f} - b^\intercal\X{f}]^2, \quad \forall e,f\in \Ec\right\}.
    \label{assump: invariance}
\end{equation}
A prediction model $b\in \Rb^p$ is \textbf{risk-invariant} if it belongs to $\Bc_{\rm inv}$. Clearly, given \eqref{eq: causal outcome model identical noise level}, $\beta^*\in \Bc_{\rm inv}$ is risk-invariant.

Note that there are other invariance principles.
For example, another seminal line of work~
\citep{fan2023environment, yin2024optimization} adopts the conditional-mean invariance principle:
\begin{equation}
    \textrm{The model \eqref{eq: invariant causal model} holds for all $e\in \Ec$ with $\mathbb{E}[Y^{e} |X^{e}_{S^*}]=(\beta^*)^\intercal X^{e}$}.
    \label{eq: causal outcome model identical conditional mean}
\end{equation}
This principle leads to the following invariance region:
\begin{equation}
\left\{b\in \Rb^p: \;\; \Eb[\Y{e}|\X{e}_{{\rm supp}(b)}=x] = \Eb[\Y{f}|\X{f}_{{\rm supp}(b)}=x] = b_{{\rm supp}(b)}^\intercal x, \quad \forall e,f\in \Ec\right\},
    \label{eq: conditional-mean invariance region}
\end{equation}
where ${\rm supp}(b) = \{j\in [p]: b\neq 0\}$ denotes the support of $b\in \Rb^p$. This formulation directly links the causal outcome model $\beta^*$ to the conditional expectations. %
However, such an invariance principle requires $\Eb[\eps{e}_Y|\X{e}_{S^*}]=0$ for all $e\in \Ec$,
which do not allow the presence of hidden confounding between $\X{e}_{S^*}$ and $\Y{e}$.
For instance, consider an instrumental variable regression example with unknown causal structure and possible hidden confounding~\citep{angrist1996identification, imbens2015causal}. %
Our invariance principle holds yet the conditional-mean invariance principle faces limitations. We emphasize that both of these two invariance principles generalize from the invariance principle proposed in \citet{peters2016causal} and serve as complements to one another. A key difference lies whether the invariance principle requires conditioning on $X_S$ or not. Such conditioning usually introduces combinatorial structures into the optimization developed based on it as we will see in the next subsection. %

\subsection{Existing Causal Invariance Learning: Main Idea and Computational Challenge}
\label{subsec: illu}

Given the risk-invariance principle \eqref{eq: causal outcome model identical noise level}, we use a concrete example to illustrate how existing causal invariance learning methods identify $\beta^*$. Due to the inherent combinatorial nature of these approaches, their computational costs grow exponentially with the dimensions of the covariates. For the purpose of demonstration, we focus on a simplified example without hidden confounding. 
\begin{Example}
\label{eg: exist}
    Consider two environments with the observed data $(\X{e}, \Y{e})$ that follows an unknown data-generating mechanism, 
    \begin{equation}
    \X{e}_1 = \eps{e}_1,\quad \Y{e} = \X{e}_1 + \varepsilon_Y^{e}, \quad \X{e}_2 = \Y{e}+\eps{e}_2, 
    \label{eq: illus-1}
    \end{equation}
where $(\eps{e}_1, \eps{e}_2)^\intercal \sim \Nc(0,~ \nu^{e} \mathbf{I}_2)$, $\varepsilon_Y^{e}\sim \Nc(0,1)$, and these noises are mutually independent.
Note that the covariate noise variance $\nu^e$ differs in the two environments, resulting in different joint distributions of $(\X{e}, \Y{e})$ for $e \in \{1, 2\}$. Despite the heterogeneity, the causal set $S^*=\{1\}$, the causal outcome model $\beta^* = (1,0)$, and the outcome noise level $\Eb[\eps{e}_Y]^2=1$ remain invariant across environments. Consequently, $\Eb[\Y{1} - (\beta^*)^\intercal \X{1}]^2 = \Eb[\Y{2} - (\beta^*)^\intercal \X{2}]^2$, satisfying the risk-invariance principle \eqref{eq: causal outcome model identical noise level}.
\end{Example}
Recall that the central challenge in identifying $\beta^*$ lies in that the true causal set $S^*$ is unknown. A natural strategy is to enumerate all possible subsets of covariates and check whether each subset yields a risk-invariant prediction model. If a subset passes this invariance check, it becomes a candidate for $S^*$. This idea is precisely adopted by the pioneering ICP framework \citep{peters2016causal}. The procedures are as follows: 
\begin{enumerate}
    \item Pool the data from the two environments into a single dataset $(X, Y)$. For clarity, we assume that both environments contribute the same proportion of data samples, so that each environment has equal weight in the pooled dataset.%
    \item Enumerate the subsets $S\in \{\emptyset, \{1\}, \{2\}, \{1,2\}\}$ with the hypothesis that $S$ is the true causal set.
    \item  Set $\bar{b}_{S^\complement} = 0$ and regress $Y$ on $X_S$ to obtain $\bar{b}_S$, thereby forming the prediction model $\bar{b}$ with the support $S$.
    \item Compute $\Eb[\Y{e} - \bar{b}^\intercal \X{e}]^2$ for $e=\{1,2\}$ to  check whether the prediction risks are equal.
\end{enumerate}
For demonstration, we carry out computations at the population level, with results summarized in Table \ref{tab: illus eg}.

\begin{table}[!ht]
\centering
\resizebox{0.7\textwidth}{!}{%
\renewcommand{\arraystretch}{1.6}%
\begin{tabular}{|c|c|c|c|c|}
\hline
$S$ & \(\emptyset\) & \(\textcolor{blue}{\{1\}}\) & \(\{2\}\) & \(\{1,2\}\) \\ \hline
\(\bar{b}\) & 
\((0, 0)^\intercal\) & 
\(\textcolor{blue}{(1, 0)^\intercal}\) & 
\(\left(0, \frac{\bar{\nu}+1}{2\bar{\nu} + 1}\right)^\intercal\) & 
\(\left(\frac{\bar{\nu}}{\bar{\nu}+1}, \frac{1}{\bar{\nu}+1}\right)^\intercal\) \\ \hline
\(\Eb[\Y{e} - \bar{b}^\intercal\X{e}]^2\) & 
\(\textcolor{red}{\nu^{e}} + 1\) & 
\(\textcolor{blue}{1}\) & 
\(\frac{\bar{\nu}^2(\textcolor{red}{\nu^{e}} + 1) + (\bar{\nu} +1)^2\textcolor{red}{\nu^{e}}}{(2\bar{\nu} + 1)^2}\) & 
\(\frac{\textcolor{red}{\nu^{e}}+\bar{\nu}^2}{(\bar{\nu} + 1)^2}\) \\ \hline
\textrm{Independence of $e$} & \textrm{No} & \textcolor{blue}{\textrm{Yes}} & \textrm{No} & \textrm{No} \\ \hline
\end{tabular}%
}
\caption{Best  population prediction model $\bar{b}$ for a given $S$; 
with $\bar{\nu} = (\nu^{1}+\nu^{2})/2$ denoting the average noise variance. The environment-dependent noise variance $\textcolor{red}{\nu^e}$ is highlighted in red.}
\label{tab: illus eg}
\end{table}

The subset $S = \{1\}$ produces a prediction model $\bar{b}=(1,0)^\intercal$, with an invariant prediction risk $\Eb[\Y{e}-\bar{b}\X{e}]^2=1$ across both environments. 
In contrast, all other subsets lead to a violation of the risk invariance, since their corresponding prediction risks involve the environment-specific noise variance $\nu^{e}$ that varies across $e\in \Ec$. Thus, they cannot be the causal set $S^*$. Hence, only $S=\{1\}$ qualifies as the causal set $S^*$, and the corresponding risk-invariant prediction model $\bar{b}$ identifies the causal outcome model $\beta^*=(1,0)^\intercal$. 
While this procedure recovers $\beta^*$, it fundamentally relies on exhaustively enumerating all subsets of covariates, which admits a computational complexity that grows exponentially with the dimension $p$.

In addition to the subset-enumeration strategy introduced above, the recent work \citet{fan2023environment} proposed a regularized least squares formulation for estimating $\beta^*$ based on the conditional-mean invariance principle. Although framed differently, their method still relies on subset enumeration, mainly due to their invariance principle requires conditioning on $X_S$. Specifically, given a regularization parameter $\gamma>0$, they consider the following problem:
\begin{equation*}
    \min_{b\in \Rb^p, S\subseteq [p]} ~~\sum_{e\in \Ec}\Eb[\Y{e} - b^\intercal\X{e}]^2 + \gamma \sum_{e\in \Ec}\sum_{j\in [p]} 1\{b_j=0\}\left|\Eb[(\Y{e} - b^\intercal\X{e})\X{e}_j]\right|^2,\quad {\rm s.t.}\quad  S = {\rm supp}(b).
\end{equation*}
where ${\rm supp}(b)=\{j\in [p]: b_j=0\}$ denotes the support of $b\in \Rb^p$, and the decision variable $S$ indicates which covariates are hypothesized to be causal. The second term in its objective acts as a regularizer that encourages solutions to fall inside the conditional-mean invariance region defined in \eqref{eq: conditional-mean invariance region}; see \citet{fan2023environment} for details. Due to the nonconvexity, solving this problem to global optimality still requires enumerating all subsets $S\subseteq [p]$. \footnote{To avoid high computational cost in the subset enumeration, they also propose to solve a surrogate objective using the Gumbel trick to approximate the indicator function, i.e., building a softmax relaxation. However, it sacrifices theoretical guarantees about identification.}

Both existing approaches rely on enumerating all subsets of covariates, which essentially treat the causal invariance learning as a \emph{mixed integer programming} problem, resulting in a computational complexity that scales exponentially with the dimension $p$. This raises a question: can we instead formulate the causal invariance learning as a \emph{continuous optimization} problem, and solve it efficiently? We provide an affirmative answer in the remaining sections.

\subsection{Nonconvex Continuous Formulation: NegDRO}
\label{subsec: negdro formulation}
Motivated by the risk-invariance principle \eqref{eq: causal outcome model identical noise level}, we frame causal invariance learning as a constrained optimization problem:
\begin{equation}
\min_{b\in \Rb^p}\Eb[\Y{1}- b^\intercal\X{1}]^2 \quad \textrm{s.t.}~~ \Eb[\Y{e}-b^\intercal \X{e}]^2= \Eb[\Y{f}-b^\intercal \X{f}]^2 ~~\forall e,f\in \Ec.
    \label{eq: bneg gamma infty}
\end{equation}
Here, the constraint enforces risk invariance across environments, while the objective ensures that the solution achieves a small prediction risk. Note that the constraint is nonconvex. Since the constraint holds for all pairs of environments, the objective can equivalently be taken from any environment; we take $e=1$ for simplicity. Note that the feasible set of \eqref{eq: bneg gamma infty} coincides with the invariant prediction model set $\Bc_{\rm inv}$ in \eqref{assump: invariance}. Since $\beta^* \in \Bc_{\rm inv}$, $\beta^*$ is always a feasible solution to \eqref{eq: bneg gamma infty}. Intuitively, if $\Bc_{\rm inv}$ collapses to a singleton, then the optimal solution of \eqref{eq: bneg gamma infty} must recover $\beta^*$ exactly. In Section \ref{subsec: ident conditions}, we establish the sufficient and nearly necessary condition for $\Bc_{\rm inv}$ to be a singleton. In Section \ref{sec: minimization}, we further provide the weaker condition showing that, even when $\Bc_{\rm inv}$ is not a singleton, the constrained problem \eqref{eq: bneg gamma infty} can still identify $\beta^*$. %

Notice that, instead of enforcing risk equality for every pair of environments as in \eqref{eq: bneg gamma infty}, it is equivalent to require that the maximum risk matches the average risk across environments. This leads to the following equivalent formulation of \eqref{eq: bneg gamma infty} with a single constraint:
\[
\min_{b\in \Rb^p}\max_{e\in \Ec} \Eb[\Y{e}-b^\intercal\X{e}]^2, \quad\textrm{s.t.}~~\max_{e\in \Ec}\Eb[\Y{e}-b^\intercal \X{e}]^2 = \frac{1}{|\Ec|}\sum_{f\in \Ec}\Eb[\Y{f}-b^\intercal\X{f}]^2.
\]
As all the environments share the same risk, we replace the objective function in \eqref{eq: bneg gamma infty} with the maximum risk. 

Although this reformulation simplifies the original constraint \eqref{eq: bneg gamma infty}, the strict equality still defines a nonconvex feasible set, which causes methods such as projected gradient descent to fail to implement, since projections onto a nonconvex set are not well defined. To address this, we penalize the equality constraint with a regularization parameter $\gamma\geq 0$, yielding
\begin{equation}
    \min_{b\in \Rb^p}\left\{\max_{e\in \Ec} \Eb[\Y{e}-b^\intercal \X{e}]^2 + \gamma |\Ec| \left(\max_{e\in \Ec}\Eb[\Y{e}-b^\intercal \X{e}]^2 - \frac{1}{|\Ec|}\sum_{f\in \Ec}\Eb[\Y{f}-b^\intercal \X{f}]^2 \right) \right\}.
    \label{eq: obj equiv}
\end{equation}
The regularization parameter $\gamma$ balances between (i) the maximum prediction risk among environments and (ii) the discrepancy between the maximum and average risks.
As $\gamma$ increases, the optimal solution of \eqref{eq: obj equiv} strives for risk parity across environments, reducing the discrepancy among environments, while minimizing the maximum risk meanwhile. 
In the limit $\gamma = \infty$, the above optimization problem recovers \eqref{eq: bneg gamma infty}.

Interestingly, \eqref{eq: obj equiv} admits an equivalent minimax form similar to Group DRO \citep{sagawa2019distributionally, hashimoto2018fairness}:
\begin{equation}
\begin{aligned}
     b_{\rm Neg}^\gamma \in 
    \argmin_{b\in \Rb^p}&\max_{{w} \in \Uc(\gamma)} \sum_{e\in \Ec} {w}_e \Eb[\Y{e}-b^\intercal\X{e}]^2,\\
    {\rm s.t.}\;\; &\Uc(\gamma) := \left\{{w}\in \Rb^{|\Ec|}: \sum_{e\in \Ec} {w}_e = 1, \min_{e\in \Ec}{w}_e\geq -\gamma\right\}.
\end{aligned}
    \label{eq: obj original}
\end{equation}
The equivalence between \eqref{eq: obj original} and \eqref{eq: obj equiv} follows from the observation that the maximization over the linear weight $w$ within $\Uc(\gamma)$ in \eqref{eq: obj original} is equivalent to assigning the weight $1+\gamma(|\Ec|-1)$ to the worst-case environment with the maximum prediction risk, and assigning the weights $-\gamma$ to all other environments.

When $\gamma=0$, the uncertainty set $\Uc(\gamma)$ becomes the simplex $\Delta^{|\Ec|}$, and then \eqref{eq: obj original} reduces to the classical Group DRO \citep{sagawa2019distributionally}, which guarantees prediction robustness in the worst environment. 
However, when $\gamma>0$, it differs fundamentally from Group DRO, as some environments will carry negative weights. These negative weights are the key to enforcing risk-invariance among environments.%
From an optimization perspective, these negative weights introduce difference-of-convex objectives, leading to a nonconvex optimization. For this reason, we term the proposed minimax optimization in \eqref{eq: obj original} as Negative weighted DRO (NegDRO). This also provides a new optimization paradigm.

NegDRO constitutes a nonconvex-linear minimax optimization. One could implement all sorts of gradient-based methods to solve it; yet, its global solution is generally challenging to obtain due to nonconvexity. Remarkably, under the causal identification conditions specified in Section \ref{sec: additive intervention}, we show that NegDRO exhibits a benign nonconvex landscape such that all stationary points of NegDRO are close to the causal outcome model $\beta^*$. This structural property allows us to design a gradient-based algorithm that converges efficiently to the causal outcome model $\beta^*$, as developed in Section \ref{subsec: computation algorithm}.

\section{Additive Intervention and Causal Identification Conditions}
\label{sec: additive intervention}
In the previous section, we introduced the constrained problem \eqref{eq: bneg gamma infty} and its relaxation, the NegDRO formulation \eqref{eq: obj original}. This section addresses two follow-up questions: (i) what conditions guarantee that the constrained problem \eqref{eq: bneg gamma infty} identifies $\beta^*$ exactly, i.e., solving the constrained problem ensures causal identification and (ii) how close is the optimal solution of NegDRO, $b_{\rm Neg}^\gamma$, to $\beta^*$ for a finite regularization parameter $\gamma\geq 0$. 
To answer these questions, we focus on the additive intervention regime, a commonly used multi-environment data generation mechanism in causal invariance learning.

\subsection{Additive Intervention}
\label{subsec: additive intervention regime}

We focus on the \emph{additive intervention regime}, a structured framework generating heterogeneous multi-source data that has been widely adopted in the literature 
\citep{peters2014causal, ghassami2017learning, rothenhausler2019causal, rothenhausler2021anchor, shen2023causality, taeb2024learning}. The key characteristic of this regime is that the causal relationships among variables remain invariant across environments, while the heterogeneity arises from additive, environment-specific interventions to the noise terms of the covariates $X$, which may induce both mean shifts and covariance shifts on $X$. %

Specifically, for each environment $e\in \Ec$, the data-generating process of $(\X{e}, \Y{e})$ follows the structural equation models (SEMs) in Definition \ref{def: SEM}, as given by:
\begin{equation}
\begin{pmatrix}\Y{e} \\ \X{e}\end{pmatrix}= \Bbf \begin{pmatrix}\Y{e} \\ \X{e}\end{pmatrix}+\begin{pmatrix}
    \varepsilon_Y^{e} \\ \varepsilon_X^{e}
\end{pmatrix},\quad \textrm{with}~~\Bbf = \begin{pmatrix}
    0 & (\beta^*)^\intercal \\
    \Bbf_{YX} & \Bbf_{XX}
\end{pmatrix},
\label{eq: SCM}
\end{equation}
where $\Bbf\in \Rb^{(p+1)\times (p+1)}$ encodes the causal relationships among $(X,Y)$, and $(\eps{e}_Y, \eps{e}_X)\in \Rb^{p+1}$ denotes the environment-specific noises.
The above SEMs indicate that (i) a subset of covariates in $X$ causally affects the outcome $Y$ through $\beta^*$, as assumed in \eqref{eq: invariant causal model}; (ii) the outcome $Y$ may causally affect some other covariates in $X$ through $\Bbf_{YX}\in \Rb^p$; (iii) covariates in $X$ may affect each other internally via $\Bbf_{XX}\in \Rb^{p\times p}$. To further elucidate each component of $\Bbf$, we provide several concrete examples in Appendix \ref{appendix: SEMs}.

We consider the SEMs \eqref{eq: SCM} that generate an acyclic causal graph, meaning that there is no direct path from one variable to itself. This guarantees that the matrix $\Ibf - \Bbf$ is invertible \citep{spirtes2001causation, pearl2009causality}
, allowing us to rewrite the SEMs \eqref{eq: SCM} equivalently as:
\begin{equation}
    \begin{pmatrix}
        \Y{e}\\ \X{e}
    \end{pmatrix} = (\Ibf - \Bbf)^{-1}\begin{pmatrix}
        \varepsilon_Y^{e} \\ \varepsilon_X^{e}
    \end{pmatrix}.
    \label{eq: SCM invert}
\end{equation}

In addition to the additive noise model in \eqref{eq: SCM invert}, the additive intervention regime further requires 
that the environment-specific noises $(\varepsilon_Y^{e},\varepsilon_X^{e})$ admit the following decomposition \citep{rothenhausler2019causal}, 
\begin{equation}
    \begin{pmatrix}
    \varepsilon_Y^{e} \\ \varepsilon_X^{e}
    \end{pmatrix} \stackrel{d}{=} \begin{pmatrix}
        \eta_Y\\ \eta_X
    \end{pmatrix} + \begin{pmatrix}
        0 \\ \delta^{e}
    \end{pmatrix} \quad \text{with} \quad \Eb[\eta (\del{e})^\intercal] = 0,
    \label{eq: intervention noise}
\end{equation}
where $\eta = (\eta_Y, \eta_X)^\intercal$, with $\eta_Y\in \Rb$ and $\eta_X\in \Rb^p$, denotes the systematic random noise shared across all environments, while $\delta^{e}\in \Rb^p$ stands for the environment-specific interventions applied only to the environment $e$. The notation $\stackrel{d}{=}$ indicates that two random vectors share the same distribution.

This noise decomposition in \eqref{eq: intervention noise} highlights three key properties. First, the outcome noise $\eps{e}_Y\stackrel{d}{=} \eta_Y$ is never  intervened, ensuring that the risk-invariance principle holds for $\beta^*$, as in \eqref{eq: causal outcome model identical noise level}, with a constant noise level $\Eb[\varepsilon_Y^{e}]^2=\Eb[\eta_Y^2]$ across all environments. Second, $\Eb[\eta(\del{e})^\intercal]=0$ ensures that the systematic noise $\eta$ and the environment-specific interventions $\del{e}$ are uncorrelated.
Third, the systematic random noise $\eta$ may capture hidden confounders that systematically affect both covariates $X$ and the outcome $Y$ with $\Eb[\eta_Y\eta_X]\neq 0$.
Consequently, the distributional heterogeneity across environments arises solely from the additive interventions $\delta^{e}$, which justifies the name. %

In practice, when there are only interventions on the covariate noise as in \eqref{eq: intervention noise}, while the underlying causal mechanisms remain intact as in \eqref{eq: SCM invert}, the setting is the additive intervention regime.
Returning to the ads-and-sales Example \ref{eg: concrete}, this corresponds to scenarios where the causal relationships between ad engagements $X=(X_1,X_2)$ and sales $Y$ remain invariant across environments, while the data heterogeneity is driven by environment-specific interventions $\delta^{e}$. Such interventions may arise, for instance, from differences in measurement tools or ad-tracking systems across distinct markets or time periods.

\subsection{Causal Identification Condition}
\label{subsec: ident conditions}
Since the causal outcome model $\beta^*$ is always feasible for \eqref{eq: bneg gamma infty}, intuitively, if the constraint set $\Bc_{\rm inv}$ in \eqref{assump: invariance} collapses to a singleton, then the optimal solution of~\eqref{eq: bneg gamma infty} recovers $\beta^*$. Following this intuition, we propose two equivalent conditions that ensure the singleton of $\Bc_{\rm inv}$. 
We emphasize that these conditions serve as \emph{sufficient} guarantees for causal identification. In later Section \ref{sec: minimization}, we establish weaker causal identification conditions when $\Bc_{\rm inv}$ is not a singleton. 

The proposed two conditions come from two complementary perspectives. 
\begin{itemize}
    \item From a \emph{statistical viewpoint}, collecting increasingly heterogeneous environments gradually shrinks the invariance region, which eventually collapses to the causal outcome model $\beta^*$; see Figure \ref{fig:illu-envs_for_causal}.  This motivates us to characterize the condition over the data generation process under which these environments contain enough heterogeneity for $\Bc_{\rm inv}$ \eqref{assump: invariance} collapses to $\beta^*$ exactly. This condition provides guidance on how multi-environment data should be collected.
    \item From an \emph{optimization viewpoint}, the invariant prediction model set $\Bc_{\rm inv}$ admits a singleton, which means that 
    \begin{equation}
    \min_{b\in \Rb^p}\left\{\max_{e\in \Ec}\Eb[\Y{e}-b^\intercal \X{e}]^2 - \frac{1}{|\Ec|}\sum_{f\in \Ec}\Eb[\Y{f}-b^\intercal \X{f}]^2\right\}
    \label{lambda_infinite}
    \end{equation}
    also admits a unique solution. This motivates us to characterize the condition directly in the optimization formulation. Such a condition will greatly help us characterize the landscape of the nonconvex NegDRO problem.
\end{itemize}

We begin with the statistical perspective.
\stepcounter{Condition}
\begin{intCondition}
\label{cond: strict positive}
    There exist two nonempty and disjoint collections of environments $\Ec_1,\Ec_2\subseteq \Ec$, and some weights $w\in \Delta^{|\Ec_1|}$ and $w'\in \Delta^{|\Ec_2|}$ such that 
\begin{equation}
\label{eq: cond - weak - equiv}
    \sum_{e \in \Ec_1} w_e \Eb[X^{e} X^{e\intercal}] \succ \sum_{f \in \Ec_2} w'_f \Eb[X^{f} X^{f\intercal}].
\end{equation}
\end{intCondition}
It states that there exist two collections of environments where the weighted average of the second-moment matrices from one collection, $\Ec_1$, strictly dominates that of another collection, $\Ec_2$. Importantly, it only requires the existence of such collections and weights, without assuming they are known in advance, which contrasts with DRIG~\citep{shen2023causality} that requires access to a reference dominated environment. 
The following theorem confirms that under this condition, the invariant prediction model set $\Bc_{\rm inv}$ in \eqref{assump: invariance} collapses to $\beta^*$.
\begin{Theorem}
Under the additive intervention regime, suppose that Condition \ref{cond: strict positive} holds. Then the causal outcome model $\beta^*$ is the unique risk-invariant prediction model such that $\Bc_{\rm inv} = \{\beta^*\}$.
\label{thm: identification infty}
\end{Theorem}

In contrast to the abstract conditions in the works \citep{fan2023environment, yin2024optimization}, Condition \ref{cond: strict positive} is formulated explicitly in terms of the covariate second-moment matrices. Moreover, unlike the works \citet{rojas2018invariant} and \citet{arjovsky2019invariant} that require a prohibitively large number of environments, our Condition \ref{cond: strict positive} holds with as few as two environments, for instance, if they satisfy $\Eb[X^{1} X^{1\intercal}] \succ \Eb[X^{2} X^{2\intercal}]$.

To further interpret Condition \ref{cond: strict positive}, recall that under the additive intervention regime, environmental heterogeneity arises from environment-specific interventions $\delta^{e}$ for $e\in \Ec$. Specifically, one can show that the second moment matrix of covariates admits the expression $\Eb[\X{e}X^{e\intercal}] = \Gbf (\Hbf + \Eb[\del{e}\delT{e}] )\Gbf$, where $\Hbf,\Gbf\in \Rb^{p\times p}$ are full-rank matrices; see Appendix Lemma \ref{lemma: expression of risk} for their explicit expressions. Substituting this into \eqref{eq: cond - weak - equiv}, Condition \ref{cond: strict positive} is equivalent to:
\[
\sum_{e \in \Ec_1} w_e \Eb[\delta^{e} \delta^{e\intercal}] \succ \sum_{f \in \Ec_2} w'_f \Eb[\delta^{f} \delta^{f\intercal}].
\]
It provides a concrete way to construct heterogeneous environments in experimental settings.
For example, one may start with a baseline environment $f$ without any interventions such that $\Eb[\del{f}\delT{f}]=0$, and then apply interventions across covariates to obtain another environment $e$ with $\Eb[\del{e}\delT{e}]\succ 0$. In this scenario, the inequality $\Eb[\delta^e\delta^{e\intercal}]\succ \Eb[\delta^f\delta^{f\intercal}]$ directly satisfies Condition \ref{cond: strict positive}. The equivalent form of Condition \ref{cond: strict positive} from the intervention perspective further provides concrete and practical guidance on collecting data, i.e., how to add intervention to build datasets satisfying environmental heterogeneity for causal identification. %

In fact, Condition \ref{cond: strict positive} is not only sufficient, but also nearly necessary to ensure $\Bc_{\rm inv} = \{\beta^*\}$. %
\begin{Theorem}
    Under the additive intervention regime, suppose that only one coordinate of the noise is intervened such that $\left|\{j\in [p]: \delta^{e}_j\neq 0\}\right| =1$ for all $e\in \Ec$. If $\Bc_{\rm inv} = \{\beta^*\}$, that is, $\beta^*$ is the unique risk-invariant prediction model, then Condition \ref{cond: strict positive} must hold.
    \label{thm: necessary condition}
\end{Theorem}

To establish necessity, this theorem focuses on a specific pattern of additive interventions, where each environment can intervene on only one coordinate of the covariate noise.
Returning to the ads-and-sales Example \ref{eg: concrete}, this corresponds to settings where each environment changes only the tracking tool for a single ad channel, either search ads or social media ads, while keeping the other unchanged.  
Within this context, Theorem \ref{thm: identification infty} and \ref{thm: necessary condition} together suggest that Condition \ref{cond: strict positive - A} is both sufficient and necessary for ensuring that the invariant prediction model set $\Bc_{\rm inv}$ collapses to $\beta^*$. 

Next, we derive the condition from an optimization perspective.  
If the feasible set  of the constrained problem \eqref{eq: bneg gamma infty} admits a unique feasible point, it has to be the desired $\beta^*$, meaning that solving the constrained optimization problem is sufficient for causal identification. Thus it remains to understand when the feasible set becomes singleton. Notice that the feasible set admits the following forms:
\begin{align*}
& \left\{b\in \Rb^p: \quad \max_{e\in \Ec}\Eb[\Y{e}-b^\intercal \X{e}]^2 = \frac{1}{|\Ec|}\sum_{f\in \Ec}\Eb[\Y{f}-b^\intercal\X{f}]^2\right\}\\
= &
\left\{b\in \Rb^p:\quad \sum_{e\in \Ec}\left(w_e - \frac{1}{|\Ec|}\right) \Eb[\Y{e}-b^\intercal\X{e}]^2 = 0, \;\;\forall w\in \Delta^{|\Ec|}\right\}.    
\end{align*}
Under the additive intervention regime, after decomposing the environmental risk $\Eb[\Y{e}-b^\intercal\X{e}]^2$ as shown in the following \eqref{eq: main risk expression}, one could show that this feasible set admits the following equivalent representation:
\begin{equation}
    \left\{b\in \Rb^p:\quad (b-\beta^*)^\intercal \left[\sum_{e\in \Ec}\left(w_e - \frac{1}{|\Ec|}\right) \Eb[X^{e}X^{e\intercal}]\right] (b-\beta^*) = 0, \;\;\forall w\in \Delta^{|\Ec|}\right\}.
    \label{eq: feasibility set}
\end{equation}
Denote $\Abm(w)\in \Rb^{p\times p}$ for any $w\in \Delta^{|\Ec|}$ as:
    \begin{equation}
    \Abm(w) := \sum_{e\in \Ec} \left(w_e - \frac{1}{|\Ec|}\right)\Eb[X^{e}X^{e\intercal}].
    \label{eq: A inf def}
\end{equation}
Clearly, the feasible set \eqref{eq: feasibility set} collapses to the singleton of $\beta^*$ if there exists some weight vector $w^0\in \Delta^{|\Ec|}$ such that $\Abm(w^0)\succ 0$. Note that the matrix $\Abm(w)$ quantifies the discrepancy between the $w$-weighted average of second-moment matrices across all environments and the simple average over all environments.
\begin{intCondition}
    There exists some $w^0\in \Delta^{|\Ec|}$ such that $\lambda = \lambda_{\rm min}(\Abm(w^0)) > 0$.
    \label{cond: strict positive - A}
\end{intCondition}
This condition requires that some convex combination of second-moment matrices strictly dominates the simple average of second-moment matrices, such that $\sum_{e\in \Ec}w^0_e\Eb[X^{e}X^{e\intercal}]\succ \sum_{e\in \Ec}\frac{1}{|\Ec|}\Eb[X^{f}X^{f\intercal}]$ for some $w^0\in \Delta^{|\Ec|}.$ 
Furthermore, $\lambda$ reflects the degree of environmental heterogeneity, in the sense that a larger $\lambda$ indicates a greater discrepancy between the $w^0$-weighted average and the simple average, signaling more pronounced heterogeneity.
The following theorem shows that given Condition \ref{cond: strict positive - A}, the constrained problem \eqref{eq: bneg gamma infty} achieves causal identification. 
\begin{Theorem}
    Under the additive intervention regime, suppose that Condition \ref{cond: strict positive - A} holds. Then the constrained problem \eqref{eq: bneg gamma infty} identifies $\beta^*$ exactly, as its feasible set reduces to $\{\beta^*\}$.
    \label{thm: optimization perspective}
\end{Theorem}

The following proposition confirms that the conditions motivated by both the statistical and the optimization perspectives are equivalent. 
\begin{Proposition}
    Conditions \ref{cond: strict positive} and \ref{cond: strict positive - A} are equivalent. That is, for any $\lambda>0$, Condition \ref{cond: strict positive - A} implies Condition \ref{cond: strict positive}, and conversely, Condition \ref{cond: strict positive} implies the existence of a $\lambda>0$ such that Condition \ref{cond: strict positive - A} holds.
\label{prop: equiv ass}
\end{Proposition} 
In the following discussions, we focus on Condition \ref{cond: strict positive - A}, as it provides a concise characterization of environmental heterogeneity via the value $\lambda$, facilitating the landscape analysis of NegDRO \eqref{eq: obj original}.

\subsection{Causality Discovery with Finite $\gamma$}
\label{subsec: finite gamma bneg}
Thus far, Condition \ref{cond: strict positive - A} ensures that solving the constrained problem \eqref{eq: bneg gamma infty} exactly recovers the causal outcome model $\beta^*$. However, it is hard to design an algorithm to directly solve the constrained problem due to its nonconvex constraints; thus, we turn to the proposed NegDRO formulation \eqref{eq: obj original}, which relaxes the strict equality constraint from the constrained problem \eqref{eq: bneg gamma infty}. Note that when $\lambda =\infty$, NegDRO is equivalent to the constrained problem. This subsection quantifies how close the optimal solution of the NegDRO problem $b_{\rm Neg}^\gamma$ in \eqref{eq: obj original} is to $\beta^*$ for a finite regularization parameter $\gamma\geq 0$.

To facilitate the discussion, we perform a reparameterization of the weights in the NegDRO formulation \eqref{eq: obj original}. 
\[
b_{\rm Neg}^\gamma \in \argmin_{b\in \Rb^p}\max_{{\Tilde{w}}\in \Delta^{|\Ec|}}\sum_{e\in \Ec}\left[(1+\gamma|\Ec|){\Tilde{w}}_e - \gamma\right]\cdot \Eb[\Y{e}-b^\intercal\X{e}]^2.
\]
where $\Tilde{w}_e = \frac{w_e+\gamma}{1+\gamma |\Ec|}$ for each $e\in \Ec$ and $w\in \Uc(\gamma)$. It is clear that the new weights $\Tilde{w}$ belong to a simplex. For the ease of notation, we denote the new weight $\tilde w_e$ as $w$. 
Further $1+\gamma|\Ec|$, we obtain
\begin{equation}
    b_{\rm Neg}^\gamma \in \argmin_{b\in \Rb^p}\Phi(b),\quad \textrm{where}\quad \Phi(b) = \max_{w\in \Delta^{|\Ec|}} \sum_{e\in \Ec}\left(w_e - \frac{\gamma}{1+\gamma |\Ec|}\right)\Eb[\Y{e}-b^\intercal\X{e}]^2.
    \label{eq: obj trans - population}
\end{equation}
We now establish the dependence of $\|b_{\rm Neg}^\gamma-\beta^*\|_2$ on $\gamma$ in the following proposition.
\begin{Proposition}[Model Error of NegDRO]
    Under the additive intervention regime, suppose that Condition \ref{cond: strict positive - A} holds. Then any global optimal solution $b_{\rm Neg}^\gamma$ of NegDRO, defined in \eqref{eq: obj trans - population}, satisfies 
    \begin{equation}
    \begin{aligned}
    \|b_{\rm Neg}^\gamma - \beta^*\|_2\lesssim \frac{1}{\lambda(1+\gamma {|\Ec|})},
    \end{aligned}
    \label{eq: bneg upper bound}
    \end{equation}
    where $\lambda>0$ denotes the degree of environmental heterogeneity, as defined in Condition \ref{cond: strict positive - A}.
    \label{prop: bneg upper bound}
\end{Proposition}
This proposition shows that if we could solve NegDRO globally, any global solution $b_{\rm Neg}^\gamma$ is $\mathcal{O}((\gamma|\Ec|)^{-1})$-close to the causal outcome model $\beta^*$. Several insights follow from this upper bound. First, increasing the regularization parameter $\gamma$ drives the solution $b_{\rm Neg}^\gamma$ closer to $\beta^*$. In the extreme case where $\gamma = \infty$, $b_{\rm Neg}^\gamma$ exactly matches $\beta^*$, consistent with Theorem \ref{thm: optimization perspective}.
Second, greater environmental heterogeneity with larger $\lambda$ and more environments with larger $|\Ec|$ both facilitate the identification, yielding the solution closer to $\beta^*$.

Despite these theoretical guarantees, solving NegDRO \eqref{eq: obj trans - population} globally is still far from trivial, due to the nonconvexity of $\Phi(b)$ in $b$. Notably, the effective weights $w_e - \frac{\gamma}{1+\gamma |\Ec|}$ can be either positive or negative, making $\sum_{e\in \Ec}(w_e - \frac{\gamma}{1+\gamma|\Ec|})\Eb[\ell(\X{e},\Y{e}; b)]$ a difference-of-convex function in $b$.
Consequently, after maximizing over $w\in \Delta^{|\Ec|}$, the objective $\Phi(b)$ remains a nonconvex function in $b$. 
Classic optimization techniques, such as gradient descent and its variants (e.g., sub-gradient descent), may get trapped in stationary points, such as local minima or saddle points, due to the nonconvex landscape \citep{boyd2004convex, bonnans2006numerical, dauphin2014identifying}. 
\section{Computationally Efficient Causal Invariance Learning}
\label{sec: alg}

In this section, we address the nonconvexity of the NegDRO problem. In Subsection \ref{subsec: stationary causal}, we establish a benign landscape of NegDRO such that any stationary point of NegDRO is guaranteed to lie close to  $\beta^*$. This insight implies that causal identification can be achieved efficiently by seeking stationary points of NegDRO rather than global optimal solutions. Building on this, Subsection \ref{subsec: computation algorithm}  presents an efficient gradient-based algorithm, with the convergence rate established in terms of both the sample size and the number of iterations in the finite-sample setting.

\subsection{Benign Landscape and Causal Identification with Stationary Points}
\label{subsec: stationary causal}

Recall in Definition \ref{def: stationary point} that stationary points refer to points where the gradient of the objective function is equal to zero. 
As the inner maximization over the simplex could have multiple solutions, the NegDRO objective $\Phi(b)$, as defined in \eqref{eq: obj trans - population}, may not be differentiable everywhere.
To simplify the notation and focus on the core insights for handling nonconvexity, we add a ridge penalty parameter into the inner maximization. This penalization ensures that the inner maximization problem has a unique solution, thereby ensuring the differentiability of the penalized objective. By properly adjusting the penalty parameter, one can show that the convergence rate for the penalized problem is the same as that for the original (nonpenalized and nonsmooth) NegDRO. The treatment of the original NegDRO and its theoretical results, which involve technical challenges of nonsmoothness analysis, is provided in Appendix \ref{subsec: without penalty}.

Given a small penalty parameter $\mu > 0$, we define the penalized version of $\Phi(b)$ in \eqref{eq: obj trans - population} as follows:
\begin{equation}
    \Phi_\mu(b) := \max_{w\in \Delta^{|\Ec|}} \left\{\sum_{e\in \Ec}\left(w_e - \frac{\gamma}{1+\gamma|\Ec|}\right)\Eb[\Y{e}-b^\intercal\X{e}]^2 - \mu \|w\|_2^2\right\}.
    \label{eq: obj trans - penal - population}
\end{equation}
The penalty term $\|w\|_2^2$ ensures the strong concavity of the inner maximization problem and thus ensures a unique maximizer $\Bar{w}(b)$ for any $b\in \Rb^p$. We use $\Bar{w}$ to denote $\Bar{w}(b)$ when its meaning is clear.
\begin{equation}
    \Bar{w} \coloneq \argmax_{w\in \Delta^{|\Ec|}}\sum_{e\in \Ec} \left\{\left(w_e - \frac{\gamma}{1+\gamma|\Ec|}\right)\Eb[\Y{e}-b^\intercal\X{e}]^2 - \mu \|w\|_2^2\right\}.
    \label{eq: maximizer w - penal - population}
\end{equation}
By Danskin's theorem \citep{danskin1966theory}\citep[Proposition~B.25]{bertsekas1997nonlinear}, $\Phi_\mu(b)$ is differentiable everywhere, with the gradient defined as
\[
\nabla \Phi_\mu(b) = \sum_{e\in \Ec}\left(\bar{w}_e - \frac{\gamma}{1+\gamma|\Ec|}\right) \Eb[(b^\intercal\X{e}-\Y{e})\X{e}].
\]
We establish the benign nonconvex landscape of the penalized NegDRO via the following theorem.
\begin{Theorem}[Landscape of NegDRO]
\label{thm: convergence theorem - penal - population - any b}
    Under the additive intervention regime, suppose that Condition \ref{cond: strict positive - A} holds. The following inequality holds for any prediction model $b\in \Rb^p$:
    \begin{equation}
    \label{eq: upper bound - penal - population - any b}
        \|b - \beta^*\|_2\lesssim \frac{1}{\lambda}\left(\frac{1}{1+\gamma|\Ec|} + \left\|\nabla\Phi_{\mu}(b)\right\|_2\right) + \sqrt{\frac{\mu}{\lambda}},
    \end{equation}
    where $\lambda>0$ denotes the degree of environmental heterogeneity as defined in Condition \ref{cond: strict positive - A}.
\end{Theorem}
This theorem indicates that the distance of $\|b - \beta^*\|_2$, for any prediction model $b\in \Rb^p$, is controlled by the gradient norm $\|\nabla\Phi_\mu(b)\|_2$. Consequently, for sufficiently large regularization $\gamma \geq 0$ and small penalty $\mu > 0$, any stationary point of the penalized NegDRO with $\|\nabla \Phi_\mu(b)\|_2 = 0$ serves as a close approximation to the causal outcome model $\beta^*$. Therefore, it suffices to find a stationary point of $\Phi_\mu(b)$ to approximate $\beta^*$.
To our knowledge, it is the first result demonstrating a benign nonconvex landscape established in the context of causal invariance learning, and we shall present the main idea for its proof within Section \ref{sec: main idea of proof}.

The parameters $\gamma\geq  0$ and $\mu>0$ play key roles in this bound. As $\gamma$ increases, the first term, $(1+\gamma|\Ec|)^{-1}$ diminishes to zero. Meanwhile, the choice of $\mu$ presents a tradeoff between $\left\|\nabla\Phi_{\mu}(b)\right\|_2$ and  $\sqrt{{\mu}/{\lambda}}$.
Specifically, while a smaller value of $\mu$ reduces $\sqrt{\mu/\lambda}$, it increases the number of iterations required for algorithms to locate a prediction model $b$ with a sufficiently small $\|\nabla\Phi_\mu(b)\|_2$. In the discussion right after the following Theorem \ref{thm: convergence theorem - penal - empirical - alg output}, we provide guidance on choosing the optimal value of $\mu$.

Theorem \ref{thm: convergence theorem - penal - population - any b} characterizes a benign landscape for NegDRO through an error-bound type inequality. We emphasize two essential distinctions from the classical error bound condition \citep{karimi2016linear}, which relates the distance from a point to the optimal solution to its gradient norm. First, in our characterization, the reference point is the causal outcome model $\beta^*$, instead of the globally optimal solution of NegDRO. Second, in addition to the gradient norm $\|\nabla \Phi_\mu(b)\|_2$, the bound in \eqref{eq: upper bound - penal - population - any b} includes other terms, $(\lambda(1+\gamma|\Ec|))^{-1}$ and $\sqrt{\mu/\lambda}$, that do not appear in the standard error bound conditions. These differences mean that one cannot directly borrow algorithms or convergence analysis from problems that satisfy the classical error bound conditions. %

\subsection{Algorithm Design}
\label{subsec: computation algorithm}
As established in Theorem \ref{thm: convergence theorem - penal - population - any b}, it suffices to find a stationary point of NegDRO to approximate $\beta^*$. Building on this insight, we now move from the population setting to the finite-sample setting. Our goal is to design a computationally efficient algorithm that converges to such stationary points. We further establish the convergence rate of the proposed algorithm in terms of both the sample size and the number of iterations.

In the finite-sample regime, we observe i.i.d. data samples $\{\x{e}_i, \y{e}_i\}_{i=1}^{n_e}$ drawn from the distribution of $(\X{e},\Y{e})$ for each environment $e\in \Ec$. 
We substitute the risk $\Eb[\Y{e}-b^\intercal\X{e}]^2$ in the population NegDRO problem \eqref{eq: obj trans - population} with its empirical counterpart, and formulate the following empirical NegDRO problem: 
\begin{equation}
\begin{aligned}
     \min_{b\in \Rb^p}\widehat{\Phi}_\mu(b),\quad &\textrm{with}\;\; \widehat{\Phi}_\mu(b) := \max_{w\in \Delta^{|\Ec|}}\sum_{e\in \Ec} \left\{\left(w_e - \frac{\gamma}{1+\gamma |\Ec|}\right)\widehat{\Eb}[\Y{e}-b^\intercal\X{e}]^2 - \mu \|w\|_2^2\right\},\\
     &\textrm{and}\;\; \widehat{\Eb}[\Y{e}-b^\intercal\X{e}]^2 := \frac{1}{n_e}\sum_{i=1}^{n_e}(\y{e}_i-b^\intercal\x{e}_i)^2.
\end{aligned}
    \label{eq: obj-raw penal empirical}
\end{equation}
We now present a gradient descent-maximization algorithm for locating a stationary point of $\widehat{\Phi}_\mu(b)$, as detailed in Algorithm \ref{alg: GDmax penal}. At each iteration $t$, given the current prediction model $b_t$, the weight vector $w_{t+1}$ is chosen to maximize the penalized objective as in \eqref{eq: alg penal step w}, which admits a closed-form solution. Since the objective is strongly concave in $w$, we apply Danskin's theorem to compute the gradient $\nabla\widehat{\Phi}_\mu(b_t)$ using the optimal weight $w_{t+1}$. We then perform a gradient descent step to update the prediction model $b_{t+1}$, as specified in \eqref{eq: alg penal step b}.

\begin{algorithm}[!ht]

    \DontPrintSemicolon
    \SetAlgoLined
    \SetNoFillComment
    \LinesNotNumbered 
    \caption{Gradient Descent-Maximization Algorithm for the Penalized NegDRO in \eqref{eq: obj-raw penal empirical}}
    \SetKwInOut{Input}{Input}
    \SetKwInOut{Output}{Output}
    \Input{The number of iterations $T$, step size $\alpha$, initial point $b_0\in \Rb^p$, parameters $\gamma\geq0$ and $\mu>0$}
    \Output{$\hat{b}^\gamma$}
    \For{$t = 0,1,...,T-1$}{
        Update the maximizer weight associated with $b_{t}$:
        {
        \begin{equation}
            w_{t+1} = \argmax_{w\in \Delta^{|\Ec|}} \left\{\sum_{e\in \Ec}\left({w}_e- \frac{\gamma}{1+\gamma |\Ec|}\right) \widehat{\Eb}[\Y{e}-b_t^\intercal\X{e}]^2 - \mu \|w\|_2^2\right\};
            \label{eq: alg penal step w}
        \end{equation}
        }
        
        Update $b_{t+1}$ corresponding to the weight $w_{t+1}$:
        {
        \begin{equation}
            b_{t+1} = b_t - \alpha \nabla\widehat{\Phi}_\mu(b_t),~~ \textrm{with} ~~\nabla \widehat{\Phi}_\mu(b_t) = \sum_{e\in \Ec}\left([{w}_{t+1}]_e- \frac{\gamma}{1+\gamma |\Ec|}\right) \nabla \widehat{\Eb}[\Y{e}-b_t^\intercal\X{e}]^2;
   \label{eq: alg penal step b}
\end{equation}
        }
    }
    
    Define $\hat{b}^\gamma \in \argmin_{\{b_t\}_{t=0}^{T}}\|\nabla \widehat{\Phi}_\mu(b_t)\|_2$.
    \label{alg: GDmax penal}
\end{algorithm}

To facilitate the theoretical analysis of Algorithm \ref{alg: GDmax penal}, we impose the following regularity conditions on the noise and empirical risk functions.
\begin{Condition}
    The random vectors $\eta\in \Rb^{p+1}$ and $\del{e}\in \Rb^p$ for all $e\in \Ec$ in \eqref{eq: SCM} are sub-Gaussian. 
    \label{cond: subgaussian}
\end{Condition}
Condition \ref{cond: subgaussian} imposes standard tail assumptions on the noise distributions that are commonly used in the analysis of empirical causal invariance learning \citep{rothenhausler2019causal, fan2023environment}. 
\begin{Condition}[Smoothness and Gradient Boundedness]
\label{cond: smooth}
    For each environment $e\in \Ec$, the empirical prediction risk  $\widehat{\Eb}[\Y{e}-b^\intercal\X{e}]^2$ defined in \eqref{eq: obj-raw penal empirical} is $\kappa_1$-smooth for some $\kappa_1>0$; that is, for any $b, b'\in \Rb^p$, 
    \begin{equation}
        \left\|\nabla \widehat{\Eb}[\Y{e}-b^\intercal\X{e}]^2 - \nabla \widehat{\Eb}[\Y{e}-(b')^\intercal\X{e}]^2\right\|_2 \leq \kappa_1 \|b - b'\|_2.
        \label{eq: smooth}
    \end{equation}
    Furthermore, all iterates $\{b_t\}_{t=0}^T$ produced by Algorithm \ref{alg: GDmax penal} remain within a bounded set $\mathcal{B}\subset \Rb^p$, and there exists $\kappa_2>0$ such that
    \begin{equation}
        \kappa_2:= \sup_{e\in \Ec, b\in \mathcal{B}}\|\nabla \widehat{\Eb}[Y^e - b^\intercal \X{e}]^2 \|_2 <\infty.
        \label{eq: bounded gradient main}
    \end{equation}
\end{Condition}
Condition \ref{cond: smooth} encodes two regularity properties of the empirical risk function that are commonly used for the convergence analysis of gradient-based algorithms \citep{ghadimi2013stochastic, lee2016gradient}. First, inequality \eqref{eq: smooth} requires the empirical risk to be smooth in $b$. For the squared loss, it holds whenever the sampled second moment matrix $\frac{1}{n_e}\sum_{i=1}^{n_e}\x{e}_ix^{e\intercal}_i$ has a bounded spectral norm. Second, inequality \eqref{eq: bounded gradient main} ensures that the gradient norms remain uniformly bounded across all environments and iterations. It holds when both $\frac{1}{n_e}\sum_{i=1}^{n_e} x_i^e x_i^{e\intercal}$ and $\frac{1}{n_e}\sum_{i=1}^{n_e} x_i^e y_i^e$ are bounded.

The following theorem establishes the convergence rate of Algorithm \ref{alg: GDmax penal} to the causal outcome model $\beta^*$. We denote $n = \min_{e\in \Ec}n_e$ as the smallest sample size across all environments.
\begin{Theorem}
    Under the additive intervention regime, suppose Conditions \ref{cond: strict positive - A}, \ref{cond: subgaussian}, \ref{cond: smooth} hold, and $n\geq (1\vee (c_1/\lambda^2)) p$. Set the step size in Algorithm \ref{alg: GDmax penal} to $\alpha = ({\mu+(\kappa_1+\kappa_2\sqrt{|\Ec|})^2/\mu})^{-1}$. Then, for any $0\leq t\leq (1\wedge (c_2\lambda^2)) n - p$, with probability at least $1-2|\Ec| e^{-t}$, the output $\hat{b}^\gamma$ of Algorithm \ref{alg: GDmax penal} satisfies:
    \begin{equation*}
        \|\hat{b}^\gamma - \beta^*\|_2 \lesssim \frac{1}{\lambda}\left(\frac{1}{1+\gamma|\Ec|} + \sqrt{\frac{1}{T}\left(\frac{\kappa_1^2+\kappa_2^2 |\Ec|}{\mu} +\mu\right)}\right) +  \sqrt{\frac{\mu}{\lambda}}  + \frac{1}{\sqrt{\lambda}}\left(\frac{p+t}{{n}}\right)^{1/4},
        \label{eq: upper bound - penal - empirical - alg output}
    \end{equation*}
    where $\lambda>0$ measures the degree of environmental heterogeneity, and $c_1,c_2$ are absolute constants.
    \label{thm: convergence theorem - penal - empirical - alg output}
\end{Theorem}
Unlike existing methods that enumerate all subsets of covariates described in Subsection \ref{subsec: illu}, Algorithm \ref{alg: GDmax penal} avoids the combinatorial computational cost that grows exponentially with the dimension $p$ and remains efficient even when $p$ is relatively large. In contrast to heuristic approaches such as the Gumbel relaxation \citep{fan2023environment}, which approximate integer programming for efficiency,  our gradient-based methods provide a rigorous finite-sample convergence guarantee to the causal outcome model $\beta^*$. 

The upper bound above explicitly quantifies the impact of the penalty parameter $\mu>0$, which presents a tradeoff between statistical accuracy and optimization stability. Specifically, with a smaller $\mu$, the term $\sqrt{\mu/\lambda}$ decreases, pulling the output $\hat{b}^\gamma$ closer to $\beta^*$. However, the smaller $\mu$ enlarges the stationary point error term $\sqrt{\frac{1}{T}\left(\frac{\kappa_1^2+\kappa_2^2 |\Ec|}{\mu} +\mu\right)}$, yielding slower convergence to the stationary point of $\widehat{\Phi}_\mu(b)$. 
To balance these effects, we select the optimal penalty parameter $\mu\asymp T^{-1/2}$, then the upper bound in the theorem is simplified as follows:
\begin{equation}
    \|\hat{b}^\gamma - \beta^*\|_2 \lesssim \frac{1}{1+\gamma |\Ec|} + T^{-1/4} + n^{-1/4}.
    \label{eq: simplified rate}
\end{equation}
If we further treat the number of environments $|\Ec|$ as constant, to achieve an $\epsilon$-accurate solution using Algorithm \ref{alg: GDmax penal}, i.e. $\|\hat{b}^\gamma - \beta^*\|_2\leq \epsilon$, it is sufficient to set the regularization parameter as 
$\gamma=\Omega(\epsilon^{-1})$, collect a sample size of $n=\Omega(\epsilon^{-4})$, and conduct  $T=\Omega(\epsilon^{-4})$ number of iterations in Algorithm \ref{alg: GDmax penal}. 

In numerical experiments within Section \ref{sec: numerical}, we demonstrate the computational efficiency and effectiveness of our method in recovering $\beta^*$. In particular, Figure \ref{fig: converge with gamma} shows the distance of our proposed NegDRO estimator to $\beta^*$ against the sample size $n$ and the number of iterations $T$ established in \eqref{eq: simplified rate}. For such a nonconvex optimization problem, it remains an interesting and open question whether one could achieve $\Oc(n^{-1/2})$ sample complexity and $\Oc(T^{-1/2})$ iteration complexity,  which could match the lower bounds of first-order methods for solving stochastic convex optimization \citep{agarwal2009information}.

\subsection{Proof Sketch of Benign Landscape in Theorem \ref{thm: convergence theorem - penal - population - any b}}
\label{sec: main idea of proof}
We sketch the main ideas for the proof of Theorem \ref{thm: convergence theorem - penal - population - any b}. The argument relies on the following expression for the population prediction risk $\Eb[\Y{e}-b^\intercal \X{e}]^2$ under the additive intervention regime:
\begin{equation}
    \Eb[\Y{e}-b^\intercal \X{e}]^2 = \sigma_Y^2 + 2v^\intercal(b-\beta^*) + (b-\beta^*)^\intercal\Eb[\X{e}X^{e\intercal}](b-\beta^*),
    \label{eq: main risk expression}
\end{equation}
where $v\in \mathbb{R}^{p}$ is a constant vector whose explicit expression is given in Appendix Lemma \ref{lemma: expression of risk}. The risk expression \eqref{eq: main risk expression} reveals that the linear(first-order) term involving $b-\beta^*$ remains invariant across environments, while the quadratic (second-order) terms depend on the environment-specific $\Eb[\X{e}X^{e\intercal}]$.
The proof sketch proceeds in three steps.

\paragraph{Step-1: Expressing $\Phi_\mu(b)$ through its gradient $\nabla \Phi_\mu(b)$.}  Given the quadratic nature of $\Phi_\mu(b)$ with respect to $b$ as shown in \eqref{eq: obj trans - penal - population} , the second-order term of $\Phi_\mu(b)$ coincides with those of $\frac{1}{2}(b - \beta^*)^\intercal\nabla\Phi_\mu(b)$, differing only by lower-order terms. Thus, we could express $\Phi_\mu(b)$ by its gradient,
    \begin{equation*}
        \Phi_\mu(b) = \frac{1}{2} (b-\beta^*)^\intercal \nabla\Phi_\mu(b) + \frac{1}{1+\gamma|\Ec|}\left[\sigma_Y^2 + v^\intercal(b - \beta^*)\right] - \mu\|\bar{w}\|_2^2,
    \end{equation*}
    where $\bar{w}$ is the unique maximizer weight defined in \eqref{eq: maximizer w - penal - population}. This reformulation links the objective value to the gradient, which will later serve to characterize the optimization landscape.

\paragraph{Step-2: Establishing a lower bound of $\Phi_\mu(b)$.} According to \eqref{eq: maximizer w - penal - population}, $\Phi_\mu(b)$ is attained with the maximizer weight $\bar{w}$, and thus any other weight $w$ provides a lower bound for $\Phi_\mu(b)$. Choosing the specific $w^0$ from Condition \ref{cond: strict positive - A}, we have
    \begin{align*}
        \Phi_\mu(b) & = \max_{w\in \Delta^{|\Ec|}} \left\{\sum_{e\in \Ec}\left(w_e - \frac{\gamma}{1+\gamma|\Ec|}\right)\Eb[\Y{e}-b^\intercal\X{e}]^2 - \mu \|w\|_2^2\right\}\\
        &\geq
        \sum_{e\in \Ec}(w^0_e - \frac{\gamma}{1+\gamma|\Ec|})\Eb[\Y{e}-b^\intercal\X{e}]^2 - \mu \|w^0\|_2^2\notag\\
        &\geq \frac{1}{1+\gamma|\Ec|}\left[\sigma_{Y}^2 + 2v^\intercal (b-\beta^*)\right] + (b-\beta^*)^\intercal \Abm(w^0)(b - \beta^*) - \mu\|w^0\|^2 
        \notag \\ 
        &\geq \frac{1}{1+\gamma|\Ec|}\left[\sigma_{Y}^2 + 2v^\intercal(b-\beta^*)\right] + \lambda \|b - \beta^*\|_2^2 - \mu\|w^0\|_2^2,\notag
    \end{align*}
where the second inequality follows from substituting the risk expression \eqref{eq: main risk expression} and the definition of $\Abm(w)$ \eqref{eq: A inf def}, and the third inequality leverages Condition \ref{cond: strict positive - A} with \(\lambda_{\rm min}(\Abm(w^0))\geq \lambda \).

\paragraph{Step-3: Bounding the error via the gradient norm.} Combining the previous two steps and applying Chebyshev's inequality, we obtain
\[
    \lambda \|b - \beta^*\|_2^2 \leq \left(\frac{1}{2}\left\|\nabla \Phi_\mu(b)\right\|_2 + \frac{\|v\|_2}{1+\gamma|\Ec|}\right)\|b - \beta^*\|_2 + \mu(\|w^0\|_2^2-\|\bar{w}\|_2^2).
    \]
Since $\|v\|_2$ is upper bounded and $\|w^0\|_2^2-\|\bar{w}\|_2^2\leq 1$, the desired result in Theorem \ref{thm: convergence theorem - penal - population - any b} follows.

\section{Limited Additive Interventions: The Importance of the Optimization Objective}
\label{sec: minimization}
The earlier Subsection \ref{subsec: ident conditions} established the conditions under which the invariant prediction model set $\Bc_{\rm inv}$ collapses to the singleton $\beta^*$, and thus the constrained problem \eqref{eq: bneg gamma infty} exactly recovers $\beta^*$.
A natural follow-up question is whether causal identification remains possible under weaker conditions. In particular, when $\Bc_{\rm inv}$ does not collapse to a singleton but instead contains multiple risk-invariant prediction models beyond $\beta^*$, how can one establish conditions under which NegDRO achieves causal identification.

The key lies in recognizing that the constrained problem \eqref{eq: bneg gamma infty} does more than just enforce the risk invariance principle. Among all risk-invariant prediction models, it also selects the one with the smallest prediction risk. This observation highlights the role of the optimization objective itself: even when the risk-invariance set is not a singleton, the objective may still single out $\beta^*$ as the unique minimizer.  
From a practical perspective, weaker conditions allow for more flexible data collection mechanisms that still guarantee causal identification as they are less restrictive. A representative scenario is the limited additive intervention regime, where only a subset of covariates is subject to interventions.
This situation frequently arises in real-world applications across diverse fields. For instance, in clinical trials, budget constraints or ethical considerations often prevent direct interventions on certain variables \citep{cohen2008likelihood, brannath2009confirmatory, banerjee2009experimental}. In these cases, Condition \ref{cond: strict positive - A} becomes too restrictive, as its strict positive definiteness requirement demands that every covariate be intervened upon at least once across all environments.

Tailored to the limited additive intervention regime, we establish a weaker condition than Condition \ref{cond: strict positive - A} that still guarantees the recovery of $\beta^*$ in Subsection \ref{subsec: descendants intervene}. 
Subsection \ref{subsec: comparison} then contrasts our approach with other causal invariance learning methods \citep{rothenhausler2019causal, shen2023causality}, highlighting the advantages of our methods under limited interventions. For clarity, the results here are presented at the population level, while the finite-sample analysis is deferred to Appendix \ref{subsec: limited negdro}.

\subsection{Causal Identification under Limited Additive Interventions}
\label{subsec: descendants intervene}

We now formalize the limited additive intervention regime. Recall from \eqref{eq: SCM} and \eqref{eq: intervention noise} that, under the additive intervention regime, the distributional heterogeneity of $(\X{e}, \Y{e})$ across environments is induced by the environment-specific interventions $\delta^{e}\in \Rb^p$ added to the covariates. The limited additive intervention regime arises when certain coordinates of $\delta^{e}$ are identically zero across all environments, meaning that the corresponding covariates are never intervened. Formally, we define the set $S_L$ for those coordinates that have been intervened upon as follows:
\begin{equation}
    S_L:= \left\{j\in [p]: \exists~e\in \Ec, \delta^{e}_j\not\equiv 0\right\},
    \label{eq: limited set}
\end{equation}
and the limited additive intervention occurs whenever $S_L\neq [p]$, i.e., at least one covariate is never intervened upon among all environments.

In this regime, the earlier Condition \ref{cond: strict positive - A} no longer holds. Indeed, one can show that $\Eb[\X{e}X^{e\intercal}] = \Gbf^\intercal (\Hbf + \Eb[\del{e}\delT{e}] )\Gbf$, where $\Hbf,\Gbf\in \Rb^{p\times p}$ are full-rank matrices. Substituting this expression, Condition \ref{cond: strict positive - A} can be equivalently rephrased in terms of the intervention second moment matrices $\Eb[\delta^e\delta^{e\intercal}]$: there exists a weight vector $w^0\in \Delta^{|\Ec|}$ such that
\begin{equation}
    \Deltabf(w^0) \succ 0, \quad \textrm{where} \quad \Deltabf(w):= \sum_{e\in \Ec}\left(w_e - \frac{1}{|\Ec|}\right)\Eb[\del{e}\delT{e}], \;\textrm{for any $w\in \Delta^{|\Ec|}$}.
    \label{eq: intervention cond}
\end{equation}
The details are deferred to Appendix Lemma \ref{lemma: expression of risk}.
With limited intervention, at least one coordinate of $\del{e}$ is always zero for all $e\in \Ec$, making $\Deltabf(w)$ singular for all $w\in \Delta^L$, and thus \eqref{eq: intervention cond} never holds. 

To weaken Condition \ref{cond: strict positive - A} tailored for the limited interventions, instead of requiring the full matrix $\Deltabf(w^0)\succ 0$, it suffices to ensure that the appropriate submatrix of $\Deltabf(w^0)$ be strictly positive definite. In particular, let $D\subseteq [p]$ denote the set of covariates that are directly affected by the outcome, i.e., those for which there is a directed causal edge from $Y$ to $X_D$. We propose the following condition:
\begin{Condition}
    \label{cond: relaxed minimization}
     $D\subseteq S_L$ and there exists a weight vector $w^0\in \Delta^{|\Ec|}$ such that both of the following hold: 
     \[
     [\Deltabf(w^0)]_{S_L, S_L}\succ 0, \quad \textrm{and}\quad \sum_{e\in \Ec} w_e^0\Eb[\X{e}\XT{e}]\succ 0.
     \]
\end{Condition}
This condition imposes three requirements. First, it requires $D\subseteq S_L$, ensuring that every child of the outcome, that is, each covariate directly affected by the outcome, is intervened upon at least once across environments. Second, it requires the submatrix $[\Deltabf(w^0)]_{S_L, S_L}$, restricted to the intervened coordinates, to be strictly positive definite, which is analogous to Condition ~\ref{cond: strict positive - A}. Finally, the requirement $\sum_{e\in \Ec} w_e^0\Eb[\X{e}\XT{e}]\succ 0$ ensures that the aggregated covariates remain nondegenerate under the weighting $w^0$.
Together, these requirements guarantee that there is sufficient environmental heterogeneity among the causal children of the outcome.

\begin{figure}[!ht]
    \centering
    \includegraphics[width=0.45\linewidth]{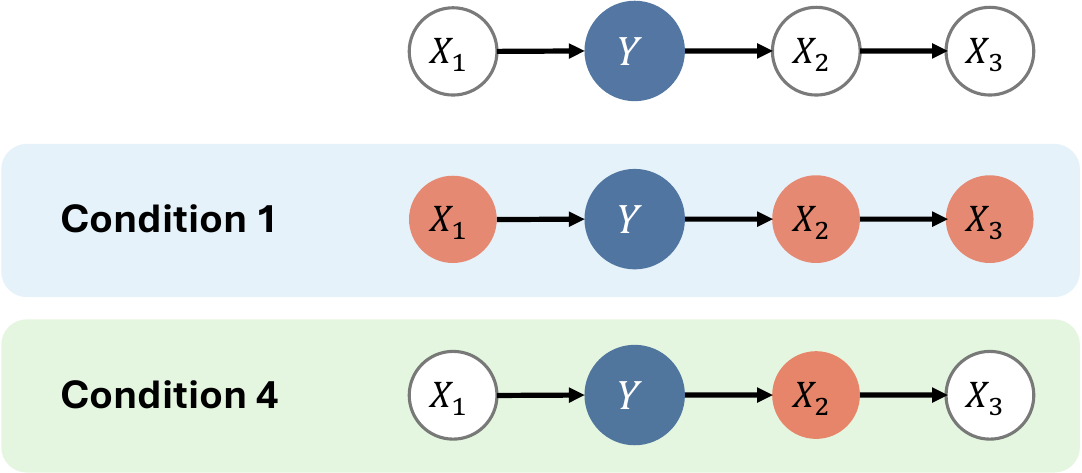}
    \caption{Comparison of Conditions \ref{cond: relaxed minimization} and \ref{cond: strict positive - A}. Orange nodes indicate intervened covariates. Condition \ref{cond: relaxed minimization} requires interventions on the covariate directly affected by the outcome ($X_2$), whereas Condition \ref{cond: strict positive - A} requires interventions on all covariates $(X_1, X_2, X_3)$.}
    \label{fig: diff between conditions}
\end{figure}

This condition is strictly weaker than Condition \ref{cond: strict positive - A}.
In fact, when all covariates are intervened with $S_L=[p]$, Condition \ref{cond: relaxed minimization} reduces exactly to Condition \ref{cond: strict positive - A}.
To illustrate the difference, consider an experiment that begins with a reference environment with no interventions, and subsequently performs a series of single-covariate interventions at a time. As depicted in Figure \ref{fig: diff between conditions}, there are $p=3$ covariates, and only $X_2$ is directly affected by the outcome $Y$, i.e., $D=\{2\}$.  Condition \ref{cond: relaxed minimization} requires intervention only on $X_2$, as it is the unique child of the outcome, whereas Condition \ref{cond: strict positive - A} requires interventions on all three covariates $(X_1, X_2, X_3)$.

The following theorem shows that Condition \ref{cond: relaxed minimization} suffices for causal identification, i.e., the optimal solution of the constrained problem \eqref{eq: bneg gamma infty} coincides with $\beta^*$.
\begin{Theorem}
    Under the additive intervention regime, suppose there is no hidden confounding between the outcome and covariates, i.e., $\Eb[\eta_Y\eta_X] = 0$. If Condition \ref{cond: relaxed minimization} holds, then the causal outcome model $\beta^*$ achieves the smallest prediction risk among all risk-invariant prediction models. Consequently, the solution to the constrained problem \eqref{eq: bneg gamma infty} exactly recovers $\beta^*$.
\label{thm: identification infty minimization - simplify}
\end{Theorem}
Compared to the earlier Theorem \ref{thm: optimization perspective}, which requires $\beta^*$ to be the unique risk-invariant prediction model, this theorem shows that even if there are multiple risk-invariant prediction models, the objective of the constrained problem \eqref{eq: bneg gamma infty} ensures that $\beta^*$ is singled out as the minimizer.

For clarity of exposition, this theorem focuses on the case without hidden confounding between the outcome and covariates, i.e., $\Eb[\eta_Y \eta_X]=0$, as assumed in prior works \citep{hauser2015jointly, peters2016causal, ghassami2017learning}. We address the more general case, where there could exist hidden confounding in Appendix \ref{Appendix: weaker condition}. Furthermore, paralleling Theorem~\ref{thm: convergence theorem - penal - population - any b},
we show in Appendix~\ref{subsec: limited negdro} that under Condition~\ref{cond: relaxed minimization},
the NegDRO objective~\eqref{eq: obj original} exhibits a benign nonconvex landscape for finite $\gamma \ge 0$ under the limited intervention regime,
ensuring that gradient-based algorithms can still reliably approach $\beta^*$.

\subsection{Comparison to Existing Methods}
\label{subsec: comparison}
In the last subsection, we showed that the constrained problem \eqref{eq: bneg gamma infty} could recover $\beta^*$ even under the challenging limited additive intervention regime. One may ask whether existing approaches can achieve the same. We focus on two representative methods, CausalDantzig \citep{rothenhausler2019causal} and DRIG \citep{shen2023causality}, and show that they fail to identify $\beta^*$ under limited interventions as their key assumptions no longer hold. Below, we provide a brief overview of these methods; a more detailed introduction is available in Appendix \ref{Appendix: CausalDantzig DRIG}.

CausalDantzig leverages the gradient invariance property inherent to the additive intervention regime, which is formalized in its Proposition 1. This property states that for any pair of environments $e,f\in \Ec$, the causal outcome model $\beta^*$ satisfies: 
\begin{equation}
\Eb[\X{e}(\Y{e} - (\beta^*)^\intercal \X{e})] = \Eb[\X{f}(\Y{f} - (\beta^*)^\intercal \X{f})].
\label{eq: grad invariance}
\end{equation}
Assuming the non-singularity of $\Eb[X^{e}X^{e\intercal} - X^{f}X^{f\intercal}]$, CausalDantzig identifies $\beta^*$ through:
\[
\beta^* = \left(\Eb[X^{e}X^{e\intercal} - X^{f}X^{f\intercal}]\right)^{-1} \Eb[X^{e}Y^{e} - X^{f}Y^{f}].
\]
However, this method fails under the limited additive intervention scenarios, since $\Eb[X^{e}X^{e\intercal} - X^{f}X^{f\intercal}]$ is singular when some covariates are never intervened upon across environments $e,f$.
In contrast, our approach remains effective in identifying $\beta^*$ under such a limited additive intervention regime, once Condition \ref{cond: relaxed minimization} holds, as established in Theorem \ref{thm: identification infty minimization - simplify}.

As for DRIG, it relies on the existence of and access to a \emph{reference environment}, denoted as $e=0$. This reference environment is assumed to be strictly dominated by all other environments $\Ec=\{1,2,...,|\Ec|\}$ such that 
$\Eb[X^{(0)}X^{(0)\intercal}]\prec \Eb[X^{e}X^{e\intercal}]$ for all $e\in \Ec$, which is a special case of our Condition \ref{cond: strict positive}. Given a regularization parameter $\gamma\geq 0$,
DRIG solves the following optimization problem:
\begin{equation}
    b_{\rm DRIG}^\gamma = \argmin_{b\in \Rb^p} \left\{\Eb[\Y{0}-b^\intercal\X{0}]^2 + \gamma\sum_{e\in \Ec} w^e\left(\Eb[\Y{e}-b^\intercal\X{e}]^2 - \Eb[\Y{0}-b^\intercal\X{0}]^2\right)\right\},
    \label{eq: DRIG formulation}
\end{equation}
where $w\in \Delta^{|\Ec|}$ is a {pre-specified} weight vector. While the primary focus of DRIG is to ensure robust generalization to unseen populations for finite $\gamma$, the method can achieve causal discovery with $b_{\rm DRIG}^\infty = \beta^*$ when $\gamma$ is set as $\infty$, but only if such a strictly dominated reference environment exists and is accessible.

We emphasize that the reference environment plays a critical role in DRIG’s formulation. With access to such an environment $e=0$, it follows that $\Eb[\X{e}(\X{e})^\intercal] - \Eb[\X{0}(\X{0})^\intercal]\succ 0$ for all $e\in \Ec$. Consequently, the optimization problem in \eqref{eq: DRIG formulation} becomes convex, since its Hessian $\Eb[\X{0}(\X{0})^\intercal] + \gamma\sum_{e\in \Ec}w^e (\Eb[\X{e}(\X{e})^\intercal] - \Eb[\X{0}(\X{0})^\intercal]) \succ 0$. 
This convexity significantly simplifies the computational effort, as DRIG requires solving only a convex problem. However, it is challenging to identify a reference environment in practice. In fact, such a reference environment may not exist, limiting the practical application of the DRIG method. This is a key distinction from our proposed NegDRO in \eqref{eq: obj original}, which involves solving a nonconvex problem to an approximate global optimality without assuming the existence of, nor access to, a reference environment.  

Next, we use the following Example \ref{eg: illus-eg2} to compare the performance of NegDRO, CausalDantzig and DRIG under three variants of the additive intervention regimes:
\begin{itemize}[itemsep=1pt, topsep=2pt, parsep=0pt]
    \item[(i)] Limited regime: only the outcome's child covariates are intervened upon, while all non-child covariates remain untouched;
    \item[(ii)] Weak regime: non-child covariates are also intervened upon, but only with very small strength;
    \item[(iii)] Strong regime: all covariates are subject to sufficiently large interventions.
\end{itemize}
\begin{figure}[!ht]
        \centering
        \includegraphics[width=0.5\linewidth]{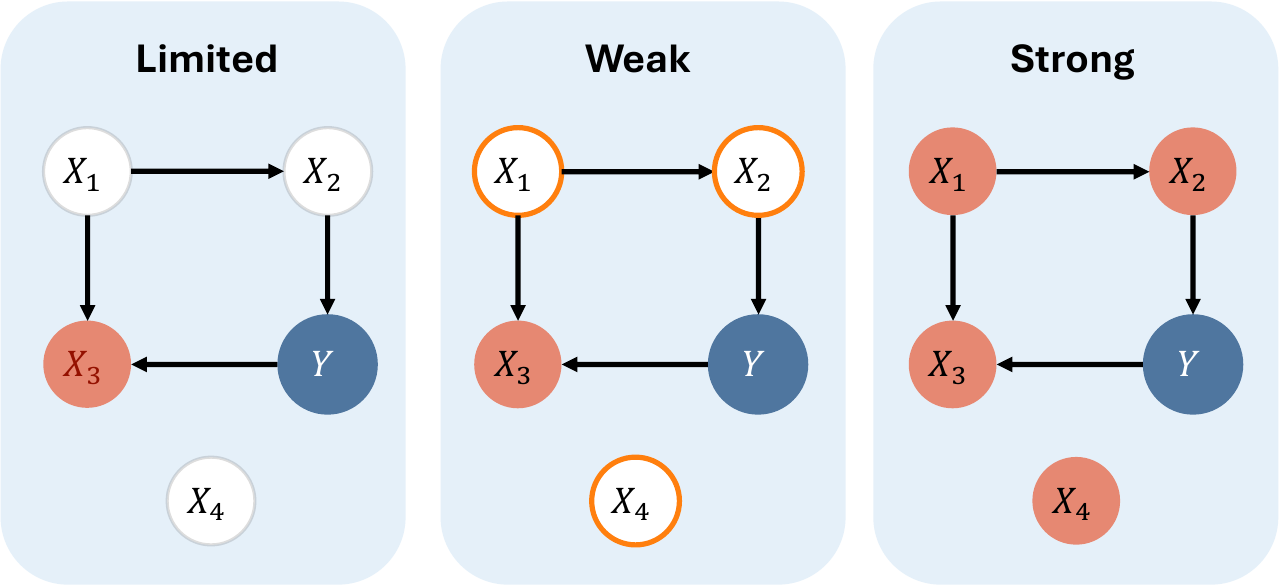}
\caption{Illustration of Example \ref{eg: illus-eg2} under the three additive intervention regimes. Solid orange nodes represent covariates with large-scale interventions, hollow orange nodes represent covariates with small-scale interventions, and white nodes indicate covariates with no interventions.}
        \label{fig:illus-scenarios}
    \end{figure}

\begin{figure}[ht]
\centering
\includegraphics[width=0.8\linewidth]{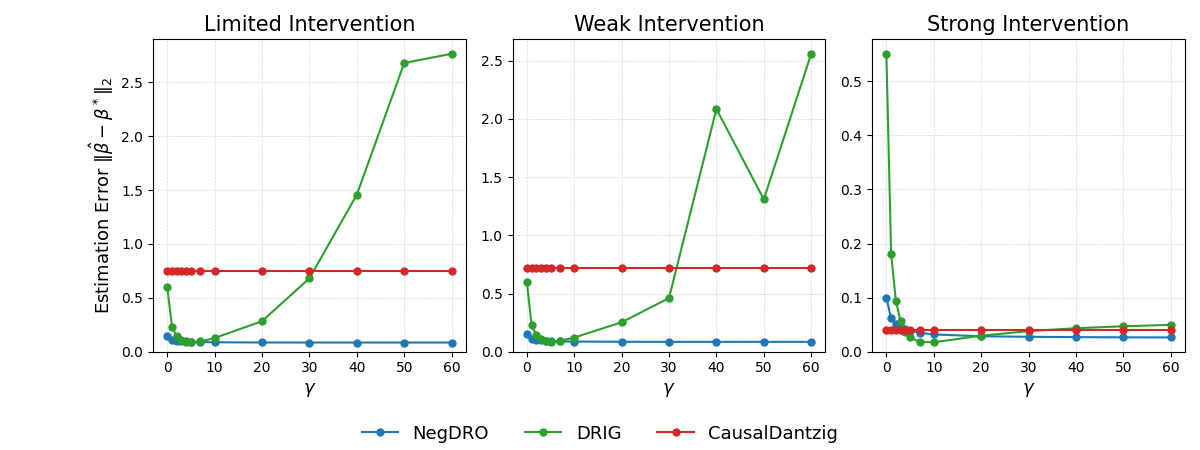}
\caption{Comparison of NegDRO with CausalDantzig and DRIG in terms of  $\ell_2$ distance of estimators computed by these methods to the causal outcome model $\beta^*$.   Sample size $n_e=10,000$ for $e\in \Ec$. For NegDRO and DRIG, we vary the regularization parameter $\gamma$ in the range of $[0,60]$, while CausalDantzig does not require it. The results are averaged over 200 simulations.
}
\label{fig: gradient fail} 
\end{figure}

\begin{Example}
    Consider two environments $\Ec = \{1,2\}$. For each  $e\in \Ec$, the variables are generated as follows:
    \[
    \X{e}_1 = \eps{e}_1,\quad \X{e}_2 = \X{e}_1 + \eps{e}_2,\quad 
     \Y{e} = 2\X{e}_2 + \eps{e}_Y,\quad \X{e}_3 = 0.5\X{e}_1 - \Y{e} +\eps{e}_3,\quad \X{e}_4 = \eps{e}_4,
    \]
    Among the four covariates, $X_3$ is the only child of $Y$, i.e. $D = \{3\}$. The noise terms $\eps{e}_Y$ and $\eps{e}_X=\eps{e}_{1:4}$ are generated according to $(\eps{e}_Y, \eps{e}_{X})^\intercal = \eta + (0, \delta^{e})^\intercal$,
    where $\eta \sim \Nc(0_4, \Ibf_4)$ denotes the systematic noise, and $\delta^{e}$ denotes the environment-specific interventions. 
    
    In the environment $e=1$, no interventions are applied, i.e., $\delta^{(1)} = 0_4$. 
    In the environment $e=2$, the outcome's child $X_3$ is intervened with \(\delta^{(2)}_3 \sim \Nc(0, 2)\). For the remaining covariates, we examine three intervention regimes, which are illustrated in Figure \ref{fig:illus-scenarios}.
    \[
    \textrm{(Limited)}~~\delta^{(2)}_{\{1,2,4\}}\equiv 0_3; ~~\textrm{(Weak)}~~\delta^{(2)}_{\{1,2,4\}}{\sim}\Nc(0, 0.01 {\bf I}_3);~~ \textrm{(Strong)}~~\delta^{(2)}_{\{1,2,4\}}{\sim}\Nc(0, 0.25{\bf I}_3).
    \]
    
\label{eg: illus-eg2} 
\end{Example}
    Figure \ref{fig: gradient fail} summarizes the numerical results across three additive intervention regimes. For NegDRO and DRIG, we evaluate the distance of the fitted estimators to the causal outcome model $\beta^*$ against the regularization parameter $\gamma$ varying within the range $[0, 60]$. Since CausalDantzig does not include a regularization parameter, its performance remains unchanged across different $\gamma$ values. 

The figure shows that under the limited and weak regimes, DRIG estimators deviate from $\beta^*$ as $\gamma$ increases. Similarly, CausalDantzig fails to approximate $\beta^*$ in these two regimes. Only in the strong intervention setting do both CausalDantzig and DRIG (with a relatively large $\gamma$) achieve an accurate approximation of $\beta^*$. In contrast, the proposed NegDRO consistently delivers estimators close to $\beta^*$ across all three regimes.

\section{Numerical Results}
\label{sec: numerical}
\begin{figure}[ht!]
    \centering
    \includegraphics[width=0.35\linewidth]{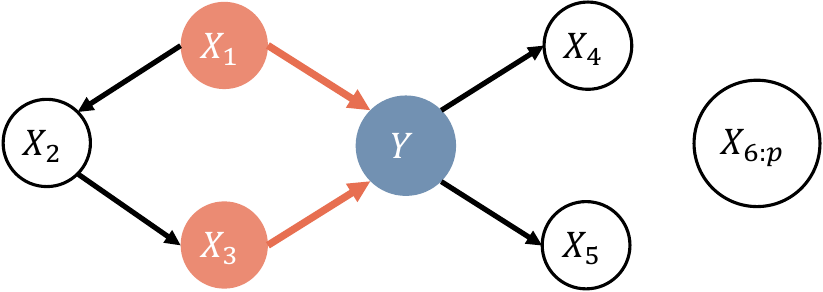}
    \caption{Causal structure of \eqref{eq: simu - SEMs}: the highlighted covariates $(X_1, X_3)$ are direct causes of $Y$.}
    \label{fig: illus simu1}
\end{figure}

In this section, we evaluate the performance of the proposed NegDRO through simulated experiments. We consider a synthetic example whose causal structure is illustrated in Figure \ref{fig: illus simu1}.

For each environment $e\in \Ec=\{1,2,3,4\}$, the covariates $\X{e}\in \Rb^p$ and outcome $\Y{e}\in \Rb$ are generated from the following structural equations:
\begin{equation}
\begin{gathered}
    \X{e}_1 = \eps{e}_1,~~\X{e}_2 = \X{e}_1 + \eps{e}_2,~~\X{e}_3 = \X{e}_2 + \eps{e}_3,\\
    \Y{e} = \frac{1}{2}\X{e}_1 -\frac{1}{2} \X{e}_3 + \eps{e}_Y,\\
    \X{e}_4 = \Y{e} + \eps{e}_4,~~\X{e}_5 = -\Y{e} + \eps{e}_5, ~~\X{e}_{6:p} = \eps{e}_{6:p}.
\end{gathered}
\label{eq: simu - SEMs}
\end{equation}
The causal set is $S^*=\{1,3\}$ with the causal outcome model $\beta^* = (\frac{1}{2},0,\frac{1}{2}, 0_{p-3})^\intercal$.

According to the relationships to $Y$, the covariates can be partitioned into three groups: (i) true causes $(X_1, X_3)$, which directly determine $Y$; (ii) correlated but non-causal covariates $(X_2, X_4, X_5)$; and (iii) irrelevant covariates $X_{6:p}$, where we vary $p$ to adjust the dimension of the covariates without changing the underlying causal structure. 

The noise terms $(\eps{e}_Y, \eps{e}_{1:p})^\intercal$ follow a structural additive form $(\eps{e}_Y, \eps{e}_{1:p})^\intercal \stackrel{d}{=} \eta + (0, \del{e})^\intercal$, where $\eta\sim \Nc(0_{p+1}, \Ibf_{p+1})$ is the systematic noise component, and $\delta^e\in \Rb^p$ are the environment-specific interventions. To introduce environmental heterogeneity, we vary the intervention strengths across environments:
\begin{equation*}
   \delta^{1}_{1:5} = 0_5;~~\delta^{2}_{1:5}\stackrel{iid}{\sim} \Nc(0, 9);~~\delta^{3}_{1:5} = (1, 1.5, 2, 2.5, 3)^\intercal;~~\delta^{4}_{1:5} \stackrel{iid}{\sim} \textrm{Unif}(-0.5, 0.5),
\end{equation*}
while the irrelevant coordinates follow $\delta^{e}_{6:p}{\sim}\Nc(0,\tfrac{e^2}{4}\cdot {\bf I}_{p-5})$ for all $e\in \Ec$. 

For each environment $e\in \Ec$, we generate i.i.d. samples $\{\x{e}_i, \y{e}_i\}_{i=1}^{n_e}$ with equal sample sizes $n_e = n$.
Note that in this setup, we focus on the scenario without hidden confounding between the covariates and the outcome, whereas additional experiments that incorporate hidden confounders are provided in Appendix \ref{appendix: subsec implementation}. 
The following numerical study consists of two parts: (i) we evaluate the accuracy of the proposed NegDRO under varying model parameters and sample sizes, and (ii) we examine both the estimation accuracy and runtime of NegDRO relative to existing approaches under varying dimensions.

\begin{figure}[!ht]
    \centering
    \includegraphics[width=0.85\linewidth]{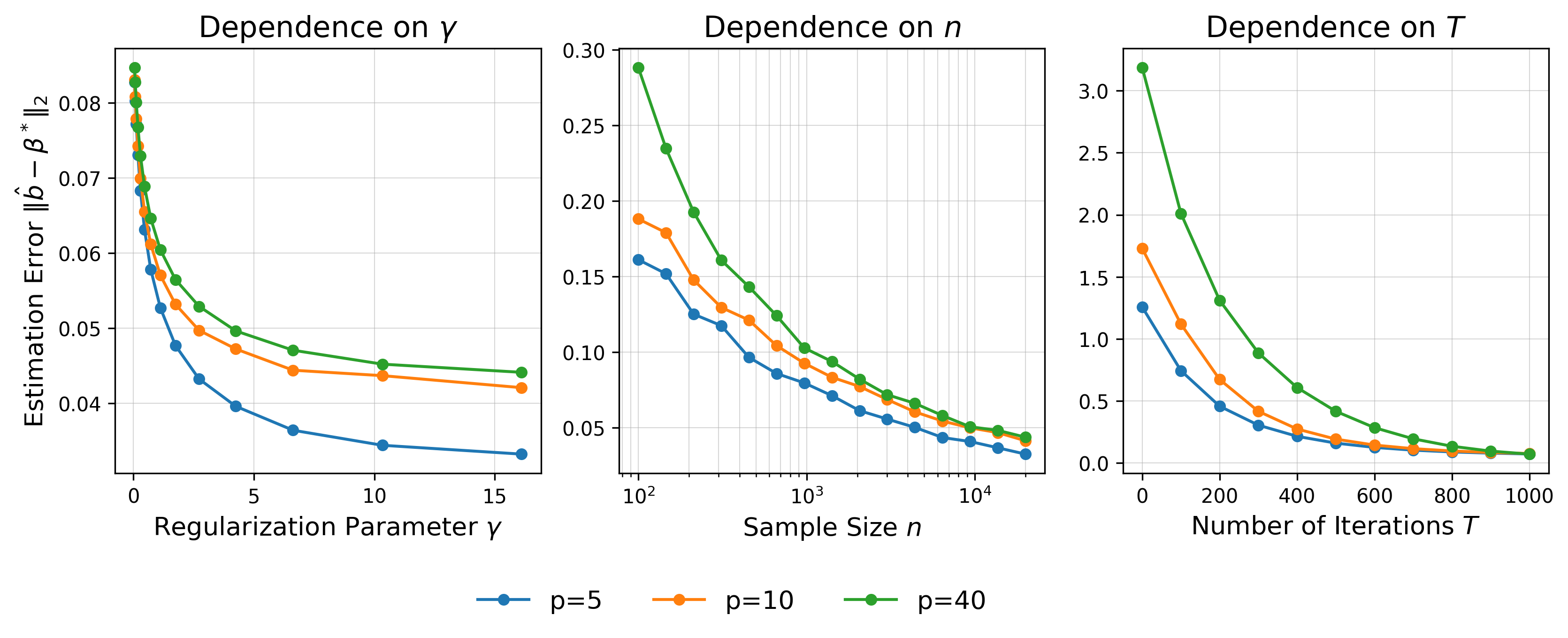}
    \caption{Dependence of $\|\hat{b} - \beta^*\|_2$ on the regularization parameter $\gamma$, the sample size $n$, and the number of iterations, where $\hat{b}$ is the output of Algorithm \ref{alg: GDmax penal}. 
    These results are averaged over 200 simulations.} 
    \label{fig: converge with gamma}
\end{figure}

\paragraph{Dependence on $\gamma, n, T$.}
Figure \ref{fig: converge with gamma} illustrates how the estimation error $\|\hat{b} - \beta^*\|_2$ varies with the regularization parameter $\gamma$, the sample size $n$, and the number of iterations $T$, where $\hat{b}$ denotes the output of Algorithm \ref{alg: GDmax penal}. The covariate dimension varies across $p \in \{5, 10, 40\}$.
In the leftmost panel, we fix the sample size at $n = 20{,}000$ and the number of iterations at $T = 1000$, while varying $\gamma$ within $[0, 17]$.
In the middle panel, $\gamma = 20$ and $T = 1000$ are fixed, and the sample size $n$ ranges from $100$ to $20{,}000$.
In the rightmost panel, we fix $n = 20{,}000$ and $\gamma = 20$, while varying $T$ from $0$ to $1000$.
The results collectively show that the proposed Algorithm \ref{alg: GDmax penal} converges toward the causal outcome model $\beta^*$ as $\gamma$, $n$, and $T$ increase, consistent with the theoretical guarantees established in Theorem \ref{thm: convergence theorem - penal - empirical - alg output}.

\paragraph{Scalability with Dimension $p$.} We compare our NegDRO with other causal invariance learning methods in terms of both estimation accuracy (measured as the $\ell_2$ distance to the causal effect $\beta^*$) and computational cost (measured in seconds), where the dimension $p$ is varied in the range of $[5,100]$. The methods included in the comparison are \texttt{ICP} \citep{peters2016causal}, \texttt{Anchor} \citep{rothenhausler2021anchor}, and \texttt{EILLS} \citep{fan2023environment}. We also include \texttt{ERM} method for comparative purposes, which pools all observed data across environments and fits a least-squares model. 
We set the regularization parameter $\gamma=20$ for our NegDRO. Following the recommendation in their original work, \texttt{EILLS} also uses a  hyperparameter $\gamma=20$. For other causal invariance learning approaches, we choose the best hyperparameters (if applicable), i.e., we enumerate a wide range of hyperparameter values and select the best one that minimizes $\ell_2$ distance to $\beta^*$. Detailed implementations for all methods are in Appendix \ref{appendix: subsec implementation}.

Figure \ref{fig: comparison with methods} summarizes the results regarding estimation error versus dimension and runtime versus dimension. 
To ensure timely execution, we impose a 30-minute time limit for each method; if a method does not finish within this limit, it terminates automatically.
In the left panel of Figure \ref{fig: comparison with methods}, among the evaluated methods, only our NegDRO closely approximates $\beta^*$, irrespective of the covariate dimension $p$. Notably, \texttt{ELLIS} performs exceptionally well in lower-dimensional settings when $p\leq 20$, yielding the smallest $\ell_2$ distance compared to all other methods. However, its computational time reaches the 30-minute time limit when $p\geq 25$, since its complexity scales exponentially with dimension due to enumeration. We also observe \texttt{ELLIS} deteriorates significantly in the presence of hidden confounders; see Appendix \ref{appendix: subsec implementation}.
The \texttt{ICP} method, which prioritizes selecting direct causes through hypothesis testing, demonstrates conservative behavior in this setting, resulting in poor accuracy for estimating $\beta^*$. Additionally, \texttt{Anchor} provides results nearly identical to those of \texttt{ERM}, and both have significant divergence from $\beta^*$.

The right panel of Figure \ref{fig: comparison with methods} depicts the average runtime for each method, with a black dashed horizontal line marking the 30-minute time limit. Methods exceeding this limit are terminated. It is evident that \texttt{ICP} and \texttt{EILLS}, which rely on exhaustive search enumeration, quickly exceed the time limit as the covariate dimension $p$ increases to $15$ and $25$, respectively. In contrast, all other methods, including our proposed NegDRO, demonstrate computational costs that scale polynomially with the dimension $p$. 

\begin{figure}[ht!]
    \centering
    \includegraphics[width=0.85\linewidth]{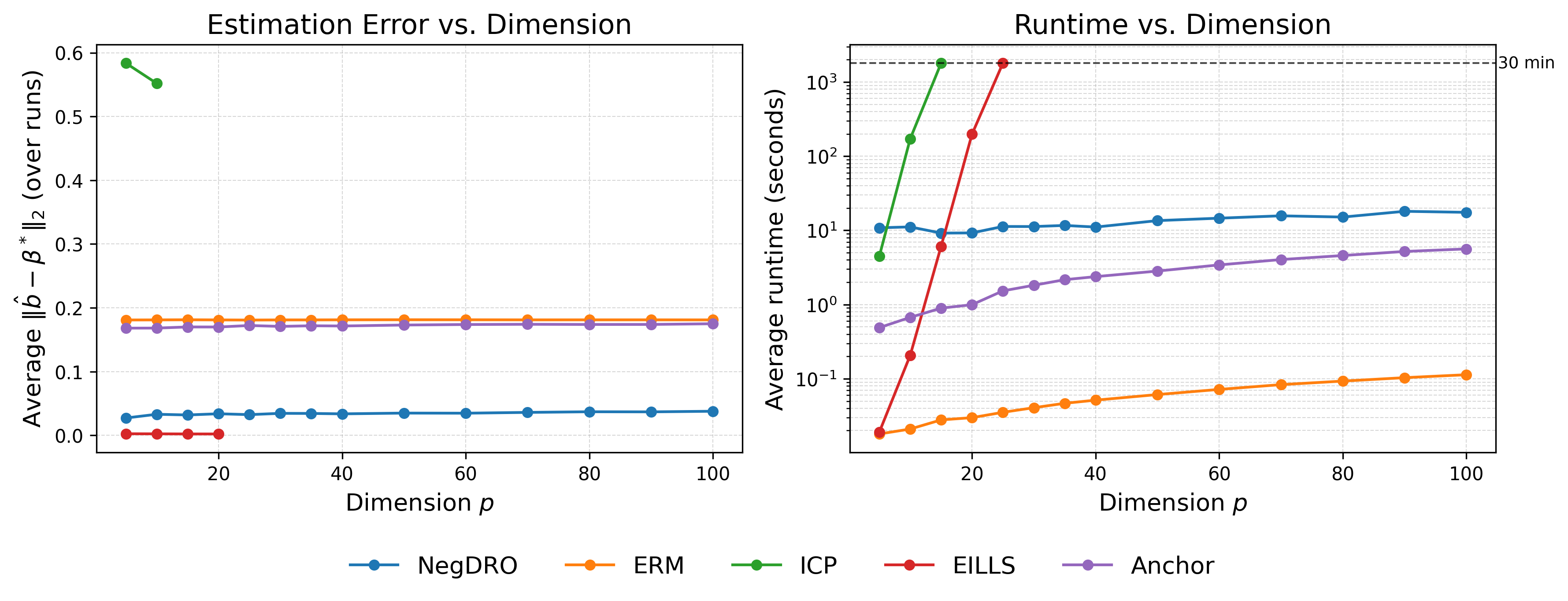}
    \caption{Comparison of different causal invariance learning methods. The dimension $p$ varies within the range of $[5, 100]$, while the sample size is fixed at $n = 20,000$. Results are averaged over 200 simulations.}
    \label{fig: comparison with methods}
\end{figure}

\section{Conclusion}
\label{sec: conclusion}
We propose the NegDRO formulation for causal invariance learning, which frames it as a nonconvex continuous minimax problem. Building on this formulation, our work addresses two fundamental questions: (i) under what conditions on environmental heterogeneity can we achieve causal identification, and (ii) how can $\beta^*$ be efficiently achieved despite the inherent nonconvexity of NegDRO. Unlike existing causal invariance learning approaches that rely on enumeration and thus face exponential computational costs in the covariate dimension $p$, our method scales mildly to higher dimensions. 

This work lies at the intersection of statistics and optimization. From the optimization perspective, the proposed NegDRO reformulates the statistical causal identification problem, transforming the combinatorial search for invariant subsets into a smooth minimax problem amenable to gradient-based methods. From a statistical perspective, the identification conditions reveal the geometric structure of the nonconvex optimization landscape, ensuring that the stationary points are statistically meaningful. Together, these insights demonstrate how optimization redefines and solves statistical causal inference, while statistical principles, in turn, endow optimization with an interpretable geometric structure.

\section*{Acknowledgment}
The authors extend their gratitude to Nicolai Meinshausen, Yihong Gu, and Menglong Li for their valuable discussions and insights. 
A large proportion of this research was conducted when Z. Guo was the associate professor at Rutgers University and when Y. Hu was the postdoc researcher at EPFL and ETH Zurich. During this period, the research of Z. Wang and Z. Guo was partly supported by the NSF grant DMS 2015373 and NIH grants R01GM140463 and R01LM013614, Y. Hu was supported by NCCR Automation, a National Centre of Competence in Research, funded by the Swiss National Science Foundation (grant number 51NF40\_225155).
Z. Guo also acknowledges financial support for visiting the Institute of Mathematical Research (FIM) at ETH Zurich. 

\bibliography{ref}
\bibliographystyle{abbrvnat}
\appendix

\newpage
\appendix
\clearpage
\addcontentsline{toc}{section}{Appendix} %
\begin{center}
    \textbf{\large Appendix to ``Causal Invariance Learning via Efficient Nonconvex Optimization''.}
\end{center}
\vspace{0.5em}
The appendix is structured as follows:
\begin{itemize}
    \item Appendix \ref{appendix: secA} provides the proofs of the main theoretical results presented in the main article.
    \item Appendix \ref{subsec: without penalty} analyzes the unpenalized NegDRO formulation, which is nonconvex and nonsmooth, providing results parallel to those in Section~\ref{sec: alg}.
    \item Appendix \ref{appendix: minimization helps} presents additional discussions and theoretical insights on the role of the optimization objective, complementing Section~\ref{sec: minimization} of the main text.
    \item Appendix \ref{appendix: secB} includes the full proofs of the results introduced in Appendices~\ref{appendix: secA}, \ref{subsec: without penalty}, and~\ref{appendix: minimization helps}.
    \item Appendix \ref{appendix: secD} contains the proofs of technical lemmas.
    \item Appendix~\ref{appendix: secE} offers additional illustrations of the structural equation models (SEMs) and further discussions on the simulation studies.
\end{itemize}
We provide a detailed Table of Contents for the Appendix below to facilitate navigation of the supplementary materials.

\setcounter{tocdepth}{2}  %
\startcontents[appendix]
\printcontents[appendix]{l}{1}{\setcounter{tocdepth}{2}}

\newpage

\section{Proofs of Main Results}
\label{appendix: secA}

This section contains the proofs of the theoretical results presented in the main article. It is outlined as follows:
\begin{itemize}
    \item \ref{subsec: appendix prepare} presents the expression of environmental risks within the additive intervention regime. It prepares for the theoretical analysis in later sections.
    \item \ref{subsec: appendix thm identification} - \ref{Appendix - proof - Lemma equiv ass} provide proofs of the theoretical results in Section \ref{sec: additive intervention}.
    \item \ref{appendix: proof of thm convergence theorem - penal - population - any b} - \ref{proof of thm: convergence theorem - penal - empirical - alg output} contain proofs of results within Sections \ref{subsec: stationary causal} and \ref{subsec: computation algorithm}.
    \item \ref{Appendix: proof of thm identification infty minimization - simplify} presents the proofs of results in Section \ref{sec: minimization}.
\end{itemize}

\subsection{Preparation}
\label{subsec: appendix prepare}
Under the additive intervention regime, we shall present the expressions of the population risk $\Eb[Y^e - b^\intercal X^e]^2$ and empirical risk $\widehat{\Eb}[Y^e - b^\intercal X^e]^2$, for a given predictor $b\in \Rb^p$.

We start with the population risk $\Eb[\Y{e}-b^\intercal\X{e}]^2$. The proof of the following lemma is provided in Appendix \ref{appendix: proof lemma: expression of risk}.
\begin{Lemma}
    Under the additive intervention, it holds that
    \begin{equation*}
    \Eb[\eps{e}_Y\X{e}] = -\Gbf^\intercal h, \quad \Eb[\X{e}X^{e\intercal}] = \Gbf^\intercal \left(\Hbf + \Eb[\del{e}\delT{e}]\right)\Gbf.
\end{equation*}
    Then, for a given predictor $b\in \Rb^p$, the population risk $\Eb[\Y{e}-b^\intercal \X{e}]^2$ is expressed as follows:
    {
    \[
    \Eb[\Y{e}-b^\intercal \X{e}]^2 = \sigma_Y^2 +2(\Gbf^\intercal h)^\intercal (b - \beta^*) + (b - \beta^*)^\intercal \left[\Gbf^\intercal(\Hbf+ \Eb[\delta^e\delta^{e\intercal}])\Gbf\right] (b - \beta^*),
    \]
    }
    where 
    {
    \begin{equation*}
    \begin{gathered}
        \Gbf^\intercal = \left[(\Ibf - \Bbf_{XX}) - \Bbf_{YX} (\beta^*)^\intercal\right]^{-1},\quad h = \sigma_Y^2\Bbf_{YX} - \Eb[\eta_Y\eta_X],\quad \Hbf = \begin{pmatrix}
    \Bbf_{YX}^\intercal \\ \Ibf
\end{pmatrix}^\intercal\Eb[\eta \eta^\intercal]\begin{pmatrix}
    \Bbf_{YX}^\intercal \\ \Ibf
\end{pmatrix}. 
    \end{gathered}
    \end{equation*}
    }
    Here, the matrix $\Gbf$ is a non-singular matrix, and $\Hbf\succeq 0.$
    \label{lemma: expression of risk}
\end{Lemma}
Through this lemma, we observe that under the additive intervention regime, the first-order term of the population risk is invariant across environments, i.e., $(\Gbf^\intercal h)^\intercal (b - \beta^*)$ is independent of $e$. Indeed, the only difference in environmental risks lies in the second-order term, which purely depends on the additive component $\Eb[\delta^e\delta^{e\intercal}]$.

Now we turn to studying the empirical environmental risk. For each environment $e\in \Ec$, we have observations $\{\x{e}_i, \y{e}_i\}_{i=1}^{n_e}$ i.i.d drawn from the distribution of $(\X{e}, \Y{e}).$ 
The following lemma presents the expression of $\widehat{\Eb}[\Y{e}-b^\intercal \X{e}]^2 = \frac{1}{n_e}\sum_{i=1}^{n_e}[y^e_i - b^\intercal x^e_i]^2$, for a given predictor $b\in \Rb^p$. The proof of the following lemma is provided in Appendix \ref{appendix: proof lemma: expression of empirical risk}.
\begin{Lemma}
    Under the additive intervention, the empirical risk $\widehat{\Eb}[\Y{e}-b^\intercal \X{e}]^2$, for a given predictor $b\in \Rb^p$, is expressed as follows:
    {
    \[
    \begin{aligned}
        \widehat{\Eb}[\Y{e}-b^\intercal \X{e}]^2 &= (\sigma_Y^2 + \hat{r}^{e}) + 2[\Gbf^\intercal(h - \hat{q}^{e})]^\intercal (b-\beta^*) \\
        &\quad\quad + (b-\beta^*)^\intercal \Gbf^\intercal \left(\Hbf + \Eb[\del{e}\delT{e}] + \hat{\Pbf}^{e}\right)\Gbf (b-\beta^*),
    \end{aligned}
    \]
    }
    where $h, \Gbf^\intercal, \Hbf$ are defined in Lemma \ref{lemma: expression of risk}, and $\hat{r}^{e},\hat{q}^{e},\hat{\Pbf}^{e}$ are terms capturing the finite-sample error, with expressions given in the following \eqref{eq: notations-empirical}.
    \label{lemma: expression of empirical risk}
\end{Lemma}

Notice that the terms $\hat{r}^{e},\hat{q}^{e},\hat{\Pbf}^{e}$ arise from the finite sample error, and thus shall be controlled with high probability, as demonstrated in the following Lemma. The proof is provided in Appendix \ref{subsec: proof of lemma rate of empirical norms}.
\begin{Lemma}
    Under the additive interventions, suppose Condition \ref{cond: subgaussian} holds. Then for any $s\geq 0$, with probability at least $1-2|\Ec|e^{-s}$, we have
    \[
    \max_{e\in \Ec} \max\{\|\hat{\Pbf}^{e}\|_2, \|\hat{q}^{e}\|_2, |\hat{r}^{e}|\} \lesssim {\sqrt{\frac{p+s}{{n}}}}+\frac{p+s}{{n}},
    \]
    where ${n} = \min_{e\in \Ec}n_e$ is the smallest sample size across all environments.
    \label{Lemma: rate of empirical norms}
\end{Lemma}

\subsection{Proof of Theorem \ref{thm: identification infty}}
\label{subsec: appendix thm identification}
We focus on the risk-invariant predictor $b\in \Rb^p$ satisfying $\Eb[\Y{e}-b^\intercal\X{e}]^2 = \Eb[\Y{f}-b^\intercal\X{f}]^2$ for any $e,f\in \Ec$. Then it follows from Lemma \ref{lemma: expression of risk} that, the risk-invariant predictor $b$ satisfies that: for any $e,f\in \Ec$, 
\[
(b-\beta^*)^\intercal \Eb[\X{e}X^{e\intercal}] (b-\beta^*) = (b-\beta^*)^\intercal \Eb[\X{f}X^{f\intercal}] (b-\beta^*).
\]
Since the above equation holds for any two environments $e,f\in \Ec$, the following equation must hold for any two collections of environments $\Ec_1,\Ec_2\subseteq \Ec$, for any weights $w\in \Delta^{|\Ec|_1}$ and $w'\in \Ec^{|\Ec_2|}$:
{\small
\begin{equation}
    (b-\beta^*)^\intercal \left[\sum_{e\in \Ec_1}w_e \Eb[\X{e}X^{e\intercal}]\right](b-\beta^*) = (b-\beta^*)^\intercal \left[\sum_{f\in \Ec_2}w_f' \Eb[\X{f}X^{f\intercal}]\right](b-\beta^*).
    \label{eq: invariance conclusion}
\end{equation}
}

By Condition \ref{cond: strict positive}, there exist two nonempty and disjoint collections of environments $\Ec_1,\Ec_2\subseteq \Ec$, and some weights $w\in \Delta^{|\Ec|_1}$ and $w'\in \Delta^{|\Ec|_2}$ such that
\[
\sum_{e\in \Ec_1}w_e \Eb[\X{e}X^{e\intercal}] \succ \sum_{f\in \Ec_2}w_f' \Eb[\X{f}X^{f\intercal}].
\]
If $b \neq \beta^*$, then it holds that
{\small
\[
(b-\beta^*)^\intercal \left[\sum_{e\in \Ec_1}w_e \Eb[\X{e}X^{e\intercal}]\right](b-\beta^*) > (b-\beta^*)^\intercal \left[\sum_{f\in \Ec_2}w_f' \Eb[\X{f}X^{f\intercal}]\right](b-\beta^*).
\]
}
which contradicts \eqref{eq: invariance conclusion}. Therefore, if $b\in \Rb^p$ is a risk-invariant predictor satisfying $\Eb[\Y{e}-b^\intercal\X{e}]^2 = \Eb[\Y{f}-b^\intercal\X{f}]^2$ for all $e,f\in \Ec$, then it must hold that $b=\beta^*$. Therefore,
\[
\Bc_{\rm inv} = \left\{b\in \Rb^p: \Eb[\Y{e}-b^\intercal\X{e}]^2 = \Eb[\Y{f}-b^\intercal\X{f}]^2,\;\forall e,f\in \Ec\right\} = \{\beta^*\}.
\]

\subsection{Proof of Theorem \ref{thm: necessary condition}}
\label{subsec: proof of necessary}

We prove that if at most one coordinate of the covariate noise vector is intervened in each environment and the unique risk-invariant predictor is $\beta^*$, then Condition \ref{cond: strict positive} necessarily holds.

\vspace{1em}
\noindent\textbf{Step-1: Representation of the intervention structure.}

By assumption, in each environment $e\in \Ec$, at most one coordinate of noise vector $\delta^e\in \Rb^p$ is intervened, while all other remain unchanged:
\[
\left|\{j\in [p]:\delta^{e}_j\equiv 0\}\right|\geq p-1.
\]
Consequently, the second-moment matrix of the intervention has the form:
\begin{equation}
    \Eb[\del{e}\delT{e}] = {\rm diag}(0,..., 0, c^{e}, 0, ..., 0),
    \label{eq: proof neccess thm intervention matrix}
\end{equation}
where the scalar $c^e\geq 0$ denotes the intervention strength at coordinate $s^e\in [p]$, the index of the intervened variable in environment $e$.
Let
\begin{equation}
    S_\Ec = \bigcup_{e\in \Ec}s^{e}, \quad { C}_\Ec = \left\{c^{e}\right\}_{e\in \Ec}.
    \label{eq: proof neccess thm intervention set}
\end{equation}
collect all intervened coordinates and their corresponding strengths.

\vspace{1em}
\noindent\textbf{Step-2: Characterizing risk invariance.}

For any risk-invariant predictor $b\in \Bc_{\rm inv}$, Lemma \ref{lemma: expression of risk} implies that for every pair of environments $e,f\in \Ec$:
\[
(b-\beta^*)^\intercal \Gbf^\intercal \Eb[\delta^{e}\delta^{e\intercal}]\Gbf (b-\beta^*) = (b-\beta^*)^\intercal \Gbf^\intercal \Eb[\delta^{f}\delta^{f\intercal}]\Gbf (b-\beta^*).
\]
Provided $\Bc_{\rm inv}=\{\beta^*\}$, i.e. $\beta^*$ is the unique risk-invariant predictor, we know that for any $b\neq \beta^*$, the above equation does not hold. Since $\Gbf$ is a full rank-matrix as defined in Lemma \ref{lemma: expression of risk}, this equivalently means that:
\[
\{0_p\} = \left\{u\in \Rb^p:~~ u^\intercal\Eb[\del{e}\delT{e}]u = u^\intercal\Eb[\del{f}\delT{f}]u,~~\textrm{for all $e,f\in \Ec$}\right\}.
\]
In words, the zero vector is the only $u$ whose quadratic forms are equal across all environments.

Using \eqref{eq: proof neccess thm intervention matrix}, this condition reduces to
\begin{equation}
    \{0_p\} = \left\{u\in \Rb^p:~~c^{e} [u_{s^{e}}]^2 = c^{f} [u_{s^{f}}]^2,\quad \textrm{for all $e,f\in \Ec$}\right\}.
    \label{eq: proof neccess thm intervention equality}
\end{equation}
In the remaining proofs, we will leverage \eqref{eq: proof neccess thm intervention equality} to construct two collections of environments $\Ec_1,\Ec_2$ required by Condition \ref{cond: strict positive}.

\vspace{1em}
\noindent\textbf{Step-3: Show that all coordinates must be intervened.}

Suppose that some coordinates have never been intervened, i.e. $S_\Ec\subsetneq [p]$. Define $u\in \Rb^p$ by
\[
\begin{cases}
    u_j = 0 &\textrm{if $j\in S_\Ec$} \\
    u_j = 1 & \textrm{if $j\in S_\Ec^\complement$}
\end{cases}.
\]
Then $u\neq 0_p$ and for all $e,f\in \Ec$,
\[
c^{e} [u_{s^{e}}]^2 = c^{f} [u_{s^{f}}]^2 = 0,
\]
contradicting \eqref{eq: proof neccess thm intervention equality}. Hence, every coordinate of the covariates must be intervened in at least one environment, such that 
\begin{equation}
    S_\Ec = [p].
    \label{eq: perturb all covariates}
\end{equation}
We now distinguish two situations:
\begin{itemize}
    \item[\textbf{(Case-1)}] at least one environment is unintervened.
    \item[\textbf{(Case-2)}] every environment is intervened (but possibly on different coordinates or scales).
\end{itemize}

\vspace{1em}
\noindent\textbf{Step-4: Construction for Case 1.}

Without loss of generality, let environment $1$ be unintervened such that $\del{1} \equiv 0_p$. Since $S_\Ec = [p]$ by \eqref{eq: perturb all covariates}, the remaining environments $\{2,...,|\Ec|\}$ collectively perturb all coordinates
\[
\bigcup_{e=2}^{|\Ec|} s^{e} = [p].
\]
Therefore,
\[
\sum_{e\in \{2,3,...,|\Ec|\}} \frac{1}{|\Ec|-1} \Eb[\del{e}\delT{e}]\succ \Eb[\del{1}\delT{1}]=0.
\]
Because $\Eb[X^eX^{e\intercal}] = \Gbf^\intercal (\Hbf + \Eb[\del{e}\delT{e}])\Gbf$, it follows that
\[
\sum_{e\in \{2,3,...,|\Ec|\}} \frac{1}{|\Ec|-1} \Eb[\X{e}X^{e\intercal}]\succ \Eb[X^1X^{1\intercal}].
\]
Hence, by letting 
\[
\Ec_1 = \{2,3,...,|\Ec|\},\quad \Ec_2 = \{1\},
\]
Condition \ref{cond: strict positive} is satisfied.

\vspace{1em}
\noindent\textbf{Step-5: Construction for Case 2.}

\vspace{0.5em}
\noindent\textit{Step-5(a): Establishing a key claim.}

We shall show that even if every environment intervenes some coordinate, there must exist two environments perturbing the same coordinate with different magnitudes. Specifically, we claim that if \eqref{eq: proof neccess thm intervention equality} holds, then there exists two environments $e,f\in \Ec$ and a coordinate $s\in [p]$ such that
\begin{equation}
    \exists e,f \in \Ec, \; {\rm s.t.}\; s^{e} = s^{f} = s~~\textrm{for some $s\in [p]$}, ~~\textrm{and}~~c^{e} \neq c^{f}.
    \label{eq: claim}
\end{equation}
We prove this by contradiction by considering the following two scenarios.
\begin{itemize}
    \item Scenario 1: suppose $s^e\neq s^f$ for all pairs $e\neq f$. Then define $u_{s^e} = 1/\sqrt{c^e}$ for each $e\in \Ec$. This nonzero $u$ satisfies $c^e[u_{s^e}]^2 = 1$ for all $e$, contradicting \eqref{eq: proof neccess thm intervention equality}. Thus this scenario is impossible.
    \item Scenario 2: suppose some coordinates are shared among environments but with identical strength, i.e. whenever $s^e=s^f$, we also have $c^e = c^f$. Such a pair of environments has no heterogeneity and can be aggregated, reducing the problem to Scenario 1, which we have already ruled out.
\end{itemize}
Combining these two scenarios, we conclude that \eqref{eq: claim} must hold.

\vspace{0.5em}
\noindent\textit{Step-5(b): Constructing two collections $\Ec_1,\Ec_2$.}

By \eqref{eq: claim}, there exist $e,f\in \Ec$ intervening the same coordinate but with different scales. 
Without loss of generality, we consider $s^e=s^f=1$ with $0<c^e< c^f$. Thus
\[
\Eb[\del{e}\delT{e}] = {\rm diag}(c^{e}, 0_{p-1}), \quad \textrm{and}\quad
\Eb[\del{f}\delT{f}] = {\rm diag}(c^{f}, 0_{p-1}).
\]
Let $\Ec_2 = \{e\}$ and $\Ec_1 = \Ec\setminus \{e\}$. 
Construct the weight vector $w = (w_g)_{g\in \Ec_1}\in \Delta^{|\Ec|-1}$ as
\[
w_f= \frac{t c^{e}}{c^{f}},\quad w_g=\left(\frac{c^{f} - t c^{e}}{(|\Ec|-2)c^{f}} \right)_{|\Ec|-2} \;\;\textrm{for } g\in \Ec_1\setminus \{f\}.
\]
where $t \in (1, \frac{c^{f}}{c^{e}}).$ Then 
\begin{equation}
    \begin{aligned}
    \sum_{g\in \Ec_1}w_g \Eb[\del{g}\delT{g}] &= \frac{t c^{e}}{c^{f}} \Eb[\del{f}\delT{f}] + \frac{c^{f} - t c^{e}}{(|\Ec|-2)c^{f}} \sum_{g\in \Ec_1\setminus\{f\}}\Eb[\del{g}\delT{g}] \\
    & = \frac{t c^{e}}{c^{f}} {\rm diag}(c^{f}, 0_{p-1}) + \frac{c^{f} - t c^{e}}{(|\Ec|-2)c^{f}} \sum_{g\in \Ec_1\setminus\{f\}}\Eb[\del{g}\delT{g}]\\
    &\succ {\rm diag}(c^{e}, 0_{p-1}) = \Eb[\del{e}\delT{e}].
\end{aligned}
\label{eq: proof neccess thm intervention dominance - LHS}
\end{equation}
where the inequality holds because $t>1$ increases the first diagonal and, by \eqref{eq: perturb all covariates}, the remaining environments perturb all other coordinates.

Consequently, the dominance relation in Condition \ref{cond: strict positive} holds with
\[
\sum_{g \in \Ec\setminus\{e\}}w_g\Eb[\X{g}\XT{g}] \succ \Eb[\X{e}\XT{e}].
\]

\vspace{1em}
\noindent\textbf{Step-6: Conclusion.}

Both Case 1 and Case 2 lead to the existence of two disjoint, nonempty environment collections $\Ec_1,\Ec_2\subset \Ec$ such that
\[
\sum_{e\in \Ec_1}w_e\Eb[\X{e}\XT{e}]\succ \sum_{f\in \Ec_2}w_f\Eb[\X{f}\XT{f}],
\]
which verifies Condition \ref{cond: strict positive}.

\subsection{Proof of Theorem \ref{thm: optimization perspective}}
\label{Appendix: proof of thm: optimization perspective}
It follows from Lemma \ref{lemma: expression of risk} that the constrained optimization problem in \eqref{eq: bneg gamma infty} is equivalently expressed as:
\[
\min_{b\in \Rb^p} \Eb[Y^1 - b^\intercal X^1]^2, \quad \textrm{s.t.}\;\; (b-\beta^*)^\intercal \Abm(w) (b-\beta^*)=0, \;\forall w\in \Delta^L.
\]
If Condition \ref{cond: strict positive - A} holds, then the $\beta^*$ is the only feasible solution satisfying $(b-\beta^*)^\intercal \Abm(w) (b-\beta^*)=0$ for all $w\in \Delta^L$. Therefore, it holds that the constrained optimization problem recovers $\beta^*$ exactly.

\subsection{Proof of Proposition \ref{prop: equiv ass}}
\label{Appendix - proof - Lemma equiv ass}

First, we show that Condition \ref{cond: strict positive - A} implies Condition \ref{cond: strict positive}. If Condition \ref{cond: strict positive - A} holds, then it holds that
\begin{equation}
    \Abm(w^0) = \sum_{l\in \Ec} \left(w_l^0 - \frac{1}{|\Ec|}\right) \Eb[\X{e}X^{e\intercal}] = \sum_{l\in \Ec}\left[(w_l^0 - \frac{1}{|\Ec|})_+ - (w_l^0 - \frac{1}{|\Ec|})_-\right] \Eb[\X{e}X^{e\intercal}] \succ 0,
    \label{proof lemma equiv - 1}
\end{equation}
where we use $x_+ = \max(x, 0)$ and $x_- = \max(-x, 0)$ to denote the positive part and negative part for a real number $x$.
Since $w^0\in \Delta^{|\Ec|}$, it must hold that $\sum_{l\in \Ec} (w_l^0 - \frac{1}{|\Ec|}) = 0$. Thus,
\begin{equation}
    \sum_{l\in \Ec}(w_l^0 - \frac{1}{|\Ec|})_+ = \sum_{l\in \Ec}(w_l^0 - \frac{1}{|\Ec|})_- = C,
    \label{proof lemma equiv - 2}
\end{equation}
for some constant $C>0.$
Subsequently, we define two collections of environments $\Gc, \Gc^\complement \subseteq \Ec$: 
$$\Gc = \{e\in \Ec: w_e^0> \frac{1}{|\Ec|}\},\quad \textrm{and}\quad \Gc^\complement = \{e\in \Ec: w_e^0 \leq \frac{1}{|\Ec|}\}.$$
It follows from \eqref{proof lemma equiv - 1} and \eqref{proof lemma equiv - 2} that
\[
\sum_{e\in \Gc} \frac{w_e^0 - 1/|\Ec|}{C}\Eb[\X{e}X^{e\intercal}] \succ \sum_{f\in \Gc^\complement}\frac{1/|\Ec| - w_f^0}{C}\Eb[\X{f}X^{f\intercal}].
\]
We let $\Ec_1 = \Gc$, and $\Ec_2 = \Gc^\complement$,
then the previous equation shows that:
\[
\sum_{e\in \Ec_1} w_e\Eb[\X{e}X^{e\intercal}] \succ \sum_{f\in \Ec_2} w^\prime_f\Eb[\X{f}X^{f\intercal}],
\]
where $w_e = \frac{w_e^0 - 1/|\Ec|}{C}$, and $w^\prime_f = \frac{1/|\Ec| - w_f^0}{C}.$ And it holds that $w\in \Delta^{|\Ec_1|}$ and ${w^\prime}\in \Delta^{|\Ec_2|}.$ Therefore, we show Condition \ref{cond: strict positive} holds.

Next, we prove that Condition \ref{cond: strict positive} implies Condition \ref{cond: strict positive - A}.
Given two weight vectors $w\in \Delta^{|\Ec_1|}$ and $w'\in \Delta^{|\Ec_2|}$ satisfying Condition \ref{cond: strict positive}, we define the constant $C = \min\left\{\frac{1-1/|\Ec|}{\max_{e\in \Ec_1}  w_e}, \frac{1/|\Ec|}{\max_{f\in \Ec_2}  w^\prime_f}\right\}$. Clearly, $C>0$. Next, we shall construct a weight $w^0\in \Delta^{|\Ec|}$ satisfying Condition \ref{cond: strict positive - A}.
\begin{itemize}
    \item For environment $e\in \Ec_1$, we let $w^0_e = Cw_e + 1/|\Ec|$;
    \item For environment $f\in \Ec_2$, we let $w^0_f = 1/|\Ec| - C {w^\prime_f};$
    \item For environment $g\notin \Ec_1\cup \Ec_2$, we let $w^0_g = 1/|\Ec|$.
\end{itemize}
It is evident that any $e\in \Ec$, $w_e^0\in [0,1]$ and $\sum_{e\in \Ec} w_e^0 = 1$, which implies that $w^0\in \Delta^{|\Ec|}$.
By Condition \ref{cond: strict positive}, we have
\[
\sum_{e\in \Ec_1} w_e \Eb[\X{e}X^{e\intercal}] \succ \sum_{f\in \Ec_2} {w^\prime_f}\Eb[\X{f}X^{f\intercal}].
\]
We plug in the constructed $w^0$, and it follows that
\[
\sum_{e\in \Ec_1} \left(\frac{w_e^0 - 1/|\Ec|}{C}\right) \Eb[\X{e}X^{e\intercal}] \succ \sum_{f\in \Ec_2} \left(\frac{1/|\Ec| - w_f^0}{C}\right)\Eb[\X{f}X^{f\intercal}].
\]
Multiplying the constant $C$ on both sides, and adding the components regarding $g\notin \Ec_1\cup \Ec_2$, we obtain that
{\small
\[
\sum_{e\in \Ec_1} \left(\frac{w_e^0 - 1/|\Ec|}{C}\right) \Eb[\X{e}X^{e\intercal}] + \sum_{f\in \Ec_2} \left(\frac{w_f^0 - 1/|\Ec|}{C}\right) \Eb[\X{f}X^{f\intercal}] + \sum_{g\notin \Ec_1\cup\Ec_2} \left(\frac{w_g^0 - 1/|\Ec|}{C}\right) \Eb[\X{g}X^{(g)\intercal}] \succ 0,
\]
}
which implies that
\[
\Abm(w^0) = \sum_{e\in \Ec}(w_e^ 0 -\frac{1}{|\Ec|})\Eb[\X{e}X^{e\intercal}]\succ 0.
\]
Thus, Condition \ref{cond: strict positive - A} is met.

\subsection{Proof of Theorem \ref{thm: convergence theorem - penal - population - any b}}
\label{appendix: proof of thm convergence theorem - penal - population - any b}
Although Section \ref{sec: main idea of proof} provides the main idea of the proof, this section presents the details for the proof of Theorem \ref{thm: convergence theorem - penal - population - any b}.
We first provide the expressions of the objective $\Phi_\mu(b)$ in \eqref{eq: obj trans - penal - population}.
It follows from Lemma \ref{lemma: expression of risk} that
{\small
\begin{equation}
    \begin{aligned}
   &\sum_{e\in \Ec}\left(w_e - \frac{\gamma}{1+\gamma |\Ec|}\right) \Eb[\Y{e}-b^\intercal\X{e}]^2 \\
   &=\sum_{e\in \Ec}\left(w_e - \frac{\gamma}{1+\gamma |\Ec|}\right)\left(\sigma_Y^2 - 2(b - \beta^*)^\intercal \Eb[\eps{e}_Y\X{e}] + (b - \beta^*)^\intercal \Eb[\X{e}(\X{e})^\intercal] (b - \beta^*)\right) \\
   &= \sum_{e\in \Ec}\left(w_e - \frac{\gamma}{1+\gamma |\Ec|}\right)\left(\sigma_Y^2 + 2 h^\intercal \Gbf(b-\beta^*) + (b - \beta^*)^\intercal \Eb[\X{e}(\X{e})^\intercal] (b - \beta^*)\right)\\
   &= \frac{\sigma_Y^2 + 2h^\intercal \Gbf(b-\beta^*) + (b-\beta^*)^\intercal \left[\frac{1}{|\Ec|}\sum_{e\in \Ec}\Eb[\X{e}X^{e\intercal}]\right] (b-\beta^*)}{1+\gamma|\Ec|}  + (b-\beta^*)^\intercal \Abm(w) (b - \beta^*),
\end{aligned}
\label{eq: expression of weighted average risks}
\end{equation}
}
where the vector $h\in \Rb^p$ and the full-rank matrix $\Gbf\in \Rb^{p\times p}$ are specified in Lemma \ref{lemma: expression of risk}, and the last equality holds due to $w\in \Delta^{|\Ec|}.$
Provided the maximizer $\Bar{w}$ defined in \eqref{eq: maximizer w - penal - population}, we establish that
{\small
\begin{equation}
\begin{aligned}
    \Phi_\mu(b) &= \sum_{e\in \Ec}\left(\Bar{w}_e - \frac{\gamma}{1+\gamma |\Ec|}\right) \Eb[\Y{e}-b^\intercal\X{e}]^2 - \mu \|\Bar{w}\|_2^2 \\
    &=  \frac{\sigma_Y^2 + 2h^\intercal \Gbf(b-\beta^*) + (b-\beta^*)^\intercal \left[\frac{1}{|\Ec|}\sum_{e\in \Ec}\Eb[\X{e}X^{e\intercal}]\right] (b-\beta^*)}{1+\gamma|\Ec|}  \\
   &\quad\quad + (b-\beta^*)^\intercal  \Abm(\Bar{w}) 0(b - \beta^*) - \mu\|\Bar{w}\|_2^2.
\end{aligned}
\label{proof: eq phi_mu b}
\end{equation}
}

\noindent\textbf{Step-1:} Express $\Phi_\mu(b)$ through its gradient $\nabla\Phi_\mu(b)$.

By Danskin' theorem, $\Phi_\mu(b)$ is differentiable with its gradient $\nabla\Phi_\mu(b)$ given by:
{\small
\begin{equation*}
    \begin{aligned}
        \nabla \Phi_\mu(b) = \frac{2 \Gbf^\intercal h + 2 \left[\frac{1}{|\Ec|}\sum_{e\in \Ec}\Eb[\X{e}X^{e\intercal}]\right] (b-\beta^*)}{1+\gamma|\Ec|} + 2 \Abm(\Bar{w}) (b - \beta^*).
    \end{aligned}
\end{equation*}
}
Left multiplying the vector $\frac{1}{2}(b - \beta^*)$, we establish that
{\small
\begin{equation*}
\begin{aligned}
    \frac{1}{2}(b-\beta^*)^\intercal \nabla \Phi_\mu(b) &=\frac{h^\intercal \Gbf(b - \beta^*) + (b-\beta^*)^\intercal \left[\frac{1}{|\Ec|}\sum_{e\in \Ec}\Eb[\X{e}X^{e\intercal}]\right] (b-\beta^*)}{1+\gamma|\Ec|} + (b-\beta^*)^\intercal\Abm(\Bar{w})(b - \beta^*). 
\end{aligned}
\end{equation*}
}
Combining the above equality with \eqref{proof: eq phi_mu b}, we establish that
{\small
\begin{equation}
    \Phi_\mu(b) = \frac{1}{2}(b-\beta^*)^\intercal \nabla \Phi_\mu(b) + \frac{\sigma_Y^2 + h^\intercal \Gbf(b-\beta^*)}{1+\gamma|\Ec|} - \mu\|\bar{w}\|_2^2.
    \label{proof: eq phi_mu b grad}
\end{equation}
}

\noindent\textbf{Step-2:} Establish the lower bound of $\Phi_\mu(b)$ via the weight $w^0$ specified in Condition \ref{cond: strict positive - A}. 

Since $\bar{w}$ is the maximizer weight as defined in \eqref{eq: maximizer w - penal - population}, we have
{\small
\begin{equation*}
\begin{aligned}
    \Phi_\mu(b) &\geq \sum_{e\in \Ec}\left(w^0 - \frac{\gamma}{1+\gamma |\Ec|}\right) \Eb[\Y{e}-b^\intercal\X{e}]^2 - \mu \|w^0\|_2^2 \\
    &= \frac{\sigma_Y^2 + 2h^\intercal \Gbf(b-\beta^*) + (b-\beta^*)^\intercal \left[\frac{1}{|\Ec|}\sum_{e\in \Ec}\Eb[\X{e}X^{e\intercal}]\right] \Gbf (b-\beta^*)}{1+\gamma|\Ec|} \\
   &\quad\quad + (b-\beta^*)^\intercal \Abm(w^0) (b - \beta^*) - \mu\|w^0\|_2^2.
\end{aligned}
\end{equation*}
}
Since $\Eb[\X{e}X^{e\intercal}]\succeq 0$ for all $e\in \Ec$, and $\Abm(w^0)\succeq \lambda\Ibf$ as specified in Condition \ref{cond: strict positive - A}, we further establish that
{\small
\begin{equation}
    \Phi_\mu(b)\geq \frac{\sigma_Y^2 + 2h^\intercal \Gbf(b-\beta^*)}{1+\gamma|\Ec|} + \lambda \|b - \beta^*\|_2^2 - \mu\|w^0\|_2^2 
    \label{proof: eq phi_mu b lower w^0}
\end{equation}
}

\noindent\textbf{Step-3:} Combine the results in the steps 1 and 2. It follows from \eqref{proof: eq phi_mu b grad} and \eqref{proof: eq phi_mu b lower w^0} that
{\small
\[
\frac{1}{2}(b-\beta^*)^\intercal \nabla \Phi_\mu(b) + \frac{\sigma_Y^2 + h^\intercal \Gbf(b-\beta^*)}{1+\gamma|\Ec|} - \mu\|\bar{w}\|_2^2 \geq \frac{\sigma_Y^2 + 2h^\intercal \Gbf(b-\beta^*)}{1+\gamma|\Ec|} + \lambda \|b - \beta^*\|_2^2 - \mu\|w^0\|_2^2.
\]
}
By reorganizing and applying Chebyshev's inequality, we shall establish that
\[
\lambda \|b-\beta^*\|_2^2 \leq \left(\frac{\|\Gbf\|_2 \|h\|_2}{1+\gamma|\Ec|} + \frac{1}{2}\left\|\nabla\Phi_\mu(b)\right\|_2\right) \|b-\beta^*\|_2 + \mu\|w^0\|_2^2 - \mu\|\bar{w}\|_2^2. 
\]
Since $w^0, \Bar{w}\in \Delta^{|\Ec|}$, it follows that
\[
\lambda\|b - \beta^*\|_2^2 \lesssim \left(\frac{1}{1+\gamma|\Ec|} + \left\|\nabla\Phi_\mu(b)\right\|_2\right) \|b -\beta^*\|_2 + \mu.
\]
It further implies that
\[
\|b - \beta^*\|_2 \lesssim \frac{1}{\lambda}\left(\frac{1}{1+\gamma|\Ec|} + \left\|\nabla\Phi_\mu(b)\right\|_2\right) + C\sqrt{\frac{\mu}{\lambda}}.
\]
The proof is complete.

\subsection{Proof of Theorem \ref{thm: convergence theorem - penal - empirical - alg output}}
\label{proof of thm: convergence theorem - penal - empirical - alg output}

The proof proceeds as follows.
\begin{itemize}
    \item We first establish the landscape of the empirical NegDRO, where we upper bound $\|b-\beta^*\|_2$ through the gradient norm $\|\nabla\widehat{\Phi}_\mu(b)\|_2$ for any predictor $b\in \Rb^p$.
    \item Second, we study the optimization convergence of Algorithm \ref{alg: GDmax penal}. Specifically, for the output $\hat{b}^\gamma$, we shall control the gradient norm $\|\nabla \widehat{\Phi}_\mu(\hat{b}^\gamma)\|_2$.
    \item Lastly, we combine these two results to complete the proof.
\end{itemize}

First, we introduce the following theorem to construct a finite-sample bound for $\|b-\beta^*\|_2$, for any predictor $b\in \Rb^p$. The proof is provided in Section \ref{appendix: proof of thm: convergence theorem - penal - empirical - any b}.
\begin{Theorem}[Landscape of Empirical NegDRO]
    Under the additive intervention regime, suppose Conditions \ref{cond: strict positive - A}, \ref{cond: subgaussian} hold and $n\geq (1\vee c_1/\lambda^2) p$. Then, for any $0\leq s\leq (1\wedge c_2\lambda^2) n - p$, with a probability of at least $1-2|\Ec| e^{-s}$, the following inequality holds uniformly for all predictors $b\in \Rb^p$
    \begin{equation}
        \left\|b - \beta^*\right\|_2 \lesssim \frac{1}{\lambda}\left(\frac{1}{1+\gamma|\Ec|} + \|\nabla \widehat{\Phi}_\mu(b)\|_2\right) +  \sqrt{\frac{\mu}{\lambda}} +\frac{1}{\sqrt{\lambda}}\left(\frac{p+s}{{n}}\right)^{1/4},
        \label{eq: upper bound - penal - empirical - any b}
    \end{equation}
    where $\lambda>0$ denotes the degree of environmental heterogeneity defined in Condition \ref{cond: strict positive - A}, and $c_1,c_2$ are some absolute constants.
    \label{thm: convergence theorem - penal - empirical - any b}
\end{Theorem}

Compared to the population-level result established in Theorem \ref{thm: convergence theorem - penal - population - any b}, the roles of the parameters $\gamma\geq 0$ and $\mu>0$ remain consistent. The only difference lies in the introduction of an additional finite-sample error term of order $\mathcal{O}(n^{-1/4})$ within the upper bound in \eqref{eq: upper bound - penal - empirical - any b}. 

Second, we show that Algorithm \ref{alg: GDmax penal} converges to a stationary point of $\widehat{\Phi}_\mu(\cdot)$ with an increasing number of iterations. The proof of the following proposition is provided in Section \ref{appendix: proof of prop: alg conv penal}.
\begin{Proposition}
\label{prop: alg conv penal}
    Suppose Condition \ref{cond: smooth} holds. Then $\widehat{\Phi}_\mu(b)$ is $L_{\widehat{\Phi}}$-smooth in $b$, with
    \[
    L_{\widehat{\Phi}} = 2\kappa_1+2\kappa_2\sqrt{|\Ec|} + 2\mu + \frac{(2\kappa_1+2\kappa_2\sqrt{|\Ec|} + 2\mu)^2}{2\mu}.
    \]
    By setting the step size in Algorithm \ref{alg: GDmax penal} as $\alpha = 1/L_{\widehat{\Phi}}\asymp (\mu + (\kappa_1+\kappa_2\sqrt{|\Ec|})/\mu)^{-1}$, the output $\hat{b}^\gamma$ of Algorithm \ref{alg: GDmax penal}  satisfies:
    \[
    \|\nabla\widehat{\Phi}_\mu(\hat{b}^\gamma)\|_2 \leq \sqrt{\frac{1}{T}\sum_{t=0}^{T-1}\left\|\nabla\widehat{\Phi}_\mu(b_t)\right\|_2^2} \lesssim \sqrt{\frac{1}{T}\left(\frac{\kappa_1^2+\kappa_2^2|\Ec|}{\mu} + \mu\right)}.
    \]
\end{Proposition}
The result aligns with similar findings on stationary convergence rates of gradient descent methods in the literature \citep{bottou2018optimization}. Notably, this stationary convergence rate is optimal for first-order methods, as established by \citet{carmon2020lower, nesterov2018lectures}.

Lastly, we combine Theorem \ref{thm: convergence theorem - penal - empirical - any b} and Proposition \ref{prop: alg conv penal} to conclude that:
\[
\begin{aligned}
    \|\hat{b}^\gamma - \beta^*\|_2 &\lesssim \frac{1}{\lambda}\left(\frac{1}{1+\gamma|\Ec|} + \|\nabla \widehat{\Phi}_\mu(\hat{b}^\gamma)\|_2\right) +  \sqrt{\frac{\mu}{\lambda}} +\frac{1}{\sqrt{\lambda}}\left(\frac{p+s}{{n}}\right)^{1/4}\\
    &\lesssim \frac{1}{\lambda}\left(\frac{1}{1+\gamma|\Ec|} + \sqrt{\frac{1}{T}\left(\frac{\kappa_1^2+\kappa_2^2|\Ec|}{\mu} + \mu\right)}\right) +  \sqrt{\frac{\mu}{\lambda}} +\frac{1}{\sqrt{\lambda}}\left(\frac{p+s}{{n}}\right)^{1/4}.
\end{aligned}
\]

\subsection{Proof of Theorem \ref{thm: identification infty minimization - simplify}}
\label{Appendix: proof of thm identification infty minimization - simplify}

Recall the constrained problem \eqref{eq: bneg gamma infty} as follows:
\begin{equation*}
    \argmin_{b\in \Rb^p} \Eb[\Y{1}-b^\intercal\X{1}]^2 \quad {\rm s.t}\quad \Eb[\Y{e}-b^\intercal\X{e}]^2 =\Eb[\Y{f}-b^\intercal\X{f}]^2\;\textrm{for all $e,f\in \Ec$}.
\end{equation*}
We apply the expression of the risk $\Eb[\Y{e}-b^\intercal\X{e}]^2$ in Lemma \ref{lemma: expression of risk} and establish the equivalent formulation of the constrained problem:
{
\begin{equation}
\begin{aligned}
    &\argmin_{b\in \Rb^{p}}2h^\intercal\Gbf(b-\beta^*) + (b-\beta^*)^\intercal \Eb[\X{e}X^{e\intercal}] (b-\beta^*)\\
    &{\rm s.t.}\quad (b - \beta^*)^\intercal \Eb[\X{e}X^{e\intercal}] (b -\beta^*) = (b - \beta^*)^\intercal \Eb[\X{f}X^{f\intercal}] (b -\beta^*),~~\textrm{for all $e,f\in \Ec$},
\end{aligned}
\label{eq: minimization - obj - u - 1 - simplify}
\end{equation}
}
where the explicit expressions of $h$ and $\Gbf$ are given in Lemma \ref{lemma: expression of risk}.

The remaining proof proceeds as follows:
\begin{itemize}
    \item[\textbf{Step-1}:] Under the additive intervention regime, suppose Condition \ref{cond: relaxed minimization} holds. We show that for any predictor $b\in \Rb^p$ subject to the constraint within \eqref{eq: minimization - obj - u - 1 - simplify} satisfies that $h^\intercal \Gbf (b-\beta^*) = 0$.
    \item[\textbf{Step-2}:] Based on Step-1, we simplify the objective function within \eqref{eq: minimization - obj - u - 1 - simplify}, and prove that solving \eqref{eq: minimization - obj - u - 1 - simplify} exactly identifies $\beta^*$.
\end{itemize}
We now detail the remaining proof.

\noindent\textbf{Step-1:} 
We observe that the following equation holds for any predictor $b\in \Rb^p$ subject to the constraint within \eqref{eq: minimization - obj - u - 1 - simplify}:
\[
\sum_{e\in \Ec} \left(w_e - \frac{1}{|\Ec|}\right) (b-\beta^*)^\intercal \Eb[X^{e}X^{e\intercal}] (b-\beta^*) = 0, \quad \forall w\in \Delta^L.
\]
We then apply Lemma \ref{lemma: expression of risk}, which shows that $\Eb[X^{e}X^{e\intercal}]=\Gbf^\intercal \left(\Hbf + \Eb[\del{e}\delT{e}]\right) \Gbf$ for $\Gbf, \Hbf\in \Rb^{p\times p}$ being full-rank matrices, to obtain that
\[
[\Gbf(b-\beta^*)]^\intercal \left[\sum_{e\in \Ec} \left(w_e - \frac{1}{|\Ec|}\right)\Eb[\del{e}\delT{e}] \right] \Gbf(b-\beta^*) = 0, \quad \forall w\in \Delta^L.
\]
Recall that we define $S_{L}$ as the set with interventions such that $S_L = \{j\in [p]: \; \exists e\in \Ec, \;\delta^{e}_j\not \equiv 0\}$, indicating that $\delta^{e}_j\equiv 0$ for all $j\notin S_L$ and $e\in \Ec$. Therefore, the preceding equality further implies that
\[
[\Gbf(b-\beta^*)]^\intercal_{S_L} \left[\sum_{e\in \Ec} \left(w_e - \frac{1}{|\Ec|}\right)\Eb[\delta^{e}_{S_L}\delta^{e\intercal}_{S_L}] \right] [\Gbf(b-\beta^*)]_{S_L} = 0, \quad \forall w\in \Delta^L.
\]
As Condition \ref{cond: relaxed minimization} requires the existence of a weight vector $w_0\in \Delta^L$ satisfying that
\[
[\Deltabf(w^0)]_{S_L, S_L}=\sum_{e\in \Ec} \left(w_e^0 - \frac{1}{|\Ec|}\right)\Eb[\delta^{e}_{S_L}\delta^{e\intercal}_{S_L}]\succ 0,
\]
we obtain that
\[
[\Gbf(b-\beta^*)]_{S_L} = 0.
\]
Since $S_L$ contains all direct children of the outcome with $D\subseteq S_L$, it further implies that
\[
[\Gbf(b-\beta^*)]_{D} = 0.
\]

Recall the vector $h=\sigma_Y^2\Bbf_{YX} - \Eb[\eta_Y\eta_X]$ as defined in Lemma \ref{lemma: expression of risk}. When $\Eb[\eta_Y\eta_X]=0$, it holds
\[
h = \sigma_Y^2 \Bbf_{YX}.
\]
Moreover, $\Bbf_{YX}$ denotes the coefficients from $Y$ to covariates with support ${\rm supp}(\Bbf_{YX}) = D$. Therefore, we have $[\Bbf_{YX}]_{D^\complement} = 0$, and thus
\begin{equation}
\begin{aligned}
    h^\intercal \Gbf (b-\beta^*) &= \sigma_Y^2 \Bbf_{YX}^\intercal \Gbf (b-\beta^*) \\
    &= \sigma_Y^2 [\Bbf_{YX}]_{D}^\intercal [\Gbf(b-\beta^*)]_{D} + \sigma_Y^2 [\Bbf_{YX}]_{D^\complement}^\intercal [\Gbf(b-\beta^*)]_{D^\complement}\\
    &= 0.
\end{aligned}
\label{eq: mini help thm - step 1}
\end{equation}

\noindent\textbf{Step 2:} Based on \eqref{eq: mini help thm - step 1}, we present the equivalent formulation of the problem \eqref{eq: minimization - obj - u - 1 - simplify} by removing the first-order term with respect to $b$ in its objective function: 
\begin{equation*}
\begin{aligned}
    &\argmin_{b\in \Rb^{p}} (b-\beta^*)^\intercal \Eb[\X{e}X^{e\intercal}] (b-\beta^*)\\
    &{\rm s.t.}\quad (b - \beta^*)^\intercal \Eb[\X{e}X^{e\intercal}] (b -\beta^*) = (b - \beta^*)^\intercal \Eb[\X{f}X^{f\intercal}] (b -\beta^*),~~\textrm{for all $e,f\in \Ec$},
\end{aligned}
\end{equation*}
Due to the risk-invariance constraint and $w^0\in \Delta^L$, equivalently, we have
\begin{equation*}
\begin{aligned}
    &\argmin_{b\in \Rb^{p}} (b-\beta^*)^\intercal \left(\sum_{e\in \Ec} w_e^0\Eb[\X{e}X^{e\intercal}] \right) (b-\beta^*)\\
    &{\rm s.t.}\quad (b - \beta^*)^\intercal \Eb[\X{e}X^{e\intercal}] (b -\beta^*) = (b - \beta^*)^\intercal \Eb[\X{f}X^{f\intercal}] (b -\beta^*),~~\textrm{for all $e,f\in \Ec$},
\end{aligned}
\end{equation*}

We observe that $b=\beta^*$ is a feasible point to the above optimization problem with the objective value $0$. And for any other feasible points $b\neq \beta^*$, the corresponding objective value satisfies that
\[
(b-\beta^*)^\intercal \left(\sum_{e\in \Ec} w_e^0\Eb[\X{e}X^{e\intercal}] \right)(b-\beta^*) > 0,
\]
as we consider that there exists an environment $e\in \Ec$ with $\sum_{e\in \Ec} w_e^0\Eb[\X{e}X^{e\intercal}] \succ 0$. Consequently, we obtain that the optimal solution is exactly $\beta^*$.

\section{Unpenalized Nonsmooth NegDRO}
\label{subsec: without penalty}

This section introduces the unpenalized NegDRO problem. Compared to the penalized version discussed in the main text, the unpenalized NegDRO is not only nonconvex but also nonsmooth. Specifically, recall the population NegDRO, defined in \eqref{eq: obj trans - population}, as follows:
\[
    \min_{b\in \Rb^p}\Phi(b),\quad \textrm{where}\quad \Phi(b) = \max_{w\in \Delta^{|\Ec|}} \sum_{e\in \Ec}\left(w_e - \frac{\gamma}{1+\gamma |\Ec|}\right)\Eb[\Y{e}-b^\intercal\X{e}]^2.
\]
The objective function of $\Phi(b)$ may admit multiple maximizers; thus, $\Phi(b)$ is not necessarily differentiable.
Correspondingly, the empirical NegDRO problem is defined as follows:
\begin{equation}
    \min_{b\in \Rb^p}\widehat{\Phi}(b),\quad \textrm{where}\quad \widehat{\Phi}(b) = \max_{w\in \Delta^{|\Ec|}}\sum_{e\in \Ec} \left(w_e - \frac{\gamma}{1+\gamma |\Ec|}\right)\widehat{\Eb}[\Y{e}-b^\intercal\X{e}]^2.
    \label{eq: obj unpenal empirical}
\end{equation}
Still, $\widehat{\Phi}(b)$ may not be differentiable.
While most theoretical properties for the unpenalized NegDRO remain consistent with those in Sections \ref{subsec: stationary causal} and \ref{subsec: computation algorithm}, they require a more general definition of stationary points due to the nonsmooth nature of $\Phi(b)$ and $\widehat{\Phi}(b)$.
To facilitate discussions, we introduce the concepts of subdifferential and generalized stationary points. 

\begin{Definition}[Subgradient and Subdifferential]
    Given a function $h:\Rb^p\to \Rb\cup\{\infty\}$, a vector $\zeta\in \Rb^p$ is called a subgradient of $h$ at a point $b\in \Rb^p$ if it satisfies:
    \[
    h(b')\geq h(b) + \zeta^\intercal (b' - b) + o(\|b'-b\|_2), ~~\textrm{for all $b'$ such that $b'\rightarrow b$}.
    \]
    The subdifferential of $h$ at $b$, denoted as $\partial h(b)$, is the set of all subgradients at $b$:
    \[
    \partial h(b) = \left\{\left.\zeta\in \Rb^p \right\vert h(b')\geq h(b) + \zeta^\intercal (b' - b) + o(\|b'-b\|_2), ~~\textrm{for all $b'$ such that $b'\rightarrow b$}\right\}.
    \]
    The distance of subdifferential $\partial h(b)$ to the origin is given by
    ${\rm Dist}(0, \partial h(b)) = \inf\{\|\zeta\|_2, ~~\zeta\in \partial h(b)\}.$
    \label{def: subdiff}
\end{Definition}
The subdifferential generalizes the notation of gradient.
If the function $h(\cdot)$ is differentiable at the point $b$, the subdifferential reduces to its gradient with $\partial h(b) = \{\nabla h(b)\}$, 
and its distance to the original becomes the gradient norm such that ${\rm Dist}(0, \partial h(b)) = \|\nabla h(b)\|_2$. Building upon the definition of subdifferential, we define generalized stationary points as follows.
\begin{Definition}[Generalized Stationary Point]
    A point $b\in \Rb^p$ is a (generalized) stationary point of the function $h:\Rb^p\to \Rb\cup\{\infty\}$ if and only if $0\in \partial h(b)$, that is, ${\rm Dist}(0, \partial h(b)) = 0$.
    \label{def: stationary}
\end{Definition}
We emphasize that this definition extends the classic stationary point for differentiable functions, as defined in Definition \ref{def: stationary point}, where a point $b$ is a stationary point of the differentiable function $h(\cdot)$ if and only if $\|\nabla h(b)\|_2 = 0$. 

This section is outlined as follows. 
\begin{itemize}
    \item Section \ref{sec: landscape of unpenal negdro} establishes the landscape of the unpenalized NegDRO in both the population and finite-sample regimes.
    \item Section \ref{appendix: proof of prop bneg upper bound} presents the proof of Proposition \ref{prop: bneg upper bound}, which relies on the landscape of unpenalized NegDRO.
    \item Lastly, we propose the algorithm to identify the stationary points for NegDRO in Section \ref{subsec: algo for unpenal negdro} along with its convergence guarantees towards the causal outcome model.
\end{itemize}  

\subsection{Landscape of Unpenalized NegDRO}
\label{sec: landscape of unpenal negdro}
In this section, we shall establish the benign landscapes for unpenalized NegDRO, for both population and empirical versions, parallel to Theorem \ref{thm: convergence theorem - penal - population - any b} in the main article.

The following theorem presents the landscape of unpenalized population NegDRO, where we control the distance between any predictor $b\in \Rb^p$ and the causal outcome model $\beta^*$, by leveraging the distance between subdifferential $\partial \Phi(b)$ to the origin. The proof is provided in Section \ref{appendix: proof of thm: convergence theorem - population - any b}.
\begin{Theorem}[Landscape of Unpenalized Population NegDRO]
    Under the additive intervention regime, suppose that Condition \ref{cond: strict positive - A} holds. The following inequality holds for any predictor $b\in \Rb^p$,
    \begin{equation*}
    \begin{aligned}
    \|b - \beta^*\|_2\lesssim \frac{1}{\lambda}\left(\frac{1}{1+\gamma {|\Ec|}} + {\rm Dist}(0, \partial \Phi(b))\right),
    \end{aligned}
    \end{equation*}
    where ${\rm Dist}(0, \partial \Phi(b)) = \inf\{\|\zeta\|_2, ~\zeta \in \partial \Phi(b)\}$ measures the distance of the subdifferential $\partial\Phi(b)$ to the origin, and $\lambda>0$ denotes the degree of environmental heterogeneity as defined in Condition \ref{cond: strict positive - A}.
\label{thm: convergence theorem - population - any b}
\end{Theorem}
This bound closely aligns with the penalized case in Theorem \ref{thm: convergence theorem - penal - population - any b}, except for two differences: the gradient norm is replaced with the subdifferential distance, and the term related to the ridge-penalty parameter $\mu$ is removed. Moreover, this bound directly indicates that for sufficiently large $\gamma\geq 0$, any (generalized) stationary point of NegDRO with ${\rm Dist}(0, \partial \Phi(b)) = 0$ closely approximates the causal outcome model $\beta^*$.

We next turn to the finite-sample regime, where we observe $\{x^e_i, y^e_i\}_{i=1}^{n_e}$ for each environment $e\in \Ec$. The following theorem establishes the landscape of empirical NegDRO, the proof of which is provided in Section \ref{appendix: proof of thm: convergence theorem - empirical - any b}.

\begin{Theorem}[Landscape of Unpenalized Empirical NegDRO]
    Under the additive intervention regime, suppose Conditions \ref{cond: strict positive - A}, \ref{cond: subgaussian} hold and $n\geq (1\vee c_1/\lambda^2) p$. Then, for any $0\leq s\leq (1\wedge c_2\lambda^2) n - p$, with a probability of at least $1-2|\Ec| e^{-s}$, the following inequality holds uniformly for all predictors $b\in \Rb^p$
    \begin{equation}
        \left\|b - \beta^*\right\|_2 \lesssim \frac{1}{\lambda}\left(\frac{1}{1+\gamma|\Ec|} + {\rm Dist}(0, \partial\widehat{\Phi}(b))\right) +  \frac{C}{\lambda^{1/2}} \left(\frac{p+s}{{n}}\right)^{1/4},
        \label{eq: upper bound - empirical - any b}
    \end{equation}
    where ${\rm Dist}(0, \partial\widehat{\Phi}(b)) = \inf\{\|\zeta\|_2, \zeta\in \partial\widehat{\Phi}(b)\}$ measures the distance of the subdifferential $\partial\widehat{\Phi}(b)$ to the origin, $\lambda>0$ denotes the degree of environmental heterogeneity as defined in Condition \ref{cond: strict positive - A}, and $c_1,c_2>0$ are absolute constants.
    \label{thm: convergence theorem - empirical - any b}
\end{Theorem}
This theorem implies that if $b$ is a (generalized) stationary point of the empirical objective function $\widehat{\Phi}(\cdot)$, such that ${\rm Dist}(0, \partial\widehat{\Phi}(b)) = 0$, it closely approximates the causal outcome model $\beta^*$ with an error $\|b - \beta^*\|_2 = \Oc((\gamma|\Ec|)^{-1} + n^{-1/4})$. Motivated by this insight, we present the algorithm designs to identify the (generalized) stationary points of $\widehat{\Phi}(\cdot)$ in Appendix \ref{subsec: algo for unpenal negdro}.

\subsection{Proof of Proposition \ref{prop: bneg upper bound}}
\label{appendix: proof of prop bneg upper bound}

For each fixed $w\in \Delta^{|\Ec|}$, the map
\[
b\mapsto \sum_{e\in \Ec}\left(w_e - \frac{\gamma}{1+\gamma|\Ec|}\right) \Eb[Y^e - b^\intercal X^e]^2
\]
is continuous in $b$. Since $\Delta^{|\Ec|}$ is compact, taking the maximization preserves continuity; that is,
\[
\Phi(b) = \max_{w\in \Delta^{|\Ec|}} \sum_{e\in \Ec}\left(w_e - \frac{\gamma}{1+\gamma|\Ec|}\right) \Eb[Y^e - b^\intercal X^e]^2
\]
is proper and continuous in $b$.
By Fermat's rule, every local minimizer $b$ of $\Phi(\cdot)$ satisfies the first-order necessary condition with $0\in \partial \Phi(b)$. 

Since $b_{\rm Neg}^\gamma$ is defined to be the global optimizer of $\Phi(\cdot)$, it follows $0\in \partial \Phi(b_{\rm Neg}^\gamma)$ with
\[
{\rm Dist}(0, \partial \Phi(b_{\rm Neg}^\gamma)) = 0.
\]
Now we apply Theorem \ref{thm: convergence theorem - population - any b} to complete the proof.

\subsection{Algorithm Design for Unpenalized NegDRO}
\label{subsec: algo for unpenal negdro}
Inspired by Theorem \ref{thm: convergence theorem - empirical - any b}, it suffices to compute a stationary point of $\widehat{\Phi}$ to approximate $\beta^*$.
To compute the stationary points of the nonsmooth function $\widehat{\Phi}(b)$, we introduce a subgradient-based Algorithm \ref{alg: GDmax subgrad}. 
In contrast to Algorithm \ref{alg: GDmax penal} in the main article, this new algorithm updates predictor $b^t$ using \textit{subgradient descent} in each iteration $t$, instead of gradient descent. Specifically, for each iteration, we first identify a maximizer weight vector $w_{t+1}$ as shown in \eqref{eq: alg step1}, and then update the predictor $b_{t+1}$ using subgradient descent as in \eqref{eq: alg step2}.

Moreover, this new algorithm employs a technique named proximal mapping to evaluate convergence,
where the proximal mapping is given by: for a parameter $\upsilon>0$, 
\begin{equation}
    {\rm prox}_{\upsilon\widehat{\Phi}}(b) = \argmin_{\zeta\in \Rb^p} \left\{\widehat{\Phi}(\zeta) + \frac{1}{2\upsilon}\|\zeta - b\|_2^2\right\}.
    \label{eq: prox map alg}
\end{equation}
We shall emphasize that while $\widehat{\Phi}(\cdot)$ itself is not convex, it can be shown to possess weak convexity, which implies that the optimization problem on the right-hand side of \eqref{eq: prox map alg} is strongly convex towards $\zeta$, for a sufficiently small $\upsilon$; thus, it is computationally efficient to solve the proximal mapping. We provide a formal definition of weak convexity and the detailed arguments in Appendix \ref{Appendix: proof of subsec computation algorithm}.

\begin{algorithm}[!ht]
    \DontPrintSemicolon
    \SetAlgoLined
    \SetNoFillComment
    \LinesNotNumbered 
    \caption{Subgradient Descent-Maximization Algorithm for the Original NegDRO in \eqref{eq: obj unpenal empirical}}
    \SetKwInOut{Input}{Input}
    \SetKwInOut{Output}{Output}
    \Input{The number of iterations $T$, step size $\alpha$, initial point $b_0$, regularization parameter $\gamma\geq 0$ and proximal-mapping parameter $\upsilon>0$}
    \Output{$\hat{b}^\gamma$}

    \For{$t = 0,1,...,T-1$}{
        Update the maximizer weight associated with $b_{t}$:
        {
        \begin{align}
            w_{t+1} \in \argmax_{w\in \Delta^{|\Ec|}} \sum_{e\in \Ec}\left({w}_e- \frac{\gamma}{1+\gamma |\Ec|}\right) \widehat{\Eb}[\Y{e}-(b_t)^\intercal \X{e}]^2;
            \label{eq: alg step1}
        \end{align}
        }
        
        Update $b^{t+1}$ corresponding to the weight $w_{t+1}$: %
        \begin{equation}
        b_{t+1} = b_t - \alpha \cdot \sum_{e\in \Ec}\left([{w}_{t+1}]_e- \frac{\gamma}{1+\gamma |\Ec|}\right) \nabla \widehat{\Eb}[\Y{e}-(b_t)^\intercal \X{e}]^2;
            \label{eq: alg step2}
        \end{equation}
        Compute proximal mapping point ${\rm prox}_{\upsilon\widehat{\Phi}}(b_{t+1})$ as in \eqref{eq: prox map alg}.
    }
    
    Define $\hat{b}^\gamma \in \argmin_{\{b_t\}_{t=0}^{T}}\|{\rm prox}_{\upsilon\widehat{\Phi}}(b_{t}) - b_{t}\|_2$.
    \label{alg: GDmax subgrad}
\end{algorithm}

We now provide additional discussions on the usage of proximal mapping, which is utilized in the final output selection step of the current Algorithm \ref{alg: GDmax subgrad}. Previously, for the ridge-penalized NegDRO, which is differentiable, Algorithm \ref{alg: GDmax penal} selects the output by minimizing the gradient norm $\|\nabla \widehat{\Phi}_\mu(b_t)\|_2$. This ensures that the selected $\hat{b}^\gamma$ approaches a stationary point of smooth $\widehat{\Phi}_\mu(\cdot)$.
In contrast, for the unpenalized NegDRO, which is not differentiable everywhere, we aim to minimize the subdifferential distances to the origin, ${\rm Dist}(0, \partial \widehat{\Phi}(b_t))$, for $t=1,...,T$, as implied by Theorem \ref{thm: convergence theorem - empirical - any b}. However, the quantities $\{{\rm Dist}(0, \partial \widehat{\Phi}(b_t))\}_{t}$ are not directly observable. To address this issue, the current Algorithm \ref{alg: GDmax subgrad} employs proximal mapping by selecting the smallest $\|{\rm prox}_{\upsilon\widehat{\Phi}}(b_{t}) - b_{t}\|_2$. In Appendix \ref{Appendix: proof of subsec computation algorithm}, we shall show that this quantity upper bounds the $\min_{t}{\rm Dist}(0, \partial \widehat{\Phi}(b_t))$, thereby effectively approaching a stationary point for the nonsmooth $\widehat{\Phi}(\cdot)$.

We impose a standard regularity condition on the iterates produced by algorithm \ref{alg: GDmax subgrad}, which is adapted from Condition \ref{cond: smooth} in the main text.
\begin{Condition}[Gradient Boundedness]
\label{cond: bounded gradient}
    All iterates $\{b_t\}_{t=0}^T$ produced by Algorithm \ref{alg: GDmax subgrad} remain within a bounded set $\mathcal{B}\subset \Rb^p$, and there exists a constant $C>0$ such that
    \begin{equation}
        \sup_{e\in \Ec, b\in \mathcal{B}}\|\nabla \widehat{\Eb}[Y^e - b^\intercal \X{e}]^2 \|_2 \leq C<\infty.
        \label{eq: bounded gradient}
    \end{equation}
\end{Condition}

We now analyze the output of Algorithm \ref{alg: GDmax subgrad} by establishing its convergence rate to $\beta^*$, with its proof provided in Section \ref{Appendix: proof of subsec computation algorithm}.
\begin{Theorem}
    Under the additive intervention regime, suppose that Conditions \ref{cond: strict positive - A}, \ref{cond: subgaussian}, \ref{cond: bounded gradient} hold, and
    the sample size satisfies $n\geq (1\vee c_1/\lambda^2)p$. 
    Set the step size to $\alpha \asymp T^{-1/2}$. Then for any $0\leq s\leq (1\wedge c_2\lambda^2)n-p$, with probability at least $1-2|\Ec|e^{-s}$,
    the output $\hat{b}^\gamma$ of Algorithm \ref{alg: GDmax subgrad} satisfies:
    \[
    \|\hat{b}^\gamma - \beta^*\|_2 \lesssim \frac{1}{\lambda}\left(\frac{1}{1+\gamma|\Ec|} + T^{-1/4}\right) +  \frac{1}{\lambda^{1/2}} \left(\frac{p+s}{{n}}\right)^{1/4},
    \]
    where $\lambda$ denotes the degree of environmental heterogeneity as defined in Condition \ref{cond: strict positive - A}, and $c_1,c_2>0$ are absolute constants.
\label{thm: convergence theorem - empirical - alg output}
\end{Theorem}

The above theorem shows that the output $\hat{b}^\gamma$  approximates $\beta^*$ up to an error $\mathcal{O}((\gamma|\Ec|)^{-1} + T^{-1/4} + n^{-1/4})$, aligned with the penalized case in Theorem \ref{thm: convergence theorem - penal - empirical - alg output} when the ridge-penalty parameter $\mu$ is set to $cM/\sqrt{T}$. 
The convergence rates with respect to both the sample size n and iteration times $T$ are consistent with those discussed for Theorem \ref{thm: convergence theorem - penal - empirical - alg output}. 

We further discuss the rate $\Oc(T^{-1/4})$ in the preceding theorem.
The derived convergence rate for $\|\hat{b}^\gamma - \beta^*\|_2$ is obtained by leveraging the upper bound of $\|b-\beta^*\|_2$ for any prediction model $b\in \Rb^p$ as established in Theorem \ref{thm: convergence theorem - empirical - any b}, in conjunction with identifying a nearly (generalized) stationary point $\hat{b}^\gamma$ such that ${\rm Dist}(0, \widehat{\Phi}(b))$ is sufficiently small. Notably, finding generalized stationary points for non-smooth functions, such as our NegDRO objective $\widehat{\Phi}(b)$, inherently requires a $T^{-1/4}$ convergence rate \citep{davis2018complexity}.
While this generalized stationary convergence rate is considered optimal without additional structural assumptions, it raises the intriguing possibility of directly quantifying $\|\hat{b}^\gamma - \beta^*\|_2$ without relying on stationary convergence theory, which could potentially lead to improved rates or new insights.

\section{Additional Results on the Role of the Optimization Objective}
\label{appendix: minimization helps}

In Section~\ref{subsec: limited negdro}, we establish the benign landscape of the unpenalized NegDRO under the limited additive intervention regime.

Building upon that result, this section extends the discussion to the general additive intervention regime and emphasizes the importance of the optimization objective in NegDRO.
Specifically, Section~\ref{Appendix: weaker condition} introduces a relaxed identification condition for the general additive intervention regime.

Furthermore, Section~\ref{Appendix: CausalDantzig DRIG} provides additional background and methodological comparisons, highlighting how alternative causal invariance learning frameworks, Causal Dantzig and DRIG, address additive interventions using different formulations of the underlying optimization problem.

\subsection{Benign Landscape for the Limited Additive Intervention Regime}
\label{subsec: limited negdro}
 We define the following quantities: with the specific weight $w^0$ from Condition \ref{cond: relaxed minimization},
\[
\lambda_1 := \lambda_{\rm min}\left([(\Deltabf(w^0)]_{S_L,S_L}\right), \quad \lambda_2:= \lambda_{\rm min}\left(\sum_{e\in \Ec} w^e_0 \Eb[\X{e}\XT{e}]\right).
\]
The following theorem establishes the benign landscape of the unpenalized NegDRO objective under the limited additive intervention regime.
The detailed proof is provided in Appendix~\ref{Appendix: thm: convergence - minimization help - landscape}.
\begin{Theorem}[Benign Landscape of Unpenalized NegDRO]
    Under the additive intervention regime, suppose there is no hidden confounding between the outcome and covariates, i.e., $\Eb[\eta_Y\eta_X]=0$. If Condition \ref{cond: relaxed minimization} holds, then the following inequality holds for any predictor $b\in \Rb^p$
    \[
    \|b-\beta^*\|_2 \lesssim \kappa{\rm Dist}(0, \partial \Phi(b))+ \kappa^{1/2} ({\rm Dist}(0, \partial \Phi(b)))^{1/2} + \kappa^{1/4} ({\rm Dist}(0, \partial \Phi(b)))^{1/4} + \kappa^{-1/2}, 
    \]
where $\kappa = 1+\gamma|\Ec|$, and ${\rm Dist}(0, \partial \Phi(b)) = \inf\{\|\zeta\|_2, ~\zeta \in \partial \Phi(b)\}$ measures the distance of the subdifferential $\partial\Phi(b)$ to the origin.
    \label{thm: convergence - minimization help - landscape}
\end{Theorem}

Compared with Theorem~\ref{thm: convergence theorem - population - any b}, which characterizes the full additive intervention regime, Theorem~\ref{thm: convergence - minimization help - landscape} shows that even under the more challenging limited interventions, the NegDRO landscape remains benign.

In particular, if a stationary point of $\Phi(b)$ is reached such that ${\rm Dist}(0, \partial \Phi(b))=0$, then it approximates $\beta^*$, then this stationary point is guaranteed to be close to $\beta^*$ with error bounded by $\mathcal{O}\left((1+\gamma|\Ec|)^{-1/2}\right)$. This convergence rate is slower than the $\mathcal{O}\left((1+\gamma|\Ec|)^{-1}\right)$ rate in Theorem \ref{thm: convergence theorem - population - any b} for the full intervention regime.

In practice, however, optimization algorithms like Algorithm \ref{alg: GDmax subgrad} rarely reach an exact stationary point. Suppose the algorithm attains an $\epsilon$-stationary solution satisfying ${\rm Dist}(0, \partial \Phi(b))\le \epsilon$. Then Theorem \ref{thm: convergence - minimization help - landscape} implies that
\[
\|b-\beta^*\|_2 \lesssim \kappa\epsilon + \kappa^{1/2}\epsilon^{1/2} + \kappa^{1/4}\epsilon^{1/4} + \kappa^{-1/2}.
\]
To ensure that $\|b-\beta^*\|_2$ converges to zero, it is necessary that $\epsilon \ll (1+\gamma|\Ec|)^{-1}$ and that $\gamma\to\infty$. This requirement is more stringent than in the full intervention regime, where Theorem \ref{thm: convergence theorem - population - any b} only requires $\epsilon\to0$ and $\gamma\to 0$ for consistency. Therefore, the limited intervention setting demands higher algorithmic precision, as the algorithm must locate a more accurate stationary point to ensure the resulting predictor remains close to $\beta^*$.

For brevity, we omit the discussion of the empirical benign landscape under the limited intervention regime. The empirical landscape bound follows essentially a similar structure to Theorem~\ref{thm: convergence - minimization help - landscape}, up to additional finite-sample errors.

\subsection{Relaxed Identification Conditions for General Additive Interventions}
\label{Appendix: weaker condition}
Section \ref{sec: minimization} in the main text introduces a relaxed identification condition tailored to the limited additive intervention regime, where only a subset of covariates is subject to interventions.
Here, we extend this idea and propose a more general relaxed condition that applies to the general additive intervention regime, encompassing the limited intervention case as a special instance.

Recall the definition of $\mathbf{\Delta}(w)$ in \eqref{eq: intervention cond}: for any weight vector $w\in \Delta^L$, 
\[
\mathbf{\Delta}(w) = \sum_{e\in \Ec}\left(w_e - \frac{1}{|\Ec|}\right) \Eb[\del{e}\delT{e}].
\]
We now introduce the following identification condition, which involves the vector $h$ and matrix $\Hbf\succeq 0$ defined in Lemma \ref{lemma: expression of risk}.
\begin{Condition}
\label{cond: relaxed minimization - general}
There exists some weight $w^0 \in \Delta^L$ such that both of the following conditions hold:
\begin{itemize}
    \item[(i)] The matrix $\Deltabf(w^0)$ satisfies that
    \begin{equation}
        \Deltabf(w^0) \succeq 0 \quad \textrm{and}~~ \Deltabf(w^0)\succ - {\Hbf} - \frac{1}{|\Ec|}\sum_{e\in \Ec}\Eb[\del{e}\delT{e}],
        \label{eq: cond semi-pos}
    \end{equation}
    \item[(ii)] Let $\Deltabf(w^0) = V^0 \Lambda^0 (V^0)^\intercal$ denote its spectral decomposition, where $\Lambda^0 = {\rm diag}(\lambda_1^0\cdots \lambda_p^0)$ contains eigenvalues $\lambda_j^0\geq 0$ for each $j\in[p]$, and $V^0 = (v_1^0\cdots v_p^0)$ are the corresponding eigenvectors. The vector $h$ lies in the subspace spanned by the eigenvectors associated with positive eigenvalues:
    \begin{equation}
    \label{eq: h span}
        h \in {\rm Span}(\{v_j:~~\textrm{for $j\in [p]$ satisfies } \lambda_j^0 > 0\}).
    \end{equation}
\end{itemize}     
\end{Condition}

Condition \ref{cond: relaxed minimization - general} relaxes the earlier Condition \ref{cond: strict positive - A}, which requires strict positive definiteness.
Here, we only require that $\Deltabf(w^0)$ be positive semidefinite and that the vector $h$ lies within the subspace spanned by its strictly positive eigenvectors.
To see that this condition is a strict relaxation of the earlier Condition \ref{cond: strict positive - A}, note from Lemma \ref{lemma: expression of risk} that
\[
\Abm(w^0) = \sum_{e\in \Ec} \left(w_e - \frac{1}{|\Ec|}\right) \Eb[\X{e}\XT{e}] = \sum_{e\in \Ec} \left(w_e - \frac{1}{|\Ec|}\right) \Gbf^\intercal \Eb[\del{e}\delT{e}] \Gbf = \Gbf^\intercal \Deltabf(w^0)\Gbf,
\]
where $\Gbf$ is a full-rank matrix. Thus, if Condition \ref{cond: strict positive - A} holds with $\Abm(w^0)\succ 0$, it follows that $\Deltabf(w^0)\succ 0$, and hence \eqref{eq: cond semi-pos} is automatically satisfied.
Moreover, since all eigenvalues of $\Deltabf(w^0)$ are strictly positive, we have
\[
\mathrm{Span}\left(\{v_j^0 : \lambda_j^0 > 0\}\right) = \Rb^p,
\]
so that \eqref{eq: h span} also holds.
Therefore, Condition \ref{cond: relaxed minimization - general} strictly relaxes Condition \ref{cond: strict positive - A}.

The following theorem shows that the relaxed Condition \ref{cond: relaxed minimization - general} suffices for causal identification, i.e., the optimal solution of the constrained problem \eqref{eq: bneg gamma infty} is $\beta^*$. The proof of Theorem \ref{thm: identification infty minimization - general} is given in Appendix \ref{Appendix - proof - identification infty minimization - general}.
\begin{Theorem} Under the additive intervention regime, suppose Condition \ref{cond: relaxed minimization - general} holds. Then the causal outcome model $\beta^*$ achieves the smallest prediction risk among all risk-invariant predictors. Consequently, the proposed constrained problem \eqref{eq: bneg gamma infty} exactly recovers $\beta^*$. 
\label{thm: identification infty minimization - general}
\end{Theorem}

\subsection{Discussions of Existing Approaches}
\label{Appendix: CausalDantzig DRIG}
In this subsection, we will provide the detailed analysis regarding the existing approaches tailored for the additive intervention regime: CausalDantzig \citep{meinshausen2018causality} and DRIG \citep{shen2023causality}. 

We start with the analysis of Causal Dantzig, where they consider two observed environments $e$ and $f$, with samples $\{\x{e}_i, \y{e}_i\}_{i=1}^{n_e}$ and $\{\x{f}_i, \y{f}_i\}_{i=1}^{n_f}$. They define
\[
\widehat{\mathbf{G}} = \frac{1}{n_e}\sum_{i=1}^{n_e} x^{e}_i x^{e\intercal}_i  - \frac{1}{n_f}\sum_{i=1}^{n_f} x^{f}_i  x^{f\intercal}_i, \quad \textrm{and}\quad
\widehat{\mathbf{Z}} = \frac{1}{n_e}\sum_{i=1}^{n_e} x^{e}_i y^{e}_i - \frac{1}{n_f}\sum_{i=1}^{n_f} x^{f}_i y^{f}_i.
\]
If $\widehat{\mathbf{G}}$ is invertible, the causal dantzig estimator would be
\[
\hat{\beta} = \widehat{\mathbf{G}}^{-1}\widehat{\mathbf{Z}}.
\]
When $\widehat{\mathbf{G}}$ is not invertible, as in the case where $p\gg n$, they proposed the regularized causal Dantzig estimator as the solution to the following optimization problem:
\[
\hat{\beta}^\lambda = \argmin\|\beta\|_1 \quad \textrm{s.t.}~~ \|\widehat{\mathbf{Z}} - \widehat{\mathbf{G}}\beta\|_\infty \leq \lambda,
\]
where $\lambda> 0$ is the tuning parameter. Suppose that $\mathbf{G} = \Eb[\widehat{\mathbf{G}}]$ is invertible and other regular conditions hold. let the tuning parameter $\lambda\asymp \sqrt{\log(p)/(n_1\wedge n_2)} \to 0$, \citet[Section 4]{meinshausen2018causality} establishes that with high probability,
\[
\|\hat{\beta}^\lambda - \beta^*\|_2 \lesssim \sqrt{\frac{\log(p)}{n_1\wedge n_2}}.
\]

Now, we will move on discussing the DRIG method \citep{shen2023causality}, where they assume the existence and availability of a reference environment, denoted as $e=0$. And all other environments $\Ec=\{1,2,...,|\Ec|\}$ have stronger interventions than the reference environment with 
$$\Eb[\del{0}\delT{0}]\prec \Eb[\del{e}\delT{e}] \textrm{ for any environment $e\in \Ec$.}$$
The DRIG estimator of the population version is the solution to the following optimization problem:
\[
b_{\rm DRIG}^\gamma = \argmin_b \Eb[\ell(\X{0},\Y{0};b)] + \gamma\sum_{e\in \Ec} w^e\left(\Eb[\ell(\X{e}, \Y{e}; b) - \Eb[\ell(\X{0}, \Y{0}; b)]\right),
\]
where $\gamma$ is the regularization parameter, $w\in \Delta^{|\Ec|}$ is some pre-fix weight. 
In comparison with DRIG, our proposed NegDRO as defined in \eqref{eq: obj original} does not assume the existence or availability of such a reference environment $e=0$, and we do not need to specify a weight $w$ in prior.

It can be shown that if $e=0$ is indeed a reference environment, $b_{\rm DRIG}^\gamma$ will be the unique stationary point of their optimization problem. Therefore, they solve $b_{\rm DRIG}^\gamma$ by setting the gradient being zero: 
\[
(1-\gamma)\Eb\left[\X{0}(\Y{0} - (b_{\rm DRIG}^\gamma)^\intercal \X{0})\right] + \gamma \sum_{e\in \Ec} w^e \Eb\left[\X{e}(\Y{e} - (b_{\rm DRIG}^\gamma)^\intercal \X{e})\right] = 0.
\]
When we set $\gamma$ to $\infty$, the above equation becomes
\[
\Eb\left[\X{0}(\Y{0} - (b_{\rm DRIG}^\infty)^\intercal \X{0})\right] = \sum_{e\in \Ec} w^e \Eb\left[\X{e}(\Y{e} - (b_{\rm DRIG}^\infty)^\intercal \X{e})\right].
\]
By organizing the above equation and assuming that $\Eb[\X{0}X^{(0)\intercal}] - \sum_{e\in \Ec} w^e\Eb[\X{e}X^{e\intercal}]$ is invertible, we establish that
\[
b_{\rm DRIG}^\infty = \left(\Eb[\X{0}X^{(0)\intercal}] - \sum_{e\in \Ec} w^e\Eb[\X{e}X^{e\intercal}]\right)^{-1} \left(\Eb[\X{0}Y^{(0)}] - \sum_{e\in \Ec} w^e\Eb[\X{e}Y^{e}]\right).
\]
Since the causal outcome model $\beta^*$ satisfies \eqref{eq: grad invariance}, it holds that
\[
\beta^* = \left(\Eb[\X{0}X^{(0)\intercal}] - \sum_{e\in \Ec} w^e\Eb[\X{e}X^{e\intercal}]\right)^{-1} \left(\Eb[\X{0}Y^{(0)}] - \sum_{e\in \Ec} w^e\Eb[\X{e}Y^{e}]\right).
\]
Combining the above two equations, we shall show that
\[
b_{\rm DRIG}^\infty = \beta^*,
\]
if $\Eb[\X{0}X^{(0)\intercal}] - \sum_{e\in \Ec} w^e\Eb[\X{e}X^{e\intercal}]$ is invertible with the prespecified weight $w\in \Delta^{|\Ec|}.$
However when there does not exist such a reference environment or the matrix is non-invertible, provided that some covariates are not intervened for all environments, the DRIG method fails to achieve causal identification.

To conclude, both CausalDantzig and DRIG rely on the invertibility of the difference of gram matrices for causal identification, which can be violated if some covariates have not been intervened in all environments $e\in \Ec.$

\section{Proofs for Other Results}
\label{appendix: secB}

\subsection{Proof of Theorem \ref{thm: convergence theorem - penal - empirical - any b}}
\label{appendix: proof of thm: convergence theorem - penal - empirical - any b}
It follows from Lemma \ref{lemma: expression of empirical risk} that the weighted empirical risk can be expressed as follows:
{
\begin{equation}
\begin{aligned}
    &\sum_{e\in \Ec}\left(w_e - \frac{\gamma}{1+\gamma|\Ec|}\right)\widehat{\Eb}[\Y{e}-b^\intercal \X{e}]^2\\
    &= \sum_{e\in \Ec}\left(w_e - \frac{\gamma}{1+\gamma|\Ec|}\right)\left[(\sigma_Y^2 + \hat{r}^{e}) + 2(h - \hat{q}^{e})^\intercal \Gbf(b-\beta^*) + (b-\beta^*)^\intercal \left(\Eb[\X{e}X^{e\intercal}] + \Gbf^\intercal \hat{\Pbf}^{e}\Gbf\right) (b-\beta^*)\right] \\
    &= \frac{\sigma_Y^2 + 2h^\intercal \Gbf(b-\beta^*) + (b-\beta^*)^\intercal \left[\frac{1}{|\Ec|}\sum_{e\in \Ec}\Eb[\X{e}X^{e\intercal}]\right]  (b-\beta^*)}{1+\gamma|\Ec|}  + (b-\beta^*)^\intercal \Abm(w) (b - \beta^*) \\
   &\quad\quad +  \sum_{e\in \Ec}\left(w_e - \frac{\gamma}{1+\gamma|\Ec|}\right)\left[\hat{r}^{e} - 2\hat{q}^{e\intercal}\Gbf(b-\beta^*) + (b-\beta^*)^\intercal \Gbf^\intercal \hat{\Pbf}^{e}\Gbf(b-\beta^*)\right].
\end{aligned}
\label{proof: eq weighted average empirical risks}
\end{equation}
}

Now we study the expression of $\widehat{\Phi}_\mu(b)$ defined in \eqref{eq: obj-raw penal empirical}. With slight abuse of notation, we define
\begin{equation}
    \Bar{w} = \argmax_{w\in \Delta^{|\Ec|}}\left\{\sum_{e\in \Ec}\left(w_e - \frac{\gamma}{1+\gamma|\Ec|}\right)\widehat{\Eb}[\Y{e}-b^\intercal \X{e}]^2 - \mu\|w\|_2^2\right\}.
    \label{eq: maximizer w - penal - empirical}
\end{equation}
Together with \eqref{proof: eq weighted average empirical risks}, we establish that
{
\begin{equation}
\begin{aligned}
    \widehat{\Phi}_\mu(b) &= \sum_{e\in \Ec}\left(\bar{w}_e - \frac{\gamma}{1+\gamma|\Ec|}\right)\widehat{\Eb}[\Y{e}-b^\intercal \X{e}]^2 \\
    &= \frac{\sigma_Y^2 + 2h^\intercal \Gbf(b-\beta^*) + (b-\beta^*)^\intercal \left[\frac{1}{|\Ec|}\sum_{e\in \Ec}\Eb[\X{e}X^{e\intercal}]\right]  (b-\beta^*)}{1+\gamma|\Ec|}  + (b-\beta^*)^\intercal \Abm(\bar w) (b - \beta^*) \\
   &\quad\quad +  \sum_{e\in \Ec}\left(\Bar{w}_e - \frac{\gamma}{1+\gamma|\Ec|}\right)\left[\hat{r}^{e} - 2\hat{q}^{e\intercal}\Gbf(b-\beta^*) + (b-\beta^*)^\intercal \Gbf^\intercal \hat{\Pbf}^{e}\Gbf(b-\beta^*)\right] - \mu\|\Bar{w}\|_2^2.
\end{aligned}
\end{equation}
}

\paragraph{Step-1:} Express $\widehat{\Phi}_\mu(b)$ through its gradient $\nabla\widehat{\Phi}_\mu(b)$.

The gradient $\nabla\widehat{\Phi}_\mu(b)$ is given by:
{\small
\begin{equation*}
    \begin{aligned}
        \nabla \widehat{\Phi}_\mu(b) &= \frac{2 \Gbf^\intercal h + 2 \left[\frac{1}{|\Ec|}\sum_{e\in \Ec}\Eb[\X{e}\XT{e}]\right] \Gbf (b-\beta^*)}{1+\gamma|\Ec|} + 2 \Abm(\Bar{w}) (b - \beta^*) \\
        &\quad + \sum_{e\in \Ec}\left(\Bar{w}_e - \frac{\gamma}{1+\gamma|\Ec|}\right)\left[-2\Gbf^\intercal\hat{q}^{e} + 2\Gbf^\intercal \hat{\Pbf}^{e}\Gbf(b-\beta^*)\right].
    \end{aligned}
\end{equation*}
}
Left multiplying the vector $\frac{1}{2}(b - \beta^*)$, we establish that
{\small
\begin{equation*}
\begin{aligned}
    \frac{1}{2}(b-\beta^*)^\intercal \nabla \widehat{\Phi}_\mu(b) &=\frac{h^\intercal \Gbf(b-\beta^*) + (b-\beta^*)^\intercal \left[ \frac{1}{|\Ec|}\sum_{e\in \Ec}\Eb[\X{e}\XT{e}]\right] (b-\beta^*)}{1+\gamma|\Ec|}  + (b-\beta^*)^\intercal \Abm(\Bar{w}) (b - \beta^*) \\
   &\quad\quad +  \sum_{e\in \Ec}\left(\Bar{w}_e - \frac{\gamma}{1+\gamma|\Ec|}\right)\left[-\hat{q}^{e\intercal}\Gbf(b-\beta^*) + (b-\beta^*)^\intercal \Gbf^\intercal \hat{\Pbf}^{e}\Gbf(b-\beta^*)\right].
\end{aligned}
\end{equation*}
}
Combining the above equality with \eqref{proof: eq phi_mu b}, we establish that
{\small
\begin{equation}
\begin{aligned}
    \widehat{\Phi}_\mu(b) &= \frac{1}{2}(b-\beta^*)^\intercal \nabla \Phi_\mu(b) + \frac{\sigma_Y^2 + h^\intercal \Gbf(b-\beta^*)}{1+\gamma|\Ec|} - \mu\|\bar{w}\|_2^2 \\
    &\quad\quad + \sum_{e\in \Ec}\left(\Bar{w}_e - \frac{\gamma}{1+\gamma|\Ec|}\right)\left[\hat{r}^{e} - \hat{q}^{e\intercal}\Gbf(b-\beta^*)\right].
\end{aligned}
    \label{proof: eq hat phi_mu b grad}
\end{equation}
}

\paragraph{Step-2:} Establish the lower bound of $\widehat{\Phi}_\mu(b)$ via the weight $w^0$ specified in Condition \ref{cond: strict positive - A}. 

Since $\bar{w}$ is the maximizer weight as defined in \eqref{eq: maximizer w - penal - empirical}, we have
{\small
\begin{equation*}
\begin{aligned}
    \widehat{\Phi}_\mu(b) &\geq \sum_{e\in \Ec}\left(w^0 - \frac{\gamma}{1+\gamma |\Ec|}\right) \widehat{\Eb}[\Y{e}-b^\intercal\X{e}]^2 - \mu \|w^0\|_2^2 \\
    &= \frac{\sigma_Y^2 + 2h^\intercal \Gbf(b-\beta^*) + (b-\beta^*)^\intercal \left[\frac{1}{|\Ec|}\sum_{e\in \Ec}\Eb[\X{e}\XT{e}]\right] (b-\beta^*)}{1+\gamma|\Ec|}  + (b-\beta^*)^\intercal \Abm(w^0) (b - \beta^*) \\
   &\quad\quad +  \sum_{e\in \Ec}\left(w^0_e - \frac{\gamma}{1+\gamma|\Ec|}\right)\left[\hat{r}^{e} - 2\hat{q}^{e\intercal}\Gbf(b-\beta^*) + (b-\beta^*)^\intercal \Gbf^\intercal \hat{\Pbf}^{e}\Gbf(b-\beta^*)\right] - \mu\|w^0\|_2^2.
\end{aligned}
\end{equation*}
}
Since $\Eb[\X{e}\XT{e}]\succeq 0$ for all $e\in \Ec$, and $\Abm(w^0)\succeq \lambda\Ibf$ as specified in Condition \ref{cond: strict positive - A}, we further establish that
{\small
\begin{equation}
\begin{aligned}
    \widehat{\Phi}_\mu(b) &\geq \frac{\sigma_Y^2 + 2h^\intercal \Gbf(b-\beta^*)}{1+\gamma|\Ec|} + \lambda \|b - \beta^*\|_2^2 - \mu\|w^0\|_2^2 \\
    &\quad \quad + \sum_{e\in \Ec}\left(w^0_e - \frac{\gamma}{1+\gamma|\Ec|}\right)\left[\hat{r}^{e} - 2\hat{q}^{e\intercal}\Gbf(b-\beta^*) + (b-\beta^*)^\intercal \Gbf^\intercal \hat{\Pbf}^{e}\Gbf(b-\beta^*)\right]
\end{aligned}
    \label{proof: eq hat phi_mu b lower w^0}
\end{equation}
}

\paragraph{Step-3:} Combine the results in the steps 1 and 2. It follows from \eqref{proof: eq hat phi_mu b grad} and \eqref{proof: eq hat phi_mu b lower w^0} that
{\small
\[
\begin{aligned}
    &\frac{1}{2}(b-\beta^*)^\intercal \nabla \widehat{\Phi}_\mu(b) - \frac{h^\intercal \Gbf(b-\beta^*)}{1+\gamma|\Ec|} - \mu\|\bar{w}\|_2^2 + \sum_{e\in \Ec}\left(\bar{w}_e - \frac{\gamma}{1+\gamma|\Ec|}\right)\left[\hat{r}^{e} - \hat{q}^{e\intercal}\Gbf(b-\beta^*)\right]\\
    &\quad\quad 
    + \mu\|w^0\|_2^2 - \sum_{e\in \Ec}\left(w^0_e - \frac{\gamma}{1+\gamma|\Ec|}\right)\left[\hat{r}^{e} - 2\hat{q}^{e\intercal}\Gbf(b-\beta^*) + (b-\beta^*)^\intercal \Gbf^\intercal \hat{\Pbf}^{e}\Gbf(b-\beta^*)\right]\\
&\geq \lambda \|b - \beta^*\|_2^2.
\end{aligned}
\]
}
By Chebyshev's inequality and the fact that $\bar{w},w^0\in \Delta^{|\Ec|}$, we establish that
{\small
\[
\begin{aligned}
    &\|b-\beta^*\|_2 \left(\frac{1}{2}\left\|\nabla \widehat{\Phi}_\mu(b)\right\|_2 + \frac{\|h\|_2 \|\Gbf\|_2}{1+\gamma|\Ec|}\right) +\mu + \sum_{e\in \Ec}\left(\bar{w}_e - \frac{\gamma}{1+\gamma|\Ec|}\right)\left[\hat{r}^{e} - \hat{q}^{e\intercal}\Gbf(b-\beta^*)\right]\\
    &\quad\quad 
    - \sum_{e\in \Ec}\left(w^0_e - \frac{\gamma}{1+\gamma|\Ec|}\right)\left[\hat{r}^{e} - 2\hat{q}^{e\intercal}\Gbf(b-\beta^*) + (b-\beta^*)^\intercal \Gbf^\intercal \hat{\Pbf}^{e}\Gbf(b-\beta^*)\right]\\
&\geq \lambda \|b - \beta^*\|_2^2.
\end{aligned}
\]
}
We further apply the results in Lemma \ref{Lemma: rate of empirical norms} and establish that for any $s\geq 0$, with probability at least $1-2|\Ec|e^{-s}$
{\small
\[
\begin{aligned}
    &\sum_{e\in \Ec}\left(\bar{w}_e - \frac{\gamma}{1+\gamma|\Ec|}\right)\left[\hat{r}^{e} - \hat{q}^{e\intercal}\Gbf(b-\beta^*)\right] \\
    &\quad\quad - \sum_{e\in \Ec}\left(w^0_e - \frac{\gamma}{1+\gamma|\Ec|}\right)\left[\hat{r}^{e} - 2\hat{q}^{e\intercal}\Gbf(b-\beta^*) + (b-\beta^*)^\intercal \Gbf^\intercal \hat{\Pbf}^{e}\Gbf(b-\beta^*)\right]\\
    &\leq 2\left[\max_{e\in \Ec}\|\hat{\Pbf}\|_2 \|\Gbf\|_2 \|b-\beta^*\|_2^2 + 3\max_{e\in \Ec}\|\hat{q}^{e}\|_2\|\Gbf\|_2 \|b-\beta^*\|_2 + 2\max_{e\in \Ec}|\hat{r}^{e}|\right] \\
    &\lesssim \left(\sqrt{\frac{p+s}{n}} + \frac{p+s}{n}\right)(\|b-\beta^*\|_2^2 + \|b-\beta^*\|_2 + 1).
\end{aligned}
\]
}
With $n\geq p+s$, we combine the above two inequalities and obtain
{
\small
\begin{equation*}
    \lambda\|b-\beta^*\|_2^2 \leq \frac{1}{2}\|b-\beta^*\|_2 \left(\left\|\nabla \widehat{\Phi}_\mu(b)\right\|_2 + \frac{\|h\|_2 \|\Gbf\|_2}{1+\gamma|\Ec|}\right) +\mu + C\sqrt{\frac{p+s}{n}}(\|b-\beta^*\|_2^2 + \|b-\beta^*\|_2 + 1),
\end{equation*}
}
for some constant $C>0$.
When $\lambda \geq 2C\sqrt{\frac{p+s}{n}}$, 
the above inequality further implies that
{\small
\[
\frac{1}{2}\lambda\|b-\beta^*\|_2^2\leq \|b-\beta^*\|_2 \left(\frac{1}{2}\left\|\nabla \widehat{\Phi}_\mu(b)\right\|_2 + \frac{1}{2}\frac{\|h\|_2 \|\Gbf\|_2}{1+\gamma|\Ec|} + C\sqrt{\frac{p+s}{n}}\right) + \mu + C\sqrt{\frac{p+s}{n}}.
\]

According to Lemma \ref{lemma: expression of risk}, $\|\Gbf\|_2$ and $\|h\|_2$ are upper bounded by some constants; it holds that
{\small
\[
\lambda\|b - \beta^*\|_2^2\leq C\left(\left\|\nabla \widehat{\Phi}_\mu(b)\right\|_2 + \frac{1}{1+\gamma|\Ec|} + \sqrt{\frac{p+s}{n}}\right)\|b-\beta^*\|_2 + C\left(\mu + \sqrt{\frac{p+s}{n}}\right).
\]
}

Therefore, there exist constants $c_0,c_1>0$ such that if $n\geq (1\vee c_0/\lambda^2) p$, then for any value $0\leq s\leq (1\wedge c_1\lambda^2) n - p$, the following inequality holds with probability at least $1-2|\Ec| e^{-s}$:
{\small
\[
\|b-\beta^*\|_2\lesssim \frac{1}{\lambda}\left(\left\|\nabla \widehat{\Phi}_\mu(b)\right\|_2 + \frac{1}{1+\gamma|\Ec|} + \sqrt{\frac{p+s}{n}}\right) + \frac{1}{\sqrt{\lambda}}\left[\sqrt{\mu} + \left(\frac{p+s}{n}\right)^{1/4}\right].
\]
}
Since we consider $\lambda\gtrsim \sqrt{\frac{p+s}{n}}$, which implies that
\[
\frac{1}{\lambda}\sqrt{\frac{p+s}{n}} \lesssim \frac{1}{\sqrt{\lambda}}\left(\frac{p+s}{n}\right)^{1/4},
\]
we shall further obtain that
{\small
\[
\|b-\beta^*\|_2\lesssim \frac{1}{\lambda}\left(\left\|\nabla \widehat{\Phi}_\mu(b)\right\|_2 + \frac{1}{1+\gamma|\Ec|}\right) + \frac{1}{\sqrt{\lambda}}\left[\sqrt{\mu} + \left(\frac{p+s}{n}\right)^{1/4}\right].
\]
}
The proof completes.

\subsection{Proof of Proposition \ref{prop: alg conv penal}}
\label{appendix: proof of prop: alg conv penal}

Recall that in \eqref{eq: obj-raw penal empirical}, we define
\[
\widehat{\Phi}_\mu(b) = \max_{w\in \Delta^L} \left\{\sum_{e\in \Ec} \left(w_e - \frac{\gamma}{1+\gamma|\Ec|}\right)[\Y{e}-b^\intercal\X{e}]^2 - \mu\|w\|_2^2\right\}.
\]
We first introduce the following lemma, which establishes the smoothness property of the inner objective function of $\widehat{\Phi}_\mu(b)$.
The proof is provided in Appendix \ref{proof of lemma: smoothness joint}.
\begin{Lemma}
    Suppose Condition \ref{cond: smooth} holds. Then the function
    \[
    \sum_{e\in \Ec}\left(w_e - \frac{\gamma}{1+\gamma|\Ec|}\right)\widehat{\Eb}[Y^e - b^\intercal X^e]^2 - \mu\|w\|_2^2
    \]
    is $(2\kappa_1+2\kappa_2\sqrt{|\Ec|} + 2\mu)$-smooth in the joint variable $(b, w)$.
    \label{lemma: smoothness joint}
\end{Lemma}

We next utilize the following lemma to establish the smoothness of $\widehat{\Phi}_\mu(b)$. This lemma essentially adapts from Lemma 4.3 of \citet{lin2020gradient}. For completeness, we provide its proof in Appendix \ref{proof of lemma: smoothness of maximized function}. 
\begin{Lemma}[Lemma 4.3 \citep{lin2020gradient}]
    Suppose the function $f(x,y)$ is 
    \begin{itemize}
        \item $\ell$-smooth in the joint variable $(x,y)\in \mathcal{X}\times \mathcal{Y}$ such that
        \[
        \|\nabla f(x_1, y_1) - \nabla f(x_2, y_2)\|_2 \leq \ell \sqrt{\|x_1-x_2\|_2^2 + \|y_1-y_2\|_2^2},
        \]
        \item $\mu$-strongly concave in $y\in \mathcal{Y}$ for every fixed $x$, and
        \item the feasible set $\mathcal{Y}$ is convex and bounded.
    \end{itemize}
    Then the function $\max_{y\in \mathcal{Y}} f(x,y)$ is $(\ell+\ell^2/\mu)$-smooth in $x$.
    \label{lemma: smoothness of maximized function}
\end{Lemma}

From Lemma \ref{lemma: smoothness joint}, we have established that the function
\[
\sum_{e\in \Ec}\left(w_e - \frac{\gamma}{1+\gamma|\Ec|}\right)\widehat{\Eb}[Y^e - b^\intercal X^e]^2 - \mu\|w\|_2^2
\]
is $(2\kappa_1+2\kappa_2\sqrt{|\Ec|} + 2\mu)$-smooth in the joint variable $(b, w)$. Moreover, the term $-\mu\|w\|_2^2$ ensures that this function is $2\mu$-strongly concave in $w\in \Delta^L$. 
Applying Lemma \ref{lemma: smoothness of maximized function}, we obtain that $\widehat{\Phi}_\mu(b)$ is $L_\Phi$-smooth in $b$, with
\begin{equation}
    L_{\widehat{\Phi}} = 2\kappa_1+2\kappa_2\sqrt{|\Ec|} + 2\mu + \frac{2(\kappa_1+\kappa_2\sqrt{|\Ec|} + \mu)^2}{\mu}.
    \label{eq: smooth lipschitz}
\end{equation}

Given the smoothness of $\widehat{\Phi}_\mu(b)$, the following inequality holds for any $b_1, b_2$:
\begin{equation}
    \widehat{\Phi}_\mu(b_2) \leq \widehat{\Phi}_\mu(b_1) + [\nabla\widehat{\Phi}_\mu(b_1)]^\intercal (b_2 -b_1) + \frac{L_{\widehat{\Phi}}}{2}\|b_2 - b_1\|_2^2.
    \label{proof: eq hat Phi_mu smooth}
\end{equation}
Algorithm \ref{alg: GDmax penal} performs a gradient descent step on $\widehat{\Phi}_\mu(b)$ with step size
\[
\alpha = \frac{1}{L_{\widehat{\Phi}}}, \quad b_{t+1} = b_t - \alpha \nabla \widehat{\Phi}_\mu(b_t).
\]
Substituting $b_1 = b_t$ and $b_2 = b_{t+1}$ into \eqref{proof: eq hat Phi_mu smooth} gives
\[
\widehat{\Phi}_\mu(b_{t+1}) \leq \widehat{\Phi}_\mu(b_t) - \frac{1}{2L_{\widehat{\Phi}}}\|\nabla\widehat{\Phi}_\mu (b_t)\|_2^2.
\]
This inequality shows that each gradient step strictly decreases the objective unless the gradient vanishes.

Summing the above inequality over $t=0,...,T$ yields
\[
\widehat{\Phi}_\mu(b_{T+1})-\widehat{\Phi}_\mu(b_{0})\leq - \frac{1}{2L_{\widehat{\Phi}}}\sum_{t=0}^{T}\|\nabla\widehat{\Phi}_\mu (b_t)\|_2^2.
\]
Since $\widehat{\Phi}_\mu(b)\geq -\mu$ for all $b$, the left-hand side is bounded below by $-\mu - \widehat{\Phi}_\mu(b_0)$. Hence, we have
\[
\frac{1}{T+1}\sum_{t=0}^{T} \|\nabla\widehat{\Phi}_\mu (b_t)\|_2^2 \leq \frac{2L_{\widehat{\Phi}} (\widehat{\Phi}_\mu(b_0) + \mu)}{T+1}.
\]

Recall that the output of the algorithm selects the iterate with the smallest gradient norm, that is,
$$\hat{b}^\gamma := \argmin_{\{b_t\}_{t=0}^{T}} \|\nabla\widehat{\Phi}_\mu (b_t)\|_2,$$
Then,
\[
\|\nabla\widehat{\Phi}_\mu(\hat{b}^\gamma)\|_2^2\leq \frac{1}{T+1}\sum_{t=0}^{T} \|\nabla\widehat{\Phi}_\mu (b_t)\|_2^2 \leq \frac{2L_{\widehat{\Phi}} (\widehat{\Phi}_\mu(b_0)+\mu)}{T+1} 
\]

Lastly, we simplify the results mentioned above. Let $A:= \kappa_1+\kappa_2\sqrt{|\Ec|}$. Then it follows from \eqref{eq: smooth lipschitz} that
\[
L_{\widehat{\Phi}} = 2(A+\mu) + \frac{2(A+\mu)^2}{\mu} = 2\left(\frac{A^2}{\mu} + 3A + 2\mu\right).
\]
By treating $A$ as a constant, we have
\[
\frac{A^2}{\mu}\lesssim L_{\widehat{\Phi}}\lesssim \frac{A^2}{\mu} + A +\mu \asymp \frac{A^2}{\mu} + \mu.
\]
And if $\mu \lesssim \widehat{\Phi}_\mu(b_0)$, the output gradient norm satisfies that
\[
\|\nabla\widehat{\Phi}_\mu(\hat{b}^\gamma)\|_2^2\leq \frac{1}{T+1}\sum_{t=0}^{T} \|\nabla\widehat{\Phi}_\mu (b_t)\|_2^2 \lesssim \frac{1}{T}\left(\frac{A^2}{\mu} + \mu\right)\asymp \frac{1}{T}\left(\frac{\kappa_1^2+\kappa_2^2 |\Ec|}{\mu} +\mu\right).
\]

\subsection{Proof of Theorem \ref{thm: convergence theorem - population - any b}}
\label{appendix: proof of thm: convergence theorem - population - any b}

For convenience, for a predictor $b\in \Rb^p$ and weight $w\in \Delta^{|\Ec|}$, we define
\begin{equation}
   {\phi}(b, w) = \sum_{e\in \Ec}\left(w_e - \frac{\gamma}{1+\gamma|\Ec|}\right){\Eb}[\Y{e}-b^\intercal\X{e}]^2.
    \label{eq: phi b,w}
\end{equation}
By the definition of ${\Phi}(b)$ in \eqref{eq: obj trans - population}, we have
\[
{\Phi}(b) = \max_{w\in \Delta^{|\Ec|}}{\phi}(b,w).
\]
Given a predictor $b\in \Rb^p$, we define the set $\Wc(b)$ of maximizer weights such that:
\begin{equation}
    \Wc(b) := \argmax_{w\in \Delta^{|\Ec|}} {\phi}(b,w).
    \label{proof: eq maximizing points population}
\end{equation}
Since ${\phi}(b,w)$ is continuous towards $w$ for every $b$, and $\Delta^{|\Ec|}$ is a compact set, the set $\Wc(b)$ is non-empty by Weierstrass' Theorem.

\noindent\textbf{Step-1:} Express ${\Phi}(b)$ through the partial derivative $\frac{\partial {\phi}(b,\bar{w})}{\partial b}$ for any $\bar{w}\in \Wc(b)$.

It follows from \eqref{eq: expression of weighted average risks} that
{\small
\begin{equation*}
\begin{aligned}
    {\phi}(b, w) &= \frac{\sigma_Y^2 + 2h^\intercal \Gbf(b-\beta^*) + (b-\beta^*)^\intercal \left[\frac{1}{|\Ec|}\sum_{e\in \Ec}\Eb[\X{e}\XT{e}]\right]  (b-\beta^*)}{1+\gamma|\Ec|} + (b-\beta^*)^\intercal \Abm(w) (b - \beta^*).
\end{aligned}
\end{equation*}
}
Then for any maximizer weight $\bar{w}\in \Wc(b)$, we study the partial derivative of ${\phi}(b,\bar{w})$ as follows:
{\small
\begin{equation*}
    \begin{aligned}
        \frac{\partial {\phi}(b,\bar{w})}{\partial b} &=  \frac{2 \Gbf^\intercal h + 2 \left[\frac{1}{|\Ec|}\sum_{e\in \Ec}\Eb[\X{e}\XT{e}]\right]  (b-\beta^*)}{1+\gamma|\Ec|} + 2  \Abm(\Bar{w}) (b - \beta^*).
    \end{aligned}
\end{equation*}
}
Left multiplying the vector $\frac{1}{2}(b - \beta^*)$, we establish that
{\small
\begin{equation*}
\begin{aligned}
    \frac{1}{2}(b-\beta^*)^\intercal \frac{\partial {\phi}(b,\bar{w})}{\partial b} &=\frac{h^\intercal \Gbf(b-\beta^*) + (b-\beta^*)^\intercal \left[\frac{1}{|\Ec|}\sum_{e\in \Ec}\Eb[\X{e}\XT{e}]\right] (b-\beta^*)}{1+\gamma|\Ec|} \\
   &\quad\quad + (b-\beta^*)^\intercal  \Abm(\Bar{w})  (b - \beta^*).
\end{aligned}
\end{equation*}
}
Therefore,
{\small
\begin{equation}
\begin{aligned}
    {\Phi}(b) = {\phi}(b,\bar{w}) &= \frac{1}{2}(b-\beta^*)^\intercal\frac{\partial {\phi}(b,\bar{w})}{\partial b} + \frac{\sigma_Y^2 + h^\intercal \Gbf(b-\beta^*)}{1+\gamma|\Ec|}.
\end{aligned}
    \label{proof: eq phi origin b grad}
\end{equation}
}

\noindent\textbf{Step-2:} Establish the lower bound of ${\Phi}(b)$ via the weight $w^0$ specified in Condition \ref{cond: strict positive - A}. 

By the definition of $\bar{w}$, we have
{\small
\begin{equation*}
\begin{aligned}
    {\Phi}(b) &= \phi(b,\bar{w}) \geq {\phi}(b,w^0) \\
    &= \frac{\sigma_Y^2 + 2h^\intercal \Gbf(b-\beta^*) + (b-\beta^*)^\intercal \left[\frac{1}{|\Ec|}\sum_{e\in \Ec}\Eb[\X{e}\XT{e}]\right] (b-\beta^*)}{1+\gamma|\Ec|}  + (b-\beta^*)^\intercal  \Abm(w^0) (b - \beta^*).
\end{aligned}
\end{equation*}
}
Since $\Eb[\X{e}\XT{e}]\succeq 0$ for all $e\in \Ec$, and $\Abm(w^0)\succeq \lambda\Ibf$ as specified in Condition \ref{cond: strict positive - A}, we further establish that
{\small
\begin{equation}
\begin{aligned}
    {\Phi}(b) &\geq \frac{\sigma_Y^2 + 2h^\intercal \Gbf(b-\beta^*)}{1+\gamma|\Ec|} + \lambda \|b - \beta^*\|_2^2.
\end{aligned}
    \label{proof: eq phi origin b lower w^0}
\end{equation}
}

\noindent\textbf{Step-3:} Combine the results in the steps 1 and 2. It follows from \eqref{proof: eq phi origin b grad} and \eqref{proof: eq phi origin b lower w^0} that
{\small
\[
\begin{aligned}
    \frac{1}{2}(b-\beta^*)^\intercal\frac{\partial {\phi}(b,\bar{w})}{\partial b}  - \frac{h^\intercal \Gbf(b-\beta^*)}{1+\gamma|\Ec|} \geq \lambda \|b - \beta^*\|_2^2.
\end{aligned}
\]
}
By Chebyshev's inequality and the fact that $\bar{w},w^0\in \Delta^{|\Ec|}$, we establish that
{\small
\[
\begin{aligned}
    \|b-\beta^*\|_2 \left(\frac{1}{2}\left\|\frac{\partial {\phi}(b,\bar{w})}{\partial b}\right\|_2 + \frac{\|h\|_2 \|\Gbf\|_2}{1+\gamma|\Ec|}\right) \geq \lambda \|b - \beta^*\|_2^2.
\end{aligned}
\]
}

According to Lemma \ref{lemma: expression of risk} that $\Gbf$ is a constant full-rank matrix and $\|h\|\leq c\sigma_Y^2$ for some constant $c>0$, it holds that
{\small
\begin{equation*}
    \lambda\|b - \beta^*\|_2\lesssim \left\|\frac{\partial {\phi}(b,\bar{w})}{\partial b}\right\|_2 + \frac{1}{1+\gamma|\Ec|}.
\end{equation*}
}

\noindent\textbf{Step-4:} Upper bound $\|b-\beta^*\|_2$ with subdifferential's distance ${\rm Dist}(0, \partial {\Phi}(b)).$

We know that the preceding inequality holds for any $\bar{w}\in \Wc(b)$, where $\Wc(b)$ is defined in \eqref{proof: eq maximizing points population}. Then the following inequality holds:
\begin{equation*}
   \|b-\beta^*\|_2\lesssim \frac{1}{\lambda}\left(\inf_{{w}\in \Wc(b)}\left\|\frac{\partial {\phi}(b,\bar{w})}{\partial b}\right\|_2 + \frac{1}{1+\gamma|\Ec|} \right).
\end{equation*}

We next introduce the lemma to facilitate our control of the term $\inf_{{w}\in \Wc(b)}\left\|\frac{\partial {\phi}(b,\bar{w})}{\partial b}\right\|_2$. This lemma essentially follows Danskin's theorem \citep{danskin1966theory}\citep[Proposition~B.25]{bertsekas1997nonlinear}, and its proof is provided in Section \ref{subsec: proof of lemma: danskin's theorem}.
\begin{Lemma}
\label{Lemma: Danskin's Theorem}
    Suppose $f(x,z)$ is a continuous function of two arguments, $f: \Rb^p\times \mathcal{Z}\to \Rb$, where $\mathcal{Z}$ is a compact set. Define 
    $F(x) = \max_{z\in \mathcal{Z}}f(x,z),$ and the set of maximizing points $\mathcal{Z}_0(x)$ as:
    $\mathcal{Z}_0(x) = \left\{\bar{z}: ~~f(x,\bar{z}) = \max_{z\in \mathcal{Z}}f(x,z)\right\}$.
    If $f(x,z)$ is differentiable with respect to $x$ for all $z\in \mathcal{Z}$, and if $\partial f /\partial x$ is continuous with respect to $z$ for all $x$, then the subdifferential of $F(x)$ is given by
    \[
    \partial F(x) = {\rm conv}\left\{\frac{\partial f(x,z)}{\partial x}:~~z\in \mathcal{Z}_0(x)\right\},
    \]
    where ${\rm conv}$ indicates the convex hull operation. Moreover, if $\mathcal{Z}$ is a convex set and $f(x,z)$ is a linear function with respect to $z$ for all $x\in \Rb^p$, the ${\rm conv}$ operation can be removed, and it holds that
    \[
    \partial F(x) = \left\{\frac{\partial f(x,z)}{\partial x}:~~z\in \mathcal{Z}_0(x)\right\}.
    \]
\end{Lemma}

We shall verify the conditions of the above lemma hold for our $\phi(b,w)$ with $b\in \Rb^p$ and $w\in \Delta^L$. Then we apply this lemma to obtain that:
\[
\partial{\Phi}(b) = \left\{\frac{\partial}{\partial b}{\phi}(b, \bar{w}),~~\bar{w}\in \Wc(b)\right\}.
\]
Therefore, it holds that
${\rm Dist}(0, \partial {\Phi}(b)) = \inf_{\bar{w}\in \Wc(b)}\left\|\frac{\partial}{\partial b}{\phi}(b, \bar{w})\right\|_2$, and 
\begin{equation*}
    \|b-\beta^*\|_2\lesssim \frac{1}{\lambda}\left({\rm Dist}(0, \partial {\Phi}(b)) + \frac{1}{1+\gamma|\Ec|} \right).
\end{equation*}

\subsection{Proof of Theorem \ref{thm: convergence theorem - empirical - any b}}
\label{appendix: proof of thm: convergence theorem - empirical - any b}
The proof of this theorem essentially follows that of Theorem \ref{thm: convergence theorem - population - any b}, with the additional work of controlling finite-sample error terms.

For convenience, for a predictor $b\in \Rb^p$ and weight $w\in \Delta^{|\Ec|}$, we define
\begin{equation}
    \hat{\phi}(b, w) = \sum_{e\in \Ec}\left(w_e - \frac{\gamma}{1+\gamma|\Ec|}\right)\widehat{\Eb}[\Y{e}-b^\intercal\X{e}]^2.
    \label{eq: hat phi b,w}
\end{equation}
By the definition of $\widehat{\Phi}(b)$ in \eqref{eq: obj unpenal empirical}, we have
\[
\widehat{\Phi}(b) = \max_{w\in \Delta^{|\Ec|}}\hat{\phi}(b,w).
\]
Given a predictor $b\in \Rb^p$, we define the set $\widehat\Wc(b)$ of maximizer weights such that:
\begin{equation}
    \widehat\Wc(b) := \argmax_{w\in \Delta^{|\Ec|}} \hat{\phi}(b,w).
    \label{proof: eq maximizing points empirical}
\end{equation}
Since $\hat{\phi}(b,w)$ is continuous towards $w$ for every $b$, and $\Delta^{|\Ec|}$ is a compact set, the set $\widehat\Wc(b)$ is non-empty by Weierstrass' Theorem.

\noindent\textbf{Step-1:} Express $\widehat{\Phi}(b)$ through the partial derivative $\frac{\partial \hat{\phi}(b,\bar{w})}{\partial b}$ for any $\bar{w}\in \widehat\Wc(b)$.

It follows from \eqref{proof: eq weighted average empirical risks} that
{\small
\begin{equation*}
\begin{aligned}
    \hat{\phi}(b, w) &= \frac{\sigma_Y^2 + 2h^\intercal \Gbf(b-\beta^*) + (b-\beta^*)^\intercal \left[\frac{1}{|\Ec|}\sum_{e\in \Ec}\Eb[\X{e}\XT{e}]\right] (b-\beta^*)}{1+\gamma|\Ec|}+(b-\beta^*)^\intercal  \Abm(w)  (b - \beta^*) \\
   &\quad\quad +  \sum_{e\in \Ec}\left(w_e - \frac{\gamma}{1+\gamma|\Ec|}\right)\left[\hat{r}^{e} - 2\hat{q}^{e\intercal}\Gbf(b-\beta^*) + (b-\beta^*)^\intercal \Gbf^\intercal \hat{\Pbf}^{e}\Gbf(b-\beta^*)\right].
\end{aligned}
\end{equation*}
}
Then for any maximizer weight $\bar{w}\in \widehat\Wc(b)$, we study the partial derivative of $\hat{\phi}(b,\bar{w})$ as follows:
{\small
\begin{equation*}
    \begin{aligned}
        \frac{\partial \hat{\phi}(b,\bar{w})}{\partial b} &=  \frac{2 \Gbf^\intercal h + 2 \left[ \frac{1}{|\Ec|}\sum_{e\in \Ec}\Eb[\X{e}\XT{e}]\right] (b-\beta^*)}{1+\gamma|\Ec|} + 2  \Abm(\Bar{w}) (b - \beta^*) \\
        &\quad + \sum_{e\in \Ec}\left(\Bar{w}_e - \frac{\gamma}{1+\gamma|\Ec|}\right)\left[-2\Gbf^\intercal\hat{q}^{e} + 2\Gbf^\intercal \hat{\Pbf}^{e}\Gbf(b-\beta^*)\right].
    \end{aligned}
\end{equation*}
}
Left multiplying the vector $\frac{1}{2}(b - \beta^*)$, we establish that
{\small
\begin{equation*}
\begin{aligned}
    \frac{1}{2}(b-\beta^*)^\intercal \frac{\partial \hat{\phi}(b,\bar{w})}{\partial b} &=\frac{h^\intercal \Gbf(b-\beta^*) + (b-\beta^*)^\intercal \left[ \frac{1}{|\Ec|}\sum_{e\in \Ec}\Eb[\X{e}\XT{e}]\right](b-\beta^*)}{1+\gamma|\Ec|} \\
   &\quad\quad + (b-\beta^*)^\intercal  \Abm(\Bar{w})(b - \beta^*) \\
   &\quad\quad +  \sum_{e\in \Ec}\left(\Bar{w}_e - \frac{\gamma}{1+\gamma|\Ec|}\right)\left[-\hat{q}^{e\intercal}\Gbf(b-\beta^*) + (b-\beta^*)^\intercal \Gbf^\intercal \hat{\Pbf}^{e}\Gbf(b-\beta^*)\right].
\end{aligned}
\end{equation*}
}
Therefore,
{\small
\begin{equation}
\begin{aligned}
    \widehat{\Phi}(b) = \hat{\phi}(b,\bar{w}) &= \frac{1}{2}(b-\beta^*)^\intercal\frac{\partial \hat{\phi}(b,\bar{w})}{\partial b} + \frac{\sigma_Y^2 + h^\intercal \Gbf(b-\beta^*)}{1+\gamma|\Ec|}\\
    &\quad\quad + \sum_{e\in \Ec}\left(\Bar{w}_e - \frac{\gamma}{1+\gamma|\Ec|}\right)\left[\hat{r}^{e} - \hat{q}^{e\intercal}\Gbf(b-\beta^*)\right].
\end{aligned}
    \label{proof: eq hat phi origin b grad}
\end{equation}
}

\noindent\textbf{Step-2:} Establish the lower bound of $\widehat{\Phi}(b)$ via the weight $w^0$ specified in Condition \ref{cond: strict positive - A}. 

By the definition of $\bar{w}$, we have
{\small
\begin{equation}
\begin{aligned}
    \widehat{\Phi}(b) &= \hat{\phi}(b,\bar{w}) \geq \hat{\phi}(b,w^0) \\
    &= \frac{\sigma_Y^2 + 2h^\intercal \Gbf(b-\beta^*) + (b-\beta^*)^\intercal \left[\frac{1}{|\Ec|}\sum_{e\in \Ec}\Eb[\X{e}\XT{e}]\right] (b-\beta^*)}{1+\gamma|\Ec|} \\
   &\quad\quad + (b-\beta^*)^\intercal \Abm(w^0) (b - \beta^*) \\
   &\quad\quad +  \sum_{e\in \Ec}\left(w^0_e - \frac{\gamma}{1+\gamma|\Ec|}\right)\left[\hat{r}^{e} - 2\hat{q}^{e\intercal}\Gbf(b-\beta^*) + (b-\beta^*)^\intercal \Gbf^\intercal \hat{\Pbf}^{e}\Gbf(b-\beta^*)\right].
\end{aligned}
\label{proof: eq hat phi b lower w^0 inter}
\end{equation}
}
Since $\Eb[\X{e}\XT{e}]\succeq 0$ for all $e\in \Ec$, and $\Abm(w^0)\succeq \lambda\Ibf$ as specified in Condition \ref{cond: strict positive - A}, we further establish that
{\small
\begin{equation}
\begin{aligned}
    \widehat{\Phi}(b) &\geq \frac{\sigma_Y^2 + 2h^\intercal \Gbf(b-\beta^*)}{1+\gamma|\Ec|} + \lambda \|b - \beta^*\|_2^2 \\
    &\quad \quad + \sum_{e\in \Ec}\left(w^0_e - \frac{\gamma}{1+\gamma|\Ec|}\right)\left[\hat{r}^{e} - 2\hat{q}^{e\intercal}\Gbf(b-\beta^*) + (b-\beta^*)^\intercal \Gbf^\intercal \hat{\Pbf}^{e}\Gbf(b-\beta^*)\right]
\end{aligned}
    \label{proof: eq hat phi origin b lower w^0}
\end{equation}
}

\noindent\textbf{Step-3:} Combine the results in the steps 1 and 2. It follows from \eqref{proof: eq hat phi origin b grad} and \eqref{proof: eq hat phi origin b lower w^0} that
{\small
\[
\begin{aligned}
    &\frac{1}{2}(b-\beta^*)^\intercal\frac{\partial \hat{\phi}(b,\bar{w})}{\partial b}  - \frac{h^\intercal \Gbf(b-\beta^*)}{1+\gamma|\Ec|} + \sum_{e\in \Ec}\left(\bar{w}_e - \frac{\gamma}{1+\gamma|\Ec|}\right)\left[\hat{r}^{e} - \hat{q}^{e\intercal}\Gbf(b-\beta^*)\right]\\
    &\quad\quad 
    - \sum_{e\in \Ec}\left(w^0_e - \frac{\gamma}{1+\gamma|\Ec|}\right)\left[\hat{r}^{e} - 2\hat{q}^{e\intercal}\Gbf(b-\beta^*) + (b-\beta^*)^\intercal \Gbf^\intercal \hat{\Pbf}^{e}\Gbf(b-\beta^*)\right]\\
&\geq \lambda \|b - \beta^*\|_2^2.
\end{aligned}
\]
}
By Chebyshev's inequality and the fact that $\bar{w},w^0\in \Delta^{|\Ec|}$, we establish that
{\small
\[
\begin{aligned}
    &\|b-\beta^*\|_2 \left(\frac{1}{2}\left\|\frac{\partial \hat{\phi}(b,\bar{w})}{\partial b}\right\|_2 + \frac{\|h\|_2 \|\Gbf\|_2}{1+\gamma|\Ec|}\right) + \sum_{e\in \Ec}\left(\bar{w}_e - \frac{\gamma}{1+\gamma|\Ec|}\right)\left[\hat{r}^{e} - \hat{q}^{e\intercal}\Gbf(b-\beta^*)\right]\\
    &\quad\quad 
    - \sum_{e\in \Ec}\left(w^0_e - \frac{\gamma}{1+\gamma|\Ec|}\right)\left[\hat{r}^{e} - 2\hat{q}^{e\intercal}\Gbf(b-\beta^*) + (b-\beta^*)^\intercal \Gbf^\intercal \hat{\Pbf}^{e}\Gbf(b-\beta^*)\right]\\
&\geq \lambda \|b - \beta^*\|_2^2.
\end{aligned}
\]
}
We further apply the results in Lemma \ref{Lemma: rate of empirical norms} to control the components involving the finite-sample error terms $\hat\Pbf^{e}, \hat{q}^{e}, \hat{r}^{e}$ on the left-hand side of the above inequality. We establish that for any $s\geq 0$, with probability at least $1-2|\Ec|e^{-s}$
{\small
\begin{equation}
    \begin{aligned}
    &\sum_{e\in \Ec}\left(\bar{w}_e - \frac{\gamma}{1+\gamma|\Ec|}\right)\left[\hat{r}^{e} - \hat{q}^{e\intercal}\Gbf(b-\beta^*)\right] \\
    &\quad\quad - \sum_{e\in \Ec}\left(w^0_e - \frac{\gamma}{1+\gamma|\Ec|}\right)\left[\hat{r}^{e} - 2\hat{q}^{e\intercal}\Gbf(b-\beta^*) + (b-\beta^*)^\intercal \Gbf^\intercal \hat{\Pbf}^{e}\Gbf(b-\beta^*)\right]\\
    &\leq 2\left[\max_{e\in \Ec}\|\hat{\Pbf}\|_2 \|\Gbf(b-\beta^*)\|_2^2 + 3\max_{e\in \Ec}\|\hat{q}^{e}\|_2\|\Gbf(b-\beta^*)\|_2 + 2\max_{e\in \Ec}|\hat{r}^{e}|\right] \\
    &\lesssim \left(\sqrt{\frac{p+s}{n}} + \frac{p+s}{n}\right)(\|b-\beta^*\|_2^2 + \|b-\beta^*\|_2 + 1).
\end{aligned}
\label{eq: finite sample small terms upper bound}
\end{equation}
}
With $n\geq p+s$, we combine the above two inequalities and obtain
{
\small
\begin{equation*}
    \lambda\|b-\beta^*\|_2^2 \lesssim\|b-\beta^*\|_2 \left(\left\|\frac{\partial \hat{\phi}(b,\bar{w})}{\partial b}\right\|_2 + \frac{\|h\|_2 \|\Gbf\|_2}{1+\gamma|\Ec|}\right) + \sqrt{\frac{p+s}{n}}(\|b-\beta^*\|_2^2 + \|b-\beta^*\|_2 + 1)
\end{equation*}
}
When $\sqrt{\frac{p+s}{n}} \lesssim \lambda$, together with Lemma \ref{lemma: expression of risk} which indicates that $\|\Gbf\|_2$ and $\|h\|_2$ are bounded, the above inequality further implies that
{\small
\[
\lambda\|b-\beta^*\|_2^2\lesssim \|b-\beta^*\|_2 \left( \left\|\frac{\partial \hat{\phi}(b,\bar{w})}{\partial b}\right\|_2 + \frac{1}{1+\gamma|\Ec|} + \sqrt{\frac{p+s}{n}}\right) + \sqrt{\frac{p+s}{n}}
\]
}
Then we show that
{\small
\begin{equation}
    \|b-\beta^*\|_2\lesssim\frac{1}{\lambda}\left(\left\|\frac{\partial \hat{\phi}(b,\bar{w})}{\partial b}\right\|_2 + \frac{1}{1+\gamma|\Ec|} + \sqrt{\frac{p+s}{n}}\right) + \frac{1}{\sqrt{\lambda}}\left(\frac{p+s}{n}\right)^{1/4}.
    \label{proof: eq upper bound with one wbar}
\end{equation}
}

\noindent\textbf{Step-4:} Upper bound $\|b-\beta^*\|_2$ with subdifferential's distance ${\rm Dist}(0, \partial \widehat{\Phi}(b)).$

We know that the above inequality \eqref{proof: eq upper bound with one wbar} holds for any $\bar{w}\in \widehat\Wc(b)$, where $\widehat\Wc(b)$ is defined in \eqref{proof: eq maximizing points empirical}. Then the following inequality holds:
\begin{equation*}
   \|b-\beta^*\|_2\lesssim\frac{1}{\lambda}\left(\inf_{\bar{w}\in \widehat\Wc(b)}\left\|\frac{\partial \hat{\phi}(b,\bar{w})}{\partial b}\right\|_2 + \frac{1}{1+\gamma|\Ec|} + \sqrt{\frac{p+s}{n}}\right) + \frac{1}{\sqrt{\lambda}}\left(\frac{p+s}{n}\right)^{1/4}.
\end{equation*}
Now we apply Lemma \ref{Lemma: Danskin's Theorem} to obtain that:
\[
\partial\widehat{\Phi}(b) = \left\{\frac{\partial}{\partial b}\hat{\phi}(b, \bar{w}),~~\bar{w}\in \widehat\Wc(b)\right\}.
\]
Therefore, it holds that
${\rm Dist}(0, \partial \widehat{\Phi}(b)) = \inf_{\bar{w}\in \widehat\Wc(b)}\left\|\frac{\partial}{\partial b}\hat{\phi}(b, \bar{w})\right\|_2$, and 
\begin{equation*}
    \|b-\beta^*\|_2\lesssim \frac{1}{\lambda}\left({\rm Dist}(0, \partial \widehat{\Phi}(b)) + \frac{1}{1+\gamma|\Ec|} + \sqrt{\frac{p+s}{n}}\right) + \frac{1}{\sqrt{\lambda}}\left(\frac{p+s}{n}\right)^{1/4}.
\end{equation*}
Since we consider $\lambda\gtrsim \sqrt{\frac{p+s}{n}}$, the above inequality shall be simplified as:
\[
    \|b-\beta^*\|_2\lesssim \frac{1}{\lambda}\left({\rm Dist}(0, \partial \widehat{\Phi}(b)) + \frac{1}{1+\gamma|\Ec|}\right) + \frac{1}{\sqrt{\lambda}}\left(\frac{p+s}{n}\right)^{1/4}.
\]

\subsection{Proof of Theorem \ref{thm: convergence theorem - empirical - alg output}}
\label{Appendix: proof of subsec computation algorithm}
We first introduce the concept of weak convexity. While $\widehat{\Phi}(\cdot)$ itself is not convex, it can be shown to possess weak convexity, which generalizes the idea of convexity by allowing for a certain controlled degree of curvature. This distinction is crucial, as weak convexity provides a meaningful balance between convex and highly non-convex behavior, enabling the development of more robust optimization techniques \citep{poliquin1992amenable, poliquin1996prox}.
\begin{Definition}[Weak convexity]
The function $f: \Rb^p \to \Rb$ is $\rho$-weakly convex if $f(\cdot) + \frac{\rho}{2}\|\cdot\|^2$ is convex.
\label{def: weak convex}
\end{Definition}
The following Lemma demonstrates that $\widehat{\Phi}(b)$ is weakly convex with high probability, whose proof is provided in Appendix \ref{subsec: proof of lemma Phi b weakly convex}.
\begin{Lemma}
    Under the additive intervention regime, suppose Condition \ref{cond: subgaussian} holds, and the sample size satisfies $n\geq p$. Then there exists a positive constant $\rho>0$ such that for any $s\in [0, n-p]$, with probability at least $1-2|\Ec|e^{-s}$, the function $\widehat{\Phi}(b)$ is $\rho$-weakly convex.
    \label{Lemma: hat Phi b weakly convex}
\end{Lemma}

After we generalized the convexity concept for $\widehat{\Phi}(b)$, now we will move on discussing the challenge arising from the non-smoothness landscape of $\widehat{\Phi}(b)$. To facilitate discussion, we present the definition of nearly stationary points as follows.
\begin{Definition}[Nearly Stationary Point]
    A point $b\in \Rb^p$ is an $\epsilon$-stationary point of the function $h:\Rb^p\to \Rb\cup\{\infty\}$ for some tolerance $\epsilon\geq 0$ if ${\rm Dist}(0, \partial h(b)):=\inf\{\|\zeta\|^2_2: \zeta\in \partial h(b)\}\leq \epsilon$, where $\partial h(b)$ is the subdifferential defined in Definition \ref{def: subdiff}. We call a point $b\in \Rb^p$ a nearly stationary point if the tolerance $\epsilon$ is small enough.
    \label{def: nearly stationary}
\end{Definition}

Identifying nearly stationary points, for non-smooth functions, like $\widehat{\Phi}(b)$, is particularly challenging due to the sharp transitions in the function's sub-gradient. In smooth optimization, the gradient provides a clear, continuous direction toward stationary points, making it straightforward to determine whether a point is close enough to being stationary. However, in non-smooth functions, there can be abrupt changes in the sub-gradient, and sub-gradient fails to provide insightful criterion on the closeness of a point to being stationary.
A simple example is $|x|$ where the only stationary point is $0$, but any other point $x\neq 0$ is not an $\epsilon$-stationary point with $\epsilon < 1$, no matter how close $x$ is to $0$.

To address this issue, we employ a smooth approximation technique using the Moreau envelope \citep{davis2018stochastic, davis2018complexity, davis2019proximally, zhang2018convergence, rafique2022weakly}. By smoothing $\widehat{\Phi}(\cdot)$, we transform the non-smooth landscape into a differentiable one, allowing for a more gradual transition in the gradient. This smoother approximation enables algorithms to track $\epsilon$-stationary points more effectively.

We define the \emph{Moreau envelope} of function $\widehat{\Phi}(\cdot)$ with the parameter $\upsilon > 0$ as follows:
\begin{equation}
    \widehat{\Phi}_\upsilon(b) = \min_{\zeta\in \Rb^p}\left\{\widehat{\Phi}(\zeta) + \frac{1}{2\upsilon}\|\zeta - b\|^2\right\}.
    \label{eq: moreau env def upsilon}
\end{equation}
We shall establish that the Moreau envelope $\widehat{\Phi}_{\upsilon}(b)$ is differentiable with respect to $b$, with gradient denoted as $\nabla \widehat{\Phi}_{\upsilon}(b)$ as long as the parameter $\upsilon$ is small enough in the following lemma. The lemma's proof is provided in Section \ref{subsec: Proof of lemma subdiff dist bound}. 
\begin{Lemma}
    \label{Lemma: subdiff dist bound}
    If $\widehat{\Phi}(b)$ is $\rho$-weakly convex, as long as $\upsilon < \rho^{-1}$, its Moreau envelope $\widehat{\Phi}_\upsilon(b)$, defined in \eqref{eq: moreau env def upsilon}, is differentiable with the gradient given by
    \begin{equation*}
         \nabla \widehat{\Phi}_\upsilon(b) = \upsilon^{-1}\left(b - {\rm prox}_{\upsilon \widehat{\Phi}}(b)\right),
    \end{equation*}
    where ${\rm prox}_{\upsilon\widehat{\Phi}}(b)$ is the proximal mapping defined as follows:
    \[
    {\rm prox}_{\upsilon\widehat{\Phi}}(b) = \argmin_{\zeta\in \Rb^p} \left\{\widehat{\Phi}(\zeta) + \frac{1}{2\upsilon}\|\zeta - b\|_2^2\right\}.
    \]
    Moreover, the proximal mapping ${\rm prox}_{\upsilon \widehat{\Phi}}(b)$ satisfies that
    \begin{equation}
        \|{\rm prox}_{\upsilon\widehat{\Phi}}(b) - b\| = \upsilon \|\nabla\widehat{\Phi}_{\upsilon}(b)\|,
        \label{eq: b_prox - b dist upsilon}
    \end{equation}
    and
    \begin{equation}
    \label{eq: eq: b_prox subdiff bounded upsilon}
        {\rm Dist}\left(0, \partial \widehat{\Phi}({\rm prox}_{\upsilon\widehat{\Phi}}(b))\right) \leq \|\nabla \widehat{\Phi}_{\upsilon}(b)\|.
    \end{equation}
\end{Lemma}
The results in \eqref{eq: b_prox - b dist upsilon} and \eqref{eq: eq: b_prox subdiff bounded upsilon} shed light on the effective identification of $\epsilon$-stationary points. Suppose that we have some point $\Bar{b}$, on which the gradient of the Moreau envelope is small enough such that $\|\nabla\widehat{\Phi}_{\upsilon}(\bar{b})\| \leq \epsilon$. The inequality \eqref{eq: eq: b_prox subdiff bounded upsilon} indicates that the proximal point ${\rm prox}_{\upsilon\widehat{\Phi}}(\Bar{b})$ is a $\epsilon$-stationary point with ${\rm Dist}\left(0, \partial \widehat{\Phi}({\rm prox}_{\upsilon\widehat{\Phi}}(b))\right) \leq \epsilon.$ Moreover, \eqref{eq: b_prox - b dist upsilon} establishes that the point $\bar{b}$ is $\upsilon\epsilon$-closed to the $\epsilon$-stationary ${\rm prox}_{\upsilon\widehat{\Phi}}(\Bar{b})$. To sum up, a small gradient $\|\nabla\widehat{\Phi}_{\upsilon}(\bar{b})\|$ implies that the point $\bar{b}$ is \emph{close} to some \emph{nearly stationary} point of the function $\widehat{\Phi}(\cdot)$.

The following lemma shows that the subdifferential $\partial \widehat{\Phi}(b)$ is bounded, whose proof is provided in Section \ref{subsec: proof of lemma bounded subdiff}. We emphasize that it is essential and commonly adopted to consider the bounded subdifferential when analyzing nonsmooth functions in the literature, such as those in \citet{davis2018stochastic, nemirovski2009robust}.
\begin{Lemma}
    \label{Lemma: bounded subdiff}
    Suppose Condition \ref{cond: bounded gradient} holds. Then there exists a constant $L>0$ such that
    \begin{equation}
            \label{eq: bounded subdiff}
        \sup\left\{\|\zeta\|_2^2,~~\zeta\in \partial \widehat{\Phi}(b)\right\} \leq L^2,~~\textrm{for all } b\in \mathcal{B},
    \end{equation}
    where $\mathcal{B}\subset \Rb^p$ is bounded set imposed in Condition \ref{cond: bounded gradient}.
\end{Lemma}

We now summarize the results we have established till now.
Under Condition \ref{cond: subgaussian}, we know that $\widehat{\Phi}(\cdot)$ is $\rho$-weakly convex function with high probability by applying the result in Lemma \ref{Lemma: hat Phi b weakly convex}, for some $\rho> 0$. Then we will follow Lemma \ref{Lemma: subdiff dist bound} by setting the parameter $\upsilon = \frac{1}{2\rho}$ in \eqref{eq: moreau env def upsilon} to obtain a differentiable Moreau envelope of $\widehat{\Phi}(b)$. Together with the bounded subdifferential result in Lemma \ref{Lemma: bounded subdiff}, we will demonstrate the convergence of Algorithm \ref{alg: GDmax subgrad} with vanishing norm of the Moreau envelope $\|\nabla \widehat{\Phi}_{1/(2\rho)} (b)\|_2$.
\begin{Lemma}
    Suppose $\widehat{\Phi}(b)$ is $\rho$-weakly convex, and \eqref{eq: bounded subdiff} holds. There exists some constant $R$ such that $\widehat{\Phi}_{1/(2\rho)}(b_0)\leq R$ with the initial point $b_0$.
    We set the step sizes of Algorithm \ref{alg: GDmax subgrad} as
    $\alpha = \sqrt{\frac{R}{\rho L^2 (T+1)}}$.
    Then it holds that
    \[
    \frac{1}{T+1}\sum_{t=0}^T \|\nabla \widehat{\Phi}_{1/(2\rho)} (b_t)\|_2^2 \leq 4 L\sqrt{\frac{\rho R}{T+1}}.
    \]
    \label{Lemma: moreau norm bound}
\end{Lemma}
The proof of Lemma essentially follows from \citet[Theorem 2.1]{davis2018stochastic}. For completeness, we will provide its proof in Section \ref{subsec: proof of lemma moreau norm bound}. 
In Algorithm \ref{alg: GDmax subgrad}, for each iteration $b_t$, we will compute its proximal mapping point with $\upsilon = 1/(2\rho)$:
\[
{\rm prox}_{1/(2\rho)\widehat{\Phi}}(b_t) = \argmin_{x\in \Rb^p}\left\{\widehat{\Phi}(x) + \rho\|x - b_t\|^2\right\}.
\]
And it follows from Lemma \ref{Lemma: subdiff dist bound} that
\[
\|{\rm prox}_{1/(2\rho)\widehat{\Phi}}(b_t) - b_t\| = \frac{1}{2\rho}\|\nabla \widehat{\Phi}_{1/(2\rho)}(b_t)\|.
\]
Since the algorithm output $\hat{b}^\gamma$ among $\{b_t\}_{t=0}^{T}$ with the smallest $\|{\rm prox}_{1/(2\rho)\widehat{\Phi}}(b_t) - b_t\|$ such that
\[
\hat{b}^\gamma \in \argmin_{\{b_t\}_{t=0}^T}\|{\rm prox}_{1/(2\rho)\widehat{\Phi}}(b_t) - b_t\|,
\]
it follows that $\hat{b}^\gamma$ achieves the smallest gradient of the Moreau envelope:
\[
\hat{b}^\gamma \in \argmin_{\{b_t\}_{t=0}^T}\|\nabla \widehat{\Phi}_{1/(2\rho)} (b_t)\|^2.
\]
Applying the results in Lemma \ref{Lemma: moreau norm bound}, we shall establish that
\begin{equation}
    \|\nabla \widehat{\Phi}_{1/(2\rho)} (\hat{b}^\gamma)\|^2_2 \leq \frac{1}{T+1}\sum_{t=0}^T \|\nabla \widehat{\Phi}_{1/(2\rho)} (b_t)\|^2_2 \leq 4 L\sqrt{\frac{\rho R}{T+1}}.
    \label{eq: moreu env norm result}
\end{equation}

Combining the above results in Lemmas \ref{Lemma: hat Phi b weakly convex}, \ref{Lemma: subdiff dist bound}, \ref{Lemma: bounded subdiff}, and \ref{Lemma: moreau norm bound}, together with \eqref{eq: moreu env norm result}, we have established the following proposition.
\begin{Proposition}
    Under the additive intervention regime, suppose Conditions \ref{cond: subgaussian}, \ref{cond: bounded gradient} hold, and the sample size satisfies $n\geq p$. Then there exists a positive constant $\rho>0$ such that: for any $s\in [0, n-p]$, with probability at least $1-2|\Ec| e^{-s}$, the output $\hat{b}^\gamma$ of Algorithm \ref{alg: GDmax subgrad} satisfies:
    \[
    \|\nabla \widehat{\Phi}_{\upsilon} (\hat{b}^\gamma)\|^2 \lesssim T^{-1/2}.
    \]
    where we set $\upsilon = \frac{1}{2\rho}$ and the step size of Algorithm \ref{alg: GDmax subgrad} as $\alpha \asymp T^{-1/2}$.
    \label{prop: alg convergence - 1/2rho}
\end{Proposition}

For notation convenience, we denote the proximal mapping point of the output $\hat{b}^\gamma$ as:
\[
[\hat{b}^\gamma]_{\rm prox} = {\rm prox}_{\upsilon\widehat{\Phi}}(\hat{b}^\gamma).
\]
We apply the result \eqref{eq: b_prox - b dist upsilon} in Lemma \ref{Lemma: subdiff dist bound} and Proposition \ref{prop: alg convergence - 1/2rho} to establish that

\begin{equation*}
    \|[\hat{b}^\gamma]_{\rm prox} - \hat{b}^\gamma\|_2 = \upsilon\|\nabla\widehat{\Phi}_{1/(2\rho)}(\hat{b}^\gamma)\|_2 \lesssim T^{-1/4}.
\end{equation*}
By triangle inequality, the above inequality implies that the distance between the output $\hat{b}^\gamma$ and the causal effect $\beta^*$ is controlled by
\begin{equation}
\begin{aligned}
     \|\hat{b}^\gamma - \beta^*\|_2 &= \|\hat{b}^\gamma - [\hat{b}^\gamma]_{\rm prox} + [\hat{b}^\gamma]_{\rm prox} - \beta^*\|_2 \leq \|\hat{b}^\gamma - [\hat{b}^\gamma]_{\rm prox}\|_2 + \|[\hat{b}^\gamma]_{\rm prox} - \beta^*\|_2 \\
     &\lesssim T^{-1/4} + \|[\hat{b}^\gamma]_{\rm prox} - \beta^*\|_2.
\end{aligned}
    \label{eq: output b* - beta*}
\end{equation}
Next, we are going to control the remaining term $\|[\hat{b}^\gamma]_{\rm prox} - \beta^*\|_2$. 

Applying \eqref{eq: eq: b_prox subdiff bounded upsilon} in Lemma \ref{Lemma: subdiff dist bound}, it holds that
\begin{equation}
    {\rm Dist}\left(0, \partial \widehat{\Phi}\left([\hat{b}^\gamma]_{\rm prox}\right)\right) \leq \|\nabla \widehat{\Phi}_{\upsilon} (\hat{b}^\gamma)\|_2 \lesssim T^{-1/4}.
\end{equation}
Put the above inequality to Theorem \ref{thm: convergence theorem - empirical - any b}, we establish that
\begin{equation}
    \begin{aligned}
    \|(\hat{b}^\gamma)_{\rm prox} - \beta^*\| &\lesssim \frac{1}{\lambda}\left[\frac{1}{1+\gamma|\Ec|} + {\rm Dist}\left(0, \partial\widehat{\Phi}([\hat{b}^\gamma]_{\rm prox})\right)\right] + \frac{1}{\lambda^{1/2}}\left(\frac{p+s}{n}\right)^{1/4}\\
    &\leq \frac{1}{\lambda}\left[\frac{1}{1+\gamma|\Ec|} + T^{-1/4}\right] + \frac{1}{\lambda^{1/2}}\left(\frac{p+s}{n}\right)^{1/4}
\end{aligned}
\label{eq: dist subdiff hatb*}
\end{equation}
Combining \eqref{eq: output b* - beta*} and \eqref{eq: dist subdiff hatb*}, we have completed the proof of Theorem \ref{thm: convergence theorem - empirical - alg output}.

\subsection{Proof of Theorem \ref{thm: convergence - minimization help - landscape}}
\label{Appendix: thm: convergence - minimization help - landscape}

The key idea is essentially the same as the proof of Theorem \ref{thm: convergence theorem - penal - population - any b}, where we relate the population objective $\Phi(b)$ to its subgradient and then use structural properties of the additive intervention regimes to obtain the upper bounds on $\|b-\beta^*\|_2$.

\vspace{1em}
\noindent\textbf{Step-1: Expressing $\Phi(b)$ in terms of its subgradient.}

From Step 1 of the proof for Theorem \ref{appendix: proof of thm: convergence theorem - population - any b}, recall that \eqref{proof: eq phi origin b grad} shows that
\begin{equation}
\begin{aligned}
    {\Phi}(b) = {\phi}(b,\bar{w}) &= \frac{1}{2}(b-\beta^*)^\intercal\frac{\partial {\phi}(b,\bar{w})}{\partial b} + \frac{\sigma_Y^2 + h^\intercal \Gbf(b-\beta^*)}{1+\gamma|\Ec|}.
\end{aligned}
    \label{eq:landscape-grad}
\end{equation}
where $\bar{w}$ is a maximizer weight vector with $\bar{w}\in \mathcal{W}(b)$ defined in \eqref{proof: eq maximizing points population}.

\vspace{1em}
\noindent\textbf{Step-2: Lower bounding $\Phi(b)$ through the specific weight $w^0$.}

Using the lower bound argument in Step 2 of the proof for Theorem \ref{appendix: proof of thm: convergence theorem - population - any b}, and writing
\[
\Abm(w^0) =\sum_{e\in \Ec}\left(w_e- \frac{1}{|\Ec|}\right)\Eb[X^eX^{e\intercal}] = \Gbf^\intercal\sum_{e\in \Ec}\left(w_e- \frac{1}{|\Ec|}\right)\Eb[\delta^e\delta^{e\intercal}]\Gbf = \Gbf \Deltabf(w^0) \Gbf,
\]
we have for the specific weight vector $w^0$:
\begin{equation}
\begin{aligned}
    {\Phi}(b) &\geq \frac{\sigma_Y^2 + 2h^\intercal \Gbf(b-\beta^*)}{1+\gamma|\Ec|}+ \frac{\lambda_2}{1+\gamma|\Ec|}\|b-\beta^*\|_2 + \frac{\gamma|\Ec|}{1+\gamma|\Ec|}(b-\beta^*)^\intercal  \Abm(w^0) (b - \beta^*),
\end{aligned}
\label{eq:landscape-lower}
\end{equation}
where the second term on the right-hand side arises from the condition that $\sum_{e\in \Ec}w_e^0 \Eb[X^e X^{e\intercal}]\succeq \lambda_2{\bf I}$.

\vspace{1em}
\noindent\textbf{Step-3: Restricting to limited-intervention coordinates.}

In the limited intervention regime, it holds $\delta^e_j\equiv 0$ for all $j\notin S_L$ for all $e\in \Ec$. Thus,
\begin{equation}
\begin{aligned}
    [\Gbf(b-\beta^*)]^\intercal  \Deltabf(w^0) [\Gbf(b - \beta^*)] &= [\Gbf(b-\beta^*)]_{S_L}^\intercal  [\Deltabf(w^0)]_{S_L,S_L} [\Gbf(b - \beta^*)]_{S_L} \\
    &\geq \lambda_1 \left\|[\Gbf(b - \beta^*)]_{S_L}\right\|_2^2,
\end{aligned}
    \label{eq:landscape-restrict}
\end{equation}
Furthermore, from Lemma \ref{lemma: expression of risk}, $h = \sigma_Y^2 \Bbf_{YX} + \Eb[\eta_Y \eta_X].$ When $\Eb[\eta_Y \eta_X] = 0$, $h=\sigma_Y^2\Bbf_{YX}$. Note that $\Bbf_{YX}$ captures the direct causal relationship from $Y$ to $X$ with $ D = {\rm supp}(\Bbf_{YX})\subseteq S_L$, it holds
\begin{equation}
h^\intercal [\Gbf(b-\beta^*)] = \sigma_Y^2 [\Bbf_{YX}]_{S_L}^\intercal [\Gbf(b-\beta^*)]_{S_L}.
    \label{eq:landscape-h}
\end{equation}

\vspace{1em}
\noindent\textbf{Step-4: Combining inequalities.}

Substituting \eqref{eq:landscape-restrict} and \eqref{eq:landscape-h} into
\eqref{eq:landscape-grad}–\eqref{eq:landscape-lower}, and applying Cauchy–Schwarz, we obtain
\begin{equation*}
    \frac{1}{2}\|b-\beta^*\|_2\left\|\frac{\partial {\phi}(b,\bar{w})}{\partial b}\right\|_2 + \frac{\sigma_Y^2\|\Bbf_{YX}\|_2 \|[\Gbf(b-\beta^*)]_{S_L}\|_2}{1+\gamma|\Ec|}\geq \frac{\lambda_1 \gamma|\Ec|}{1+\gamma|\Ec|} \left\|[\Gbf(b - \beta^*)]_{S_L}\right\|_2^2 + \frac{\lambda_2}{1+\gamma|\Ec|}\|b-\beta^*\|_2^2.
\end{equation*}
For notation convenience, we denote
\[
r = \|b-\beta^*\|_2,\quad x = \|[\Gbf(b-\beta^*)]_{S_L}\|_2, \quad g = \left\|\frac{\partial {\phi}(b,\bar{w})}{\partial b}\right\|_2, \quad \kappa = 1+\gamma|\Ec|.
\]
Then we simplify the preceding inequality as follows:
\begin{equation}
    \frac{\lambda_1 (\kappa-1)}{\kappa} x^2 + \frac{\lambda_2}{\kappa}r^2\lesssim r g + \frac{x}{\kappa}.
    \label{eq:landscape-main}
\end{equation}

\vspace{1em}
\noindent\textbf{Step-5: Bounding $\|b-\beta^*\|_2$ preliminarily.}

We establish the following inequality:
\[
x = \|[\Gbf(b-\beta^*)]_{S_L}\|_2 \leq \|\Gbf (b-\beta^*)\|_2 \leq \|\Gbf\|_2 r \lesssim r.
\]
Therefore, if we drop the first term $\frac{\lambda_1 (\kappa-1)}{\kappa} x^2$ on the LHS of \eqref{eq:landscape-main}, we obtain
\[
\frac{\lambda_2}{\kappa} r^2 \lesssim  gr + \frac{1}{\kappa} r.
\]
Diving both sides by $\lambda_2 r /\kappa$ gives
\begin{equation}
r \lesssim \frac{g}{\lambda_2} +\frac{1}{\lambda_2}.
\label{eq:landscape-r-1}
\end{equation}

\vspace{1em}
\noindent\textbf{Step-6: Bounding $[\Gbf(b-\beta^*)]_{S_L}$.}

We remove the $\frac{\lambda_2r^2}{\kappa}$ term on the LHS of \eqref{eq:landscape-main} to get the inequality:
\[
\frac{\lambda_1(\kappa-1)}{\kappa} x^2\lesssim rg + \frac{x}{\kappa}.
\]
Substitute \eqref{eq:landscape-r-1} back into the product $rg$:
\[
rg \lesssim \left(\frac{\kappa}{\lambda_2} g + \frac{1}{\lambda_2}\right)g = \frac{\kappa}{\lambda_2} g^2 + \frac{1}{\lambda_2} g.
\]
Therefore,
\[
\lambda_1 (\kappa-1) x^2 \lesssim x+\frac{\kappa^2}{\lambda_2} g^2 + \frac{\kappa}{\lambda_2} g.
\]
We establish that
\begin{equation}
    x\lesssim \frac{1}{\lambda_1 (\kappa-1)} + \sqrt{\frac{1}{\lambda_1(\kappa-1)}\left(\frac{\kappa^2}{\lambda_2} g^2 + \frac{\kappa}{\lambda_2}g\right)}.
    \label{eq:landscape-x}
\end{equation}

\vspace{1em}
\noindent\textbf{Step-7: Bounding $\|b-\beta^*\|_2$.}

Go back to \eqref{eq:landscape-main}, but drop the nonnegative $x^2$ term to get the following inequality:
\[
\frac{\lambda_2}{\kappa}r^2\lesssim rg + \frac{x}{\kappa}.
\]
Applying Young's inequality to the term $rg$: for any $\varepsilon\in (0,1)$,
\[
rg \leq \frac{\varepsilon\lambda_2}{\kappa} r^2 + \frac{\kappa}{\varepsilon \lambda_2} g^2.
\]
Choosing $\varepsilon=1/2$, the preceding two inequalities imply
\[
\frac{\lambda_2}{\kappa} r^2 \lesssim \frac{\kappa}{\lambda_2} g^2 + \frac{x}{\kappa}.
\]
Now substitute the bound \eqref{eq:landscape-x} for $x$:
\[
\frac{\lambda_2}{\kappa} r^2 \lesssim \frac{\kappa}{\lambda_2} g^2 + \frac{1}{\lambda_1 (\kappa-1)\kappa} +\frac{1}{\kappa} \sqrt{\frac{1}{\lambda_1(\kappa-1)}\left(\frac{\kappa^2}{\lambda_2} g^2 + \frac{\kappa}{\lambda_2}g\right)}.
\]
Multiplying both sides by $\kappa/\lambda_2$ gives:
\[
r^2\lesssim \frac{\kappa^2}{\lambda_2^2} g^2 + \frac{1}{\lambda_1\lambda_2(\kappa-1)\kappa} + \frac{1}{\lambda_2} \sqrt{\frac{\kappa}{\lambda_1(\kappa-1)}\left(\frac{\kappa}{\lambda_2} g^2 + \frac{1}{\lambda_2}g\right)}.
\]
Taking square roots and using $\sqrt{u+v+w}\leq \sqrt{u}+\sqrt{v}+\sqrt{w}$, we get
\begin{equation}
r\lesssim \frac{\kappa}{\lambda_2}g + \frac{1}{\sqrt{\lambda_1\lambda_2(\kappa-1)\kappa}} + \left(\frac{\kappa}{\lambda_2}\right)^{1/2} \left(\frac{1}{\lambda_1 (\kappa-1)}\right)^{1/4} \left(\frac{1}{\lambda_2}\right)^{1/4} g^{1/2} + \left(\frac{\kappa}{\lambda_2}\right)^{1/4} g^{1/4}.
    \label{eq:landscape-r}
\end{equation}

\vspace{1em}
\noindent\textbf{Step-8: Bounding via ${\rm Dist}(0,\partial \Phi(b))$.}

Now we apply Lemma \ref{Lemma: Danskin's Theorem}, which essentially follows the celebrated Danskin's theorem \citep{danskin1966theory}\citep[Proposition~B.25]{bertsekas1997nonlinear}, to obtain that:
\[
\partial{\Phi}(b) = \left\{\frac{\partial}{\partial b}{\phi}(b, \bar{w}),~~\bar{w}\in \Wc(b)\right\}.
\]
Therefore, it holds that
${\rm Dist}(0, \partial {\Phi}(b)) = \inf_{\bar{w}\in \Wc(b)}\left\|\frac{\partial}{\partial b}{\phi}(b, \bar{w})\right\|_2$. We denote
\[
D := {\rm Dist}(0, \partial \Phi(b)).
\]

Note that \eqref{eq:landscape-r} holds for $g=\left\|\frac{\partial \phi(b, \bar{w})}{\partial b}\right\|_2$ for any $\bar{w}\in \mathcal{W}(b)$.  Taking the infimum over $\bar{w}\in \Wc(b)$ gives:
\[
\begin{aligned}
    r&\lesssim \frac{\kappa}{\lambda_2}D + \frac{1}{\sqrt{\lambda_1\lambda_2(\kappa-1)\kappa}} + \left(\frac{\kappa}{\lambda_2}\right)^{1/2} \left(\frac{1}{\lambda_1 (\kappa-1)}\right)^{1/4} \left(\frac{1}{\lambda_2}\right)^{1/4} D^{1/2} + \left(\frac{\kappa}{\lambda_2}\right)^{1/4} D^{1/4}.
\end{aligned}
\]
We now simplify the dependence by treating $\lambda_1,\lambda_2$ as fixed constants. Using the expressions of $r=\|b-\beta^*\|_2$ and $D = {\rm Dist}(0, \partial \Phi(b))$, we achieve
\[
\|b-\beta^*\|_2 \lesssim \kappa {\rm Dist}(0, \partial \Phi(b))+ \kappa^{1/2} ({\rm Dist}(0, \partial \Phi(b)))^{1/2} + \kappa^{1/4} ({\rm Dist}(0, \partial \Phi(b)))^{1/4} + \kappa^{-1/2}, 
\]
with $\kappa = 1+\gamma|\Ec|$.

\subsection{Proof of Theorem \ref{thm: identification infty minimization - general}}
\label{Appendix - proof - identification infty minimization - general}
Recall the constrained problem in \eqref{eq: bneg gamma infty} as follows:
\begin{equation*}
    \argmin_{b\in \Rb^p} \Eb[\Y{1}-b^\intercal\X{1}]^2 \quad {\rm s.t}\quad \Eb[\Y{e}-b^\intercal\X{e}]^2 =\Eb[\Y{f}-b^\intercal\X{f}]^2\;\textrm{for all $e,f\in \Ec$}.
\end{equation*}
We obtain its equivalent formulation by leveraging Lemma \ref{lemma: expression of risk}:
{\small
\begin{equation}
\begin{aligned}
    &\argmin_{b\in \Rb^{p}} 2h^\intercal\Gbf(b-\beta^*) + (b-\beta^*)^\intercal \Gbf^\intercal \left(\Hbf + \Eb[\del{1}\delT{1}]\right)\Gbf (b-\beta^*)\\
    &{\rm s.t.}\quad (b - \beta^*)^\intercal \Gbf^\intercal \Eb[\del{e}\delT{e}]\Gbf (b -\beta^*) = (b - \beta^*)^\intercal\Gbf^\intercal\Eb[\del{f}\delT{f}]\Gbf (b -\beta^*),~~\textrm{for all $e,f\in \Ec$},
\end{aligned}
\label{eq: general thm obj}
\end{equation}
}
where the explicit expressions of $h$ and $\Gbf$ are given in Lemma \ref{lemma: expression of risk}.

The remaining proof proceeds as follows:
\begin{itemize}
    \item[\textbf{Step-1}:] Under the additive intervention, suppose Condition \ref{cond: relaxed minimization - general} holds. We show that for any predictor $b\in \Rb^p$ subject to the constraint within \eqref{eq: general thm obj} satisfies that $h^\intercal \Gbf (b-\beta^*) = 0$.
    \item[\textbf{Step-2}:] Based on Step-1, we simplify the objective function, and prove that solving the problem \eqref{eq: general thm obj} uniquely identifies $\beta^*$.
\end{itemize}
We now detail the remaining proof.

\noindent\textbf{Step-1:} 
We observe that for any predictor $b\in \Rb^p$ satisfying the constraint in \eqref{eq: general thm obj}, it must satisfy that:
\[
(b-\beta^*)^\intercal \Gbf^\intercal\Deltabf(w) \Gbf(b-\beta^*) = (b-\beta^*)^\intercal \Gbf^\intercal\left[\sum_{e\in \Ec}\left(w_e - \frac{1}{|\Ec|}\right) \Eb[\del{e}\delT{e}]\right] \Gbf(b-\beta^*) = 0, \quad \forall w\in \Delta^L.
\]
Therefore, for the specific weight vector $w^0$ in Condition \ref{cond: relaxed minimization - general}, it holds
\begin{equation}
    (b-\beta^*)^\intercal \Gbf^\intercal \Deltabf(w^0) \Gbf(b-\beta^*) = 0.
    \label{eq: general limit intervene - 1}
\end{equation}

Since we consider $\Deltabf(w^0) \succeq 0$ with spectral decomposition:
\[
\Deltabf(w^0) = V^0 \Lambda^0 (V^0)^\intercal,
\]
where $\Lambda^0 = {\rm diag}(\lambda^0_1,\lambda^0_2,..., \lambda^0_p)$ is the diagonal matrix of eigenvalues with $\Lambda^0\succeq 0$, and $V^0 = (v_1^0, v_2^0, ..., v_p^0)$ are columns of corresponding eigenvectors. 
We let $S_+$ represent the set of indices for positive eigenvalues:
\[
S_+ = \{i\in [p]:~~\lambda_i^0 > 0\}.
\]
For any vector $u\in\Rb^p$ satisfying that $u^\intercal \Deltabf(w^0) u =0$, we shall have
\[
0 = u^\intercal \Deltabf(w^0) u = u^\intercal V^0 \Lambda^0 (V^0)^\intercal = \sum_{i=1}^p \lambda_i^0 [(v_i^0)^\intercal u]^2 = \sum_{i\in S_+} \lambda_i^0 [(v_i^0)^\intercal u]^2,
\]
where the last equality holds since $\lambda_i^0 = 0$ for all $i\notin S_+.$
Moreover, considering that $\lambda_i > 0$ for $i\in S_+$, we further have
\[
(v_i^0)^\intercal u = 0, \quad \textrm{for $i\in S_+$}.
\]

In Condition \ref{cond: relaxed minimization - general}, the vector $h$ lies within the positive eigenspace of $\Deltabf(w^0)$ in the sense that
\[
h \in {\rm Span}\left\{v_i^0, ~~i\in S_+\right\}.
\]
Combining the two preceding results, we shall establish that if a vector $u\in \Rb^p$ satisfying $u^\intercal \Deltabf(w^0)u = 0$, then
\[
h^\intercal u = 0.
\]
Together with \eqref{eq: general limit intervene - 1}, we obtain that
\[
h^\intercal \Gbf (b-\beta^*) = 0.
\]
\noindent\textbf{Step-2:} Based on Step-1, we shall remove the first-order term (w.r.t. $b$) in the objective function of the problem \eqref{eq: general thm obj} as follows:
{\small
\begin{equation}
\begin{aligned}
    & \argmin_{b\in \Rb^{p}} (b-\beta^*)^\intercal \Gbf^\intercal \left(\Hbf + \Eb[\del{1}\delT{1}]\right)\Gbf (b-\beta^*)\\
    &{\rm s.t.}\quad (b - \beta^*)^\intercal \Gbf^\intercal \Eb[\del{e}\delT{e}]\Gbf (b -\beta^*) = (b - \beta^*)^\intercal\Gbf^\intercal\Eb[\del{f}\delT{f}]\Gbf (b -\beta^*),~~\textrm{for all $e,f\in \Ec$},
\end{aligned}
\label{eq: mini step1}
\end{equation}
}
Due to the risk-invariant constraint and $w^0\in \Delta^L$, we have
\[
\begin{aligned}
    (b-\beta^*)^\intercal \Gbf^\intercal\Eb[\del{1}\delT{1}]\Gbf (b-\beta^*) &= (b-\beta^*)^\intercal \Gbf^\intercal\left(\sum_{e\in \Ec} w^0_e \Eb[\del{e}\delT{e}]\right)\Gbf (b-\beta^*)\\
    &=(b-\beta^*)^\intercal \Gbf^\intercal\left(\Deltabf(w^0) + \sum_{e\in \Ec} \frac{1}{|\Ec|}\Eb[\del{e}\delT{e}]\right)\Gbf (b-\beta^*).
\end{aligned}
\]
Then we shall further write the optimization problem \eqref{eq: mini step1} equivalently:
{\small
\begin{equation*}
\begin{aligned}
    & \argmin_{b\in \Rb^{p}} (b-\beta^*)^\intercal\Gbf^\intercal  \left[\Hbf + \Deltabf(w^0) + \sum_{e\in \Ec} \frac{1}{|\Ec|}\Eb[\del{e}\delT{e}] \right]\Gbf(b-\beta^*)\\
    &{\rm s.t.}\quad (b - \beta^*)^\intercal \Gbf^\intercal \Eb[\del{e}\delT{e}]\Gbf (b -\beta^*) = (b - \beta^*)^\intercal\Gbf^\intercal\Eb[\del{f}\delT{f}]\Gbf (b -\beta^*),~~\textrm{for all $e,f\in \Ec$}.
\end{aligned}
\end{equation*}
}
We observe that $b=\beta^*$ is a feasible point to the above optimization problem with the objective value $0$. And for any other feasible points $b\neq \beta^*$, the corresponding objective value satisfies that
\[
(b-\beta^*)^\intercal\Gbf^\intercal  \left[\Hbf + \Deltabf(w^0) + \sum_{e\in \Ec} \frac{1}{|\Ec|}\Eb[\del{e}\delT{e}] \right]\Gbf(b-\beta^*) > 0,
\]
as we consider $\Deltabf(w^0)\succ -\Hbf -\sum_{e\in \Ec} \frac{1}{|\Ec|}\Eb[\del{e}\delT{e}]$ as specified in Condition \ref{cond: relaxed minimization - general}. Consequently, we show that solving \eqref{eq: mini step1} exactly recovers $\beta^*$. Equivalently, the optimal solution of the constrained problem \eqref{eq: bneg gamma infty} identifies $\beta^*$.

\section{Proof of Technical Lemmas}
\label{appendix: secD}

\subsection{Proof of Lemma \ref{lemma: expression of risk}}
\label{appendix: proof lemma: expression of risk}
    Since $\Y{e} = (\beta^*)^\intercal \X{e} + \eps{e}_Y$ with the noise term satisfying $\Eb[(\eps{e})^2]=\sigma_Y^2$, we have
\begin{equation}
    \begin{aligned}
    \Eb[\Y{e} - b^\intercal \X{e}]^2 &= \Eb[(\beta^* - b)^\intercal \X{e} + \eps{e}_Y]^2 \\
    &= \sigma_Y^2 - 2 (b - \beta^*)^\intercal\Eb[\eps{e}_Y \X{e}] + (b - \beta^*)^\intercal \Eb[\X{e}(\X{e})^\intercal] (b - \beta^*).
\end{aligned}
\label{eq: risk-raw}
\end{equation}
Now, we continue to express $\Eb[\eps{e}_Y \X{e}]$ and $\Eb[\X{e}(\X{e})^\intercal]$. It follows from \eqref{eq: SCM invert} and \eqref{eq: intervention noise}
\[
\X{e} \stackrel{d}{=} [(\Ibf - \Bbf)^{-1}]_{-1,\cdot}[\eta + \begin{pmatrix}
    0 \\ \del{e}
\end{pmatrix}],\quad \textrm{and}\quad \eps{e}_Y \stackrel{d}{=} \eta_{Y}.
\]
Moreover, as we consider that $\Eb[\del{e}\eta^\intercal] = 0$, it holds that
{
\begin{equation}
    \begin{aligned}
        \Eb[\eps{e}_Y \X{e}] &= [(\Ibf - \Bbf)^{-1}]_{-1,\cdot}\Eb[\eta_Y \eta],\\
\Eb[\X{e}X^{e\intercal}] &= [(\Ibf - \Bbf)^{-1}]_{-1,\cdot}\Eb[\eta \eta^\intercal][(\Ibf - \Bbf)^{-1}]_{-1,\cdot}^\intercal + [(\Ibf - \Bbf)^{-1}]_{-1,-1}\Eb[\del{e} \delT{e}][(\Ibf - \Bbf)^{-1}]_{-1,-1}^\intercal.
    \end{aligned}
\label{eq: expression gram and inner prod}
\end{equation}
}
Next, we study the term $(\Ibf - \Bbf)^{-1}$. It follows from \eqref{eq: SCM} that
\[
\Ibf - \Bbf = \begin{pmatrix}
    1 & - (\beta^{*})^\intercal \\ - \Bbf_{YX} & \Ibf - \Bbf_{XX}
\end{pmatrix}
\]
By the block matrix inversion formula, we establish that
\[
(\Ibf - \Bbf)^{-1} = 
\begin{pmatrix}
    1 + (\beta^*)^\intercal \left[(\Ibf - \Bbf_{XX}) - \Bbf_{YX} (\beta^*)^\intercal\right]^{-1}\Bbf_{YX} & (\beta^*)^\intercal \left[(\Ibf - \Bbf_{XX}) - \Bbf_{YX} (\beta^*)^\intercal\right]^{-1} \\
    -\left[(\Ibf - \Bbf_{XX}) - \Bbf_{YX} (\beta^*)^\intercal\right]^{-1} \Bbf_{YX} & \left[(\Ibf - \Bbf_{XX}) - \Bbf_{YX} (\beta^*)^\intercal\right]^{-1}
\end{pmatrix}
\]
We denote $\Gbf^\intercal = \left[(\Ibf - \Bbf_{XX}) - \Bbf_{YX} (\beta^*)^\intercal\right]^{-1}$, then
\begin{equation}
    [(\Ibf - \Bbf)^{-1}]_{-1,\cdot} = (-\Gbf^\intercal\Bbf_{YX}, \Gbf^\intercal)\quad\textrm{and}\quad [(\Ibf - \Bbf)^{-1}]_{-1,-1} = \Gbf^\intercal.
    \label{proof eq I-B inverse components}
\end{equation}
Due to the invertibility of $\Ibf-\Bbf$, we know $\Gbf$ is a non-singular matrix as well.
Putting the above expressions back to \eqref{eq: expression gram and inner prod}, together with the notation that $\eta = (\eta_Y, \eta_X)^\intercal$, we have

\begin{equation*}
\begin{aligned}
    \Eb[\eps{e}\X{e}] &= -\Gbf^\intercal\Bbf_{YX} \Eb[\eta_Y \eta_Y] + \Gbf^\intercal\Eb[\eta_Y \eta_{X}] = \Gbf^\intercal(-\sigma_Y^2\Bbf_{YX} + \Eb[\eta_Y \eta_{X}]) \\
    \Eb[\X{e}X^{e\intercal}] &= \Gbf^\intercal \left(\sigma_Y^2\Bbf_{YX}\Bbf_{YX}^\intercal - 2\Bbf_{YX}\Eb[\eta_Y\eta_{X}] + \Eb[\eta_{X}\eta_{X}^\intercal] + \Eb[\del{e}\delT{e}]\right)\Gbf.
\end{aligned}
\end{equation*}
We define 
$$h = \sigma_Y^2\Bbf_{YX} - \Eb[\eta_Y\eta_X],$$ and 
\[
\Hbf = \sigma_Y^2\Bbf_{YX}\Bbf_{YX}^\intercal - 2\Bbf_{YX}\Eb[\eta_Y\eta_X] + \Eb[\eta_X\eta_X^\intercal] = \begin{pmatrix}
    \Bbf_{YX}^\intercal \\ \Ibf
\end{pmatrix}^\intercal\Eb[\eta \eta^\intercal]\begin{pmatrix}
    \Bbf_{YX}^\intercal \\ \Ibf
\end{pmatrix} \succeq 0.
\]
We then establish that
{
\begin{equation*}
\begin{gathered}
    \Eb[\eps{e}_Y\X{e}] = -\Gbf^\intercal h, \quad \Eb[\X{e}X^{e\intercal}] = \Gbf^\intercal \left(\Hbf + \Eb[\del{e}\delT{e}]\right)\Gbf.
\end{gathered}
\end{equation*}
}
Putting the above expressions to \eqref{eq: risk-raw}, we complete the proof.

\subsection{Proof of Lemma \ref{lemma: expression of empirical risk}}
\label{appendix: proof lemma: expression of empirical risk}

According to SEMs in \eqref{eq: SCM}, we have
\begin{equation}
    \begin{pmatrix}
    \y{e}_i \\ \x{e}_i
\end{pmatrix} = (\Ibf - \Bbf)^{-1}\begin{pmatrix}
    \epsilon^{e}_{Y,i}\\ \epsilon^{e}_{X,i} 
\end{pmatrix},
\label{proof eq: y,x empirical noise}
\end{equation}
where
$\{\epsilon^{e}_{Y,i}, \epsilon^{e}_{X,i}\}_{i=1}^{n_e}$ are the (unobserved) noise samples drawn from the distribution of additive noise terms $(\eps{e}_Y, \eps{e}_X)$. 
Since $\y{e}_i = (\beta^*)^\intercal \x{e}_i + \epsilon^{e}_{Y,i}$, we shall express the empirical risk as follows: 
{\small
\[
\begin{aligned}
    \widehat{\Eb}[\Y{e}-b^\intercal \X{e}]^2 &= \frac{1}{n_e}\sum_{i=1}^{n_e}[\y{e}_i - b^\intercal \x{e}_i]^2 = \frac{1}{n_e}\sum_{i=1}^{n_e}[(\beta^* - b)^\intercal \x{e}_i + \epsilon^{e}_{Y, i}]^2 \\
    &=\frac{1}{n_e}\sum_{i=1}^{n_e}[\epsilon^{e}_{Y, i}]^2 - 2 (b-\beta^*)^\intercal \frac{1}{n_e}\sum_{i=1}^{n_e}\epsilon^{e}_{Y, i} \x{e}_i + (b -\beta^*)^\intercal \frac{1}{n_e}\sum_{i=1}^{n_e} \x{e}_i x^{e\intercal}_i (b-\beta^*).
\end{aligned}
\]
}
Compared with the population risk ${\Eb}[\Y{e}-b^\intercal \X{e}]^2$, whose expression presented in Lemma \ref{lemma: expression of risk}, we have
{
\begin{equation}
\begin{aligned}
     &\widehat{\Eb}[\Y{e}-b^\intercal \X{e}]^2-{\Eb}[\Y{e}-b^\intercal \X{e}]^2\\  &=  \left(\frac{1}{n_e}\sum_{i=1}^{n_e}[\epsilon^{e}_{Y, i}]^2 - \Eb[(\eps{e}_Y)^2]\right) - 2(b-\beta^*)^\intercal \left(\frac{1}{n_e}\sum_{i=1}^{n_e}\epsilon^{e}_{Y, i} \x{e}_i - \Eb[\eps{e}_Y \eps{e}_X]\right) \\
     &\quad\quad + (b-\beta^*)^\intercal \left(\frac{1}{n_e}\sum_{i=1}^{n_e} \x{e}_i x^{e\intercal}_i - \Eb[\X{e}X^{e\intercal}]\right) (b-\beta^*).
\end{aligned}
\label{proof: eq emprirical risk raw}
\end{equation}
}
We continue investigating the three components on the right-hand side of the above equation.

It follows from \eqref{eq: SCM invert} and  \eqref{proof eq I-B inverse components} that 
\[
x^{e}_i = (-\Gbf^\intercal\Bbf_{YX}, \Gbf^\intercal)\begin{pmatrix}
    \epsilon_{Y,i}^e \\ \epsilon_{X,i}^e
\end{pmatrix},\quad \textrm{and}\quad 
\X{e} \stackrel{d}{=} (-\Gbf^\intercal\Bbf_{YX}, \Gbf^\intercal)\begin{pmatrix}
    \varepsilon_Y^e \\ \varepsilon_X^e
\end{pmatrix}.
\]
We define the matrix $\hat{\mathbf{\Omega}}^{e}$ as follows:
{
\begin{equation}
    \hat{\mathbf{\Omega}}^{e} =  \frac{1}{n_e}\sum_{i=1}^{n_e}\begin{pmatrix}
     \epsilon^{e}_{Y,i} \\ \epsilon^{e}_{X,i}
\end{pmatrix}\begin{pmatrix}
     \epsilon^{e}_{Y,i} \\ \epsilon^{e}_{X,i}
\end{pmatrix}^\intercal- \Eb\left[\begin{pmatrix}
        \eps{e}_Y\\\eps{e}_X
    \end{pmatrix}\begin{pmatrix}
        \eps{e}_Y \\\eps{e}_X
    \end{pmatrix}^\intercal\right].
\end{equation}
}
Then it holds that:
{
\[
\frac{1}{n_e}\sum_{i=1}^{n_e}[\epsilon^{e}_{Y, i}]^2 - \Eb[(\eps{e}_Y)^2] = \begin{pmatrix}
        1 \\ 0
    \end{pmatrix}^\intercal \hat{\mathbf{\Omega}}^{e}\begin{pmatrix}
        1 \\ 0
    \end{pmatrix}.
\]
\[
\begin{aligned}
    \frac{1}{n_e}\sum_{i=1}^{n_e}\epsilon^{e}_{Y, i} \x{e}_i -\Eb[\eps{e}_Y \X{e}] = \Gbf^\intercal \begin{pmatrix}
        -\Bbf_{YX}^\intercal \\ \Ibf
    \end{pmatrix}^\intercal \hat{\mathbf{\Omega}}^e \begin{pmatrix}
        1 \\ 0
    \end{pmatrix}
\end{aligned}
\]
\[
\frac{1}{n_e}\sum_{i=1}^{n_e} \x{e}_i x^{e\intercal}_i - \Eb[\X{e}X^{e\intercal}] =  \Gbf^\intercal   \begin{pmatrix}
        -\Bbf_{YX}^\intercal \\ \Ibf
    \end{pmatrix}^\intercal \hat{\mathbf{\Omega}}^e \begin{pmatrix}
        -\Bbf_{YX}^\intercal \\ \Ibf
    \end{pmatrix}\Gbf.
\]
}
We define
{
\begin{equation}
    \hat{\Pbf}^{e} = \begin{pmatrix}
    -\Bbf_{YX}^\intercal \\ \Ibf
\end{pmatrix}^\intercal\hat{\mathbf{\Omega}}^{e}\begin{pmatrix}
    -\Bbf_{YX}^\intercal \\ \Ibf
\end{pmatrix},~~\hat{q}^{e} = \begin{pmatrix}
    1 \\ 0
\end{pmatrix}^\intercal\hat{\mathbf{\Omega}}^{e}\begin{pmatrix}
    -\Bbf_{YX}^\intercal \\ \Ibf
\end{pmatrix},~~\hat{r}^{e} = \begin{pmatrix}
    1 \\ 0
\end{pmatrix}^\intercal\hat{\mathbf{\Omega}}^{e}\begin{pmatrix}
    1 \\ 0
\end{pmatrix}.
\label{eq: notations-empirical}
\end{equation}
}
Then combining the above results and putting them back to \eqref{proof: eq emprirical risk raw}, we establish that
{
\begin{equation*}
\begin{aligned}
    & \widehat{\Eb}[\Y{e}-b^\intercal \X{e}]^2 = {\Eb}[\Y{e}-b^\intercal \X{e}]^2 + \hat{r}^{e} -2 \hat{q}^{e\intercal}\Gbf(b-\beta^*) + (b-\beta^*)^\intercal\Gbf^\intercal \hat{\Pbf}^{e}\Gbf(b-\beta^*) \\
    &= (\sigma_Y^2 + \hat{r}^{e}) + 2(h - \hat{q}^{e})^\intercal \Gbf(b-\beta^*) + (b-\beta^*)^\intercal \Gbf^\intercal \left(\Hbf + \Eb[\del{e}\delT{e}] + \hat{\Pbf}^{e}\right)\Gbf (b-\beta^*),
\end{aligned}
\end{equation*}
}
where the second equality holds due to Lemma \ref{lemma: expression of risk}.

\subsection{Proof of Lemma \ref{Lemma: rate of empirical norms}}
\label{subsec: proof of lemma rate of empirical norms}
Since the spectral norm is sub-multiplicative, it follows from \eqref{eq: notations-empirical} that the following inequalities hold simultaneously:
\begin{equation}
    \|\hat{\Pbf}^{e}\|_2 \leq \left\|\begin{pmatrix}
    \Bbf_{YX}^\intercal \\ \Ibf
\end{pmatrix}\right\|_2^2\|\hat{\mathbf{\Omega}}^{e}\|_2,~~ 
    \|\hat{q}^{e}\|_2 \leq \left\|\begin{pmatrix}
    \Bbf_{YX}^\intercal \\ \Ibf
\end{pmatrix}\right\|_2\|\hat{\mathbf{\Omega}}^{e}\|_2,~~
    |\hat{r}^{e}| \leq \|\hat{\mathbf{\Omega}}^{e}\|_2
    \label{eq: empirical norm upper bound -1}
\end{equation}
For the term $\hat{\mathbf{\Omega}}^{e}$, notice that it is related to the covariance estimation, and a standard concentration inequality can be applied to upper bound $\|\hat{\mathbf{\Omega}}^{e}\|_2$ \citep[Theorem 4.7.1, Example 4.7.3]{vershynin2018high}.

It follows from Condition \ref{cond: subgaussian} that $(\eps{e}_Y, (\eps{e}_X)^\intercal)^\intercal$ is a sub-gaussian random vector. Then for any $t\geq 0$, with probability at least $1-2e^{-t}$, we have
\[
\left\|\hat{\mathbf{\Omega}}^{e}\right\|_2 \leq C K^2 \left({\sqrt{\frac{p+1+t}{n_e}}}+\frac{p+1+t}{n_e}\right)\left\|\Eb\left[\begin{pmatrix}
    \eps{e}_Y\\\eps{e}_X
\end{pmatrix}\begin{pmatrix}
    \eps{e}_Y \\\eps{e}_X
\end{pmatrix}^\intercal\right]\right\|_2,
\]
where $C, K$ are some universal constants. Next, we utilize the union bound to control $\max_{e\in \Ec}\|\hat{\mathbf{\Omega}}^{e}\|_2$. We define $n = \min_{e\in \Ec}n_e$ as the smallest sample size across all environments. For any $t\geq 0$, with probability at least $1-2|\Ec|e^{-t}$, 
\begin{equation*}
    \begin{aligned}
        \max_{e\in \Ec}\left\|\hat{\mathbf{\Omega}}^{e}\right\|_2 \lesssim \left({\sqrt{\frac{p+1+t}{{n}}}}+\frac{p+1+t}{{n}}\right) \max_{e\in \Ec}\left\|\Eb\left[\begin{pmatrix}
    \eps{e}_Y\\\eps{e}_X
\end{pmatrix}\begin{pmatrix}
    \eps{e}_Y \\\eps{e}_X
\end{pmatrix}^\intercal\right]\right\|_2, 
    \end{aligned}
\end{equation*}
Since $(\eps{e}_Y, (\eps{e}_X)^\intercal)^\intercal$ is a sub-gaussian random vector with a bounded second moment for all $e\in \Ec$, we shall simplify the above inequality as follows: for any $t\geq 0$, with probability at least $1-2|\Ec|e^{-t}$,
\[
\max_{e\in \Ec}\left\|\hat{\mathbf{\Omega}}^{e}\right\|_2 \lesssim \left({\sqrt{\frac{p+t}{{n}}}}+\frac{p+t}{{n}}\right).
\]

Together with the result in the preceding result and \eqref{eq: empirical norm upper bound -1}, we establish that: with probability at least $1-2|\Ec|e^{-t}$:
\[
\max_{e\in \Ec}\max\{\|\hat{\Pbf}^{e}\|_2, \|\hat{q}^{e}\|_2, |\hat{r}^{e}|\} \lesssim \left({\sqrt{\frac{p+t}{{n}}}}+\frac{p+t}{{n}}\right).
\]

\subsection{Proof of Lemma \ref{lemma: smoothness joint}}
\label{proof of lemma: smoothness joint}

For each environment $e\in \Ec$, we define the empirical risk as
\[
g_e(b)=\widehat{\Eb}[\Y{e}-b^\intercal\X{e}]^2,
\]
and we abbreviate $g(b):= (g_e(b))_{e\in \Ec} \in \Rb^{|\Ec|}$. We define the function as
\[
f(b, w) = \sum_{e\in \Ec} \left(w_e - \frac{\gamma}{1+\gamma|\Ec|}\right)g_e(b)-\mu\|w\|_2^2.
\]
Differentiating $f$ w.r.t. $b$ and $w$ gives two gradient blocks:
\[
\nabla_b f(b,w) = \sum_{e\in \Ec}\left(w_e - \frac{\gamma}{1+\gamma|\Ec|}\right)\nabla_b g_e(b), \quad \textrm{and}\quad \nabla_w f(b,w) = g(b) - 2\mu w.
\]
Hence, the full gradient is
\[
\nabla f(b,w) = \begin{pmatrix}
    \nabla_b f(b, w) \\ \nabla_w f(b,w)
\end{pmatrix} = \begin{pmatrix}
     \sum_{e\in \Ec}\left(w_e - \frac{\gamma}{1+\gamma|\Ec|}\right)\nabla_b g_e(b) \\
     g(b) - 2\mu w
\end{pmatrix}.
\]

According to Condition \ref{cond: smooth}, $g_e(b)$ is $M$-smooth with respect to $b$ for all $e\in \Ec$ such that
\[
\|\nabla_b g_e(b_1) - \nabla_b g_e(b_2)\|_2 \leq \kappa_1\|b_1-b_2\|_2,
\]
and we work on a bounded parameter set $b\in \mathcal{B}$ with bounded gradient norm
\[
\kappa_2:= \sup_{e\in \Ec, b\in \mathcal{B}}\|\nabla g_e(b)\|_2 <\infty.
\]
Our goal is to show that for any two points $(b_1, w_1), (b_2, w_2)$,
\[
\|\nabla f(b_1,w_1) - \nabla f(b_2, w_2)\|_2 \leq L\sqrt{\|b_1-b_2\|_2^2 + \|w_1-w_2\|_2^2},
\]
for $L = 2\kappa_1+2\kappa_2\sqrt{|\Ec|} + 2\mu$.

\vspace{1em}
\noindent\textbf{Step-1: Variation of the $b$-gradient.}

Using the triangle inequality, we shall split the difference as follows:
\[
\left\|\nabla_b f(b_1,w_1) - \nabla_b f(b_2,w_2)\right\|_2\leq \left\|\nabla_b f(b_1,w_2) - \nabla_b f(b_2,w_2)\right\|_2 + \left\|\nabla_b f(b_1,w_1) - \nabla_b f(b_1,w_2)\right\|_2.
\]
Next, we upper bound the two terms on the right-hand side in order.

\begin{itemize}
    \item[(a)] \textbf{Change in $b$ with fixed $w$.}
\end{itemize}
For the first term, 
\[
\begin{aligned}
    \left\|\nabla_b f(b_1,w_2) - \nabla_b f(b_2,w_2)\right\|_2 
    &=\left\|\sum_{e\in \Ec}\left(w_{2e} -\frac{\gamma}{1+\gamma|\Ec|}\right) \left(\nabla_b g_e(b_1) - \nabla_b g_e(b_2)\right)\right\|_2 \\
    &\leq \sum_{e\in \Ec}\left|w_{2e} - \frac{\gamma}{1+\gamma|\Ec|}\right| \left\|\nabla_b g_e(b_1) - \nabla_b g_e(b_2)\right\|_2\\
    &\leq 2\kappa_1\|b_1-b_2\|_2,
\end{aligned}
\]
where the last inequality holds as $g_e(b)$ is $\kappa_1$-smooth with
\[
\|\nabla_b g_e(b_1) - \nabla_b g_e(b_2)\|_2 \leq \kappa_1\|b_1-b_2\|_2,
\]
and
\[
\sum_{e\in \Ec}\left|w_{2e} - \frac{\gamma}{1+\gamma|\Ec|}\right| \leq \sum_{e\in \Ec}w_{2e} + \frac{\gamma|\Ec|}{1+\gamma|\Ec|}\leq 2.
\]

\begin{itemize}
    \item[(b)] \textbf{Change in $w$ with fixed $b$.}
\end{itemize}
For the second term,
\[
\begin{aligned}
    \left\|\nabla_b f(b_1,w_1) - \nabla_b f(b_1,w_2)\right\|_2 &=\left\|\sum_{e\in \Ec}(w_{1e}-w_{2e}) \nabla_b g_e(b_1)\right\|_2 \\
    &\leq \kappa_2\|w_1-w_2\|_1 \leq \kappa_2\sqrt{|\Ec|}\|w_1-w_2\|_2,
\end{aligned}
\]
where the first inequality holds due to $\kappa_2:=\sup_{e\in \Ec, b\in \mathcal{B}}\|\nabla_b g_e(b)\|_2$.

Combining {\bf (a)} and {\bf (b)}, we establish that
\begin{equation}
    \left\|\nabla_b f(b_1,w_1) - \nabla_b f(b_2,w_2)\right\|_2 \leq 2\kappa_1\|b_1-b_2\|_2 + \kappa_2\sqrt{|\Ec|} \|w_1-w_2\|_2.
    \label{eq: b smooth}
\end{equation}

\vspace{1em}
\noindent\textbf{Step-2: Variation of the $w$-gradient.}

Similarly, we shall decompose:
\[
\begin{aligned}
    \left\|\nabla_w f(b_1,w_1) - \nabla_w f(b_2,w_2)\right\|_2 &= \|g(b_1) - g(b_2) -2\mu (w_1 - w_2)\|_2 \\
 &\leq \|g(b_1) - g(b_2)\|_2 + 2\mu\|w_1 - w_2\|_2.
\end{aligned}
\]
For each coordinate $e$,
\[
g_e(b_1) - g_e(b_2) = \int_0^1 \nabla_b g_e(b_2 + t(b_1 - b_2))^\intercal (b_1 - b_2) dt.
\]
Taking norms and using the bounded gradient assumption gives
\[
\|g(b_1) - g(b_2)\|_2 \leq \sqrt{|\Ec|} \kappa_2\|b_1 - b_2\|_2.
\]
Therefore,
\begin{equation}
    \left\|\nabla_w f(b_1,w_1) - \nabla_w f(b_2,w_2)\right\|_2 \leq \sqrt{|\Ec|} \kappa_2\|b_1 - b_2\|_2 + 2\mu\|w_1-w_2\|_2.
    \label{eq: w smooth}
\end{equation}

\vspace{1em}
\noindent\textbf{Step-3: Joint Smoothness.}

Stacking both blocks from \eqref{eq: b smooth} and \eqref{eq: w smooth}, 
\[
\begin{aligned}
    \|\nabla f(b_1, w_1) - \nabla f(b_2, w_2)\|_2 &\leq \|\nabla_b f(b_1,w_1) - \nabla_b f(b_2,w_2)\|_2+ \|\nabla_w f(b_1,w_1) - \nabla_w f(b_2,w_2)\|_2\\
    &\leq \left(2 \kappa_1 + \sqrt{|\Ec|}\kappa_2\right)\|b_1-b_2\|_2 + (\kappa_2\sqrt{|\Ec|}+2\mu) \|w_1 - w_2\|_2\\
    &\leq L_{\rm joint} \sqrt{\|b_1-b_2\|_2^2 + \|w_1-w_2\|_2^2},
\end{aligned}
\]
where
\[
L_{\rm joint}\leq \sqrt{\left(2 \kappa_1+ \sqrt{|\Ec|}\kappa_2\right)^2 + \left(\sqrt{|\Ec|} \kappa_2+ 2\mu\right)^2} \leq 2\kappa_1 + 2\kappa_2\sqrt{|\Ec|}+2\mu.
\]
Hence, $f(b,w)$ is $L_{\rm joint}$-smooth with $L_{\rm joint}\leq 2\kappa_1 + 2\kappa_2\sqrt{|\Ec|}+2\mu$.

\subsection{Proof of Lemma \ref{lemma: smoothness of maximized function}}
\label{proof of lemma: smoothness of maximized function}

    We denote by $y^*(x)$ the unique maximizer:
\[
y^*(x) = \argmax_{y\in \mathcal{Y}}f(x,y).
\]
The uniqueness of $y^*(x)$ follows from the strong concavity of $f(x,\cdot)$ in $y$, for each fixed $x$.

\vspace{1em}
\noindent\textbf{Step-1: Optimality conditions.}

For any $x\in \mathcal{X}$, the first-order optimality condition for the concave maximization problem gives:
\[
(y - y^*(x))^\intercal \nabla_y f(x, y^*(x)) \leq 0, \quad \forall y\in \mathcal{Y}.
\]
Applying this condition to $x_1, x_2\in \mathcal{X}$:
\begin{equation}
    \left(y - y^*(x_1)\right)^\intercal \nabla_y f(x_1, y^*(x_1))\leq 0, \quad \forall y\in \mathcal{Y},
    \label{eq: smooth lemma-1}
\end{equation}
\begin{equation}
    \left(y - y^*(x_2)\right)^\intercal \nabla_y f(x_2, y^*(x_2))\leq 0, \quad \forall y\in \mathcal{Y}.
\label{eq: smooth lemma-2}
\end{equation}
Substituting $y=y^*(x_2)$ in \eqref{eq: smooth lemma-1} and $y=y^*(x_1)$ in \eqref{eq: smooth lemma-2} and adding them yields:
\begin{equation}
    \left(y^*(x_2)-y^*(x_1)\right)^\intercal\left(\nabla_y f(x_1, y^*(x_1)) - \nabla_y f(x_2, y^*(x_2))\right) \leq 0.
    \label{eq: smooth max step-1}
\end{equation}

\vspace{1em}
\noindent\textbf{Step-2: Using strong concavity in $y$.}

Since the function $f(x_1, \cdot)$ is $\mu$-strongly concave, its gradient satisfies:
\[
\left(\nabla_y f(x_1, y^*(x_2)) - \nabla_y f(x_1, y^*(x_1))\right)^\intercal (y^*(x_2) - y^*(x_1)) \leq -\mu\|y^*(x_2) - y^*(x_1)\|_2^2.
\]
Adding the inequality above and \eqref{eq: smooth max step-1} eliminates the middle term $\nabla_y f(x_1, y^*(x_1))$ and yields
\begin{equation}
    \left(y^*(x_2)-y^*(x_1)\right)^\intercal\left(\nabla_y f(x_2, y^*(x_2))-\nabla_y f(x_1, y^*(x_2))\right) \geq \mu\|y^*(x_2) - y^*(x_1)\|_2^2.
\label{eq: smooth max step-2}
\end{equation}

\vspace{1em}
\noindent\textbf{Step-3: Bounding using joint smoothness}

Because $f$ is $\ell$-smooth jointly in $(x,y)$, varying $x$ while keeping $y$ fixed gives
\[
\left\|\nabla_y f(x_2, y^*(x_2)) - \nabla_y f(x_1, y^*(x_2))\right\| \leq \ell \|x_2 - x_1\|_2.
\]
Combining with \eqref{eq: smooth max step-2} and applying the Cauchy-Schwarz inequality,
\[
\begin{aligned}
    \mu\|y^*(x_2) - y^*(x_1)\|_2^2 &\leq \left\|y^*(x_2)-y^*(x_1)\right\|_2\left\|\nabla_y f(x_2, y^*(x_2))-\nabla_y f(x_1, y^*(x_2))\right\|\\
    &\leq \ell \left\|y^*(x_2)-y^*(x_1)\right\|_2  \|x_2 - x_1\|_2.
\end{aligned}
\]
Diving both sides by $\mu\|y^*(x_2) - y^*(x_1)\|_2$ (nonzero since $\mu>0$) yields
\begin{equation}
    \|y^*(x_2) - y^*(x_1)\|_2 \leq \frac{\ell}{\mu} \|x_2 - x_1\|_2.
\label{eq: smooth max step-3}
\end{equation}
Thus, the maximizer mapping $y^*(\cdot)$ is $\ell/\mu$-Lipschitz continuous in $x$.

\vspace{1em}
\noindent\textbf{Step-4: Smoothness of the value function.}

We define
\[
F(x) = \max_{y\in \mathcal{Y}}f(x,y).
\]
By Danskin's theorem (which applies since $\mathcal{Y}$ is convex and bounded, and $y^*(x)$ is uniquely defined),
\[
\nabla F(x) = \nabla_x f(x, y^*(x)).
\]
Then for any $x_1,x_2$,
\[
\begin{aligned}
    \left\|\nabla F(x) - \nabla F(x')\right\|_2 &= \left\|\nabla_x f(x, y^*(x)) - \nabla_x f(x', y^*(x'))\right\|_2 \\
    &\leq  \ell\left(\|x-x'\|_2 + \|y^*(x) - y^*(x')\|_2\right).
\end{aligned}
\]
where the inequality follows from the $\ell$-smoothness of $f$ jointly on $(x,y)$.

Leveraging the Lipschitz property of $y^*(\cdot)$ in \eqref{eq: smooth max step-3} gives
\[
\left\|\nabla F(x) - \nabla F(x')\right\|_2\leq (\ell + \ell^2/\mu) \|x-x'\|_2.
\]
Hence $F(x) = \max_{y\in \mathcal{Y}}f(x,y)$ is $(\ell + \ell^2/\mu)$-smooth in $x$.

\subsection{Proof of Lemma \ref{Lemma: Danskin's Theorem}}
\label{subsec: proof of lemma: danskin's theorem}
    The first part is one of the multiple versions of the celebrated Danskin's theorem \citep{danskin1966theory}, and it was proven in \citet[Proposition B.25]{bertsekas1997nonlinear}. Here, we only show the second part of removing the ${\rm conv}$ operation.

    To remove the ${\rm conv}$ operation, it suffices to show that
    $$
    A(x) = \left\{\frac{\partial f(x,z)}{\partial x}:~~z\in \mathcal{Z}_0(x)\right\}
    $$
    itself has already been a convex set. For any $a_1, a_2 \in A(x)$, it holds that
    \[
    a_1 = \frac{\partial f(x,\bar{z}_1)}{\partial x},~~a_2  = \frac{\partial f(x,\bar{z}_2)}{\partial x}
    \]
    for $\bar{z}_1, \bar{z}_2 \in \mathcal{Z}_0(x).$ Since $f(x, z)$ is linear function towards $z$, it holds that $\frac{\partial}{\partial x}f(x,z)$ is a linear function with respect to $z$ as well. Then for any $t\in [0,1]$, it holds that
    \[
    t a_1 + (1 - t) a_2 = t \frac{\partial f(x,\bar{z}_1)}{\partial x} + (1 - t) \frac{\partial f(x,\bar{z}_2)}{\partial x} =  \frac{\partial f(x, t\bar{z}_1 + (1-t)\bar{z}_2)}{\partial x}. 
    \]
    If we have $t\bar{z}_1 + (1-t)\bar{z}_2\in \mathcal{Z}_0(x)$, the preceding equation implies that $ta_1 + (1-t)a_2 \in A(x)$ for any $t\in [0,1]$, meaning that $A(x)$ is a convex set.
    Therefore, to show that $A(x)$ is a convex set, it suffices to show that $t\bar{z}_1 + (1-t)\bar{z}_2\in \mathcal{Z}_0$.
    
    By the definition of $\bar{z}_1$ and $\bar{z}_2$, we know that
    \[
    f(x,\bar{z}_1) = f(x, \bar{z}_2) = \max_{z\in \mathcal{Z}}f(x,z).
    \]
    As $\mathcal{Z}$ is a convex set, $t\bar{z}_1 + (1-t)\bar{z}_2 \in \mathcal{Z}$ as well. Moreover, since $f(x,z)$ is a linear function with respect to $z$ for every $x$, we have
    \[
    t f(x, \bar{z}_1) + (1-t) f(x, \bar{z}_2) = f(x, t\bar{z}_1 + (1-t)\bar{z}_2) = \max_{z\in \mathcal{Z}}f(x,z),
    \]
    for every $t\in [0,1].$ It implies that $t\bar{z}_1 + (1-t)\bar{z}_2 \in \mathcal{Z}_0$ as well. The proof is completed.

\subsection{Proof of Lemma \ref{Lemma: hat Phi b weakly convex}}
\label{subsec: proof of lemma Phi b weakly convex}
By the definition of $\hat{\phi}(b,w)$ in \eqref{eq: hat phi b,w}, its Hessian matrix towards $b$ is:
\[
\nabla_b^2\hat\phi(b,w) = \sum_{e\in \Ec}\left({w}_e - \frac{\gamma}{1+\gamma |\Ec|}\right) \frac{1}{n_e}\sum_{i=1}^{n_e} \x{e}_i (\x{e}_i)^\intercal.
\]
Since $w\in \Delta^{|\Ec|}$, it holds that:
\begin{equation}
    \nabla_b^2\hat\phi(b,w)\succeq -\frac{1}{|\Ec|} \sum_{e\in \Ec}\frac{1}{n_e}\sum_{i=1}^{n_e} \x{e}_i (\x{e}_i)^\intercal.
    \label{eq: lower bound Hessian phi(b,w)}
\end{equation}
We observe that
\[
\begin{aligned}
    \left\|\frac{1}{n_e}\sum_{i=1}^{n_e} \x{e}_i (\x{e}_i)^\intercal\right\|_2 
    &= \left\|\Eb[\X{e}(\X{e})^\intercal]+\frac{1}{n_e}\sum_{i=1}^{n_e} \x{e}_i (\x{e}_i)^\intercal - \Eb[\X{e}(\X{e})^\intercal]\right\|_2\\
    &\leq \left\|\Eb[\X{e}(\X{e})^\intercal]\right\|_2 + \left\|\frac{1}{n_e}\sum_{i=1}^{n_e} \x{e}_i (\x{e}_i)^\intercal - \Eb[\X{e}(\X{e})^\intercal]\right\|_2.
\end{aligned}
\]
Therefore, we establish that
{\small
\begin{equation}
    \begin{aligned}
    \left\|\sum_{e\in \Ec}\frac{1}{n_e}\sum_{i=1}^{n_e} \x{e}_i (\x{e}_i)^\intercal\right\|_2&\leq \sum_{e\in \Ec}\left\|\frac{1}{n_e}\sum_{i=1}^{n_e} \x{e}_i (\x{e}_i)^\intercal\right\|_2 \\
    &\leq |\Ec|\max_{e\in \Ec}\left\|\frac{1}{n_e}\sum_{i=1}^{n_e} \x{e}_i (\x{e}_i)^\intercal\right\|_2\\
    &\leq |\Ec|\max_{e\in \Ec}\left\|\Eb[\X{e}(\X{e})^\intercal]\right\|_2 + |\Ec|\max_{e\in \Ec}\left\|\frac{1}{n_e}\sum_{i=1}^{n_e} \x{e}_i (\x{e}_i)^\intercal - \Eb[\X{e}(\X{e})^\intercal]\right\|_2.
\end{aligned}
\label{eq: sum covariance est x spectral norm}
\end{equation}
}

By the standard concentration inequality for covariance matrix estimation \citet[Theorem 4.7.1, Example 4.7.3]{vershynin2018high}, for any $t\geq 0$, with probability at least $1-2e^{-t}$, we have
\begin{equation}
    \left\|\frac{1}{n_e}\sum_{i=1}^{n_e} \x{e}_i (\x{e}_i)^\intercal - \Eb[\X{e}(\X{e})^\intercal]\right\|_2 \leq C K^2 \left(\sqrt{\frac{p+t}{n_e}} + \frac{p+t}{n_e}\right) \left\|\Eb[\X{e}(\X{e})^\intercal]\right\|_2,
    \label{eq: covariance est x spectral norm}
\end{equation}
where $C,K$ are some universal constants.
We define ${n}=\min_{e\in \Ec} n_e$ the smallest sample size across environments. 
By union bound, it follows from \eqref{eq: covariance est x spectral norm} that for any $t\geq 0$, with probability at least $1-2|\Ec|e^{-t}$:
{\small
\begin{equation}
    \max_{e\in \Ec}\left\|\frac{1}{n_e}\sum_{i=1}^{n_e} \x{e}_i (\x{e}_i)^\intercal - \Eb[\X{e}(\X{e})^\intercal]\right\|_2 \leq C K^2 \left(\sqrt{\frac{p+t}{{n}}} + \frac{p+t}{{n}}\right) \max_{e\in \Ec}\left\|\Eb[\X{e}(\X{e})^\intercal]\right\|_2.
    \label{eq: max covariance est x spectral norm}
\end{equation}}

Putting the results in \eqref{eq: sum covariance est x spectral norm} and \eqref{eq: max covariance est x spectral norm} together, we obtain that
{\small
\[
\frac{1}{|\Ec|}\left\|\sum_{e\in \Ec}\frac{1}{n_e}\sum_{i=1}^{n_e} \x{e}_i (\x{e}_i)^\intercal\right\|_2 \leq \left[1+ C K^2 \left(\sqrt{\frac{p+t}{{n}}} + \frac{p+t}{{n}}\right)\right]\max_{e\in \Ec}\left\|\Eb[\X{e}(\X{e})^\intercal]\right\|_2.
\]
}
Together with the above inequality with \eqref{eq: lower bound Hessian phi(b,w)}, we shall establish that: for any $t> 0$, with probability at least $1 - 2|\Ec|e^{-t}$
\begin{equation*}
    \nabla_b^2\hat\phi(b,w)\succeq -\left[1+ C K^2 \left(\sqrt{\frac{p+t}{{n}}} + \frac{p+t}{{n}}\right)\right]\max_{e\in \Ec}\left\|\Eb[\X{e}(\X{e})^\intercal]\right\|_2.
\end{equation*}

According to Lemma \ref{lemma: expression of risk}, $\Eb[\X{e}(\X{e})^\intercal]$ shall be expressed as follows:
\[
\Eb[\X{e}(\X{e})^\intercal] = \Gbf^\intercal\left(\begin{pmatrix}
    \Bbf_{YX}^\intercal \\ \Ibf
\end{pmatrix}^\intercal\Eb[\eta \eta^\intercal]\begin{pmatrix}
    \Bbf_{YX}^\intercal \\ \Ibf
\end{pmatrix} + \Eb[\del{e}\delT{e}]\right)\Gbf.
\]
Together with the bounded second moments of the vector $\eta$ and $\del{e}$, for all $e\in \Ec$, as implied in Condition \ref{cond: subgaussian}, we obtain that 
$\max_{e\in \Ec}\left\|\Eb[\X{e}(\X{e})^\intercal]\right\|_2 < C$ bounded as well. We consider ${n}\geq p+t$, then
\[
\sqrt{\frac{p+t}{n}} + \frac{p+t}{n} \leq 2\frac{p+t}{n} \leq 2.
\]
The preceding results imply that
\begin{equation*}
    \nabla_b^2\hat\phi(b,w)\succeq - C,
\end{equation*}
for some universal constant $C>0.$
Therefore, for any weak convexity parameter $\rho > C$, we know that the function
\[
b\mapsto \hat\phi(b,w) + \frac{1}{2}\|b\|^2
\]  
is a convex function towards $b$, and $\hat\phi(b,w)$ is $\rho$-weakly convex towards $b$ for any $w\in \Delta^{|\Ec|}.$

Recall the definition of $\widehat{\Phi}(b) = \max_{w\in \Delta^{|\Ec|}}\hat\phi(b,w)$, it can be written as:
\[
\widehat{\Phi}(b) = \hat\phi(b,\bar{w}), \quad \textrm{with} ~~\bar{w}\in \argmax_{w\in \Delta^{|\Ec|}}\hat\phi(b,w).
\]
Therefore, according to our preceding discussions, $\widehat{\Phi}(b)$ is $\rho$-weakly convex as well.

\subsection{Proof of Lemma \ref{Lemma: subdiff dist bound}}
\label{subsec: Proof of lemma subdiff dist bound}
Since $\widehat{\Phi}(\zeta)$ is $\rho$-weakly convex towards $\zeta$, according to Definition \ref{def: weak convex}, we know that for any given $b\in \Rb^p$, the function $\zeta \mapsto \widehat\Phi(\zeta) + \frac{\rho}{2}\|\zeta-b\|_2^2$ is convex. Therefore, as long as $\upsilon < \rho^{-1}$, we shall establish that the function
\[
\widehat{\Phi}(\zeta) + \frac{1}{2\upsilon}\|\zeta-b\|^2 = \left(\widehat{\Phi}(\zeta) + \frac{\rho}{2}\|\zeta-b\|_2^2\right) + \left(\frac{1}{2\upsilon}-\frac{\rho}{2}\right) \|\zeta - b\|_2^2
\]
is $(1/\upsilon - \rho)$-strongly convex towards $\zeta$. Therefore, for any given $b\in \Rb^p$, the function $\zeta\mapsto \widehat{\Phi}(\zeta) + \frac{1}{2\upsilon}\|\zeta-b\|^2$ is strongly convex and has a unique minimizer, denoted as follows: 
\[
{\rm prox}_{\upsilon \Phi}(b) = \argmin_{\zeta\in \Rb^p}\left\{\widehat{\Phi}(\zeta) + \frac{1}{2\upsilon}\|\zeta-b\|^2_2\right\}.
\]
Then the Moreau envelope of $\widehat{\Phi}(b)$ follows that:
\[
\widehat{\Phi}_\upsilon(b) = \min_{\zeta\in \Rb^p} \left\{\widehat{\Phi}(\zeta) + \frac{1}{2\upsilon}\|\zeta - b\|_2^2\right\} = \widehat{\Phi}({\rm prox}_{\upsilon \Phi}(b)) + \frac{1}{2\upsilon}\|{\rm prox}_{\upsilon \Phi}(b) - b\|_2^2.
\]
Due to the uniqueness of the minimizer ${\rm prox}_{\upsilon \Phi}(b)$, Danskin's theorem implies that the gradient of $\widehat{\Phi}_\upsilon(b)$ towards $b$ can be expressed as:
\[
\begin{aligned}
    \nabla \widehat{\Phi}_\upsilon(b) &=\nabla \min_{\zeta\in \Rb^p}\left\{\widehat{\Phi}(\zeta) + \frac{1}{2\upsilon}\|\zeta-b\|_2^2\right\} = \upsilon^{-1}(b - {\rm prox}_{\upsilon\widehat{\Phi}}(b)).
\end{aligned}
\]
Therefore,
\[
\|b - {\rm prox}_{\upsilon\widehat{\Phi}}(b)\|_2 = \upsilon\|\nabla\widehat{\Phi}_\upsilon(b)\|_2.
\]

Moreover, the subdifferential of $\widehat{\Phi}(\zeta) + \frac{1}{2\upsilon}\|\zeta-b\|^2_2$ towards $\zeta$ is given by
\[
\partial \left\{\widehat{\Phi}(\zeta) + \frac{1}{2\upsilon}\|\zeta-b\|^2_2\right\} = \partial \widehat{\Phi}(\zeta) + \upsilon^{-1}(\zeta - b).
\]
Since ${\rm prox}_{\upsilon\widehat{\Phi}}(b) = \argmin_{\zeta\in \Rb^p}\{\widehat{\Phi}(\zeta) + \frac{1}{2\upsilon}\|\zeta-b\|_2^2\}$, we have 
$$0\in \partial \left.\left\{\widehat{\Phi}(\zeta) + \frac{1}{2\upsilon}\|\zeta-b\|^2_2\right\}\right\vert_{\zeta={\rm prox}_{\upsilon\widehat{\Phi}}(b) }.$$ 
Combining the above two equations, we obtain that
\[
\upsilon^{-1}(b - {\rm prox}_{\upsilon\widehat{\Phi}}(b) ) \in \partial \widehat{\Phi}({\rm prox}_{\upsilon\widehat{\Phi}}(b) ).
\]
Recall the distance of $\partial \widehat{\Phi}({\rm prox}_{\upsilon\widehat{\Phi}}(b) )$ towards the origin is defined as
\[
{\rm dist}(0, \partial \widehat{\Phi}({\rm prox}_{\upsilon\widehat{\Phi}}(b) ) = \inf\left\{\|x\|_2, ~x\in \partial\widehat{\Phi}({\rm prox}_{\upsilon\widehat{\Phi}}(b) )\right\},
\]
therefore,
\[
{\rm dist}(0, \partial \widehat{\Phi}({\rm prox}_{\upsilon\widehat{\Phi}}(b) )) \leq \|\upsilon^{-1}(b-{\rm prox}_{\upsilon\widehat{\Phi}}(b) )\|_2 = \|\nabla \widehat{\Phi}_\upsilon (b)\|_2.
\]

\subsection{Proof of Lemma \ref{Lemma: bounded subdiff}}
\label{subsec: proof of lemma bounded subdiff}
For any $w\in \Delta^{|\Ec|}$, it follows the expression of $\hat\phi(b,w)$ in \eqref{eq: hat phi b,w} that
\begin{equation*}
    \begin{aligned}
     \frac{\partial}{\partial b}\hat\phi(b,w) = \sum_{e\in \Ec}\left(w_e - \frac{\gamma}{1+\gamma|\Ec|}\right) \nabla\widehat{\Eb}[\ell(\X{e},\Y{e};b)] .
    \end{aligned}
\end{equation*}
Then for any $w\in \Delta^{|\Ec|}$, we observe
\begin{equation}
\begin{aligned}
    \left\|\frac{\partial}{\partial b}\hat\phi(b,w)\right\|_2 &= \left\|\sum_{e\in \Ec}\left(w_e - \frac{\gamma}{1+\gamma|\Ec|}\right) \nabla\widehat{\Eb}[\ell(\X{e},\Y{e};b)]\right\|_2 \\
    &\leq 2 \max_{e\in \Ec}\left\|\nabla\widehat{\Eb}[\ell(\X{e},\Y{e};b)]\right\|_2 \leq C.
\end{aligned}
\label{eq: lemma bounded key}
\end{equation}
where the last inequality holds by Condition \ref{cond: bounded gradient}.

Since $\hat\phi(b, w)$, as defined in \eqref{eq: hat phi b,w}, is differentiable with respect to $b$ for all $w\in \Delta^{|\Ec|}$, and $\frac{\partial \hat\phi(b,w)}{\partial b}$ is continuous with respect to $w$ for all $b$, then by Lemma \ref{Lemma: Danskin's Theorem}, the subdifferential of $\widehat{\Phi}(b)$ is given by:
\begin{equation*}
    \partial \widehat{\Phi}(b) = \left\{\frac{\partial \hat\phi(u, \bar{w})}{\partial u}, \bar{w}\in \argmax_{w\in \Delta^{|\Ec|}} \hat\phi(u,w)\right\}.
\end{equation*}
Then it follows from \eqref{eq: lemma bounded key} that
\[
\sup\left\{\|x\|, ~x\in \partial \widehat\Phi(b)\right\} \leq C.
\]

\subsection{Proof of Lemma \ref{Lemma: moreau norm bound}}
\label{subsec: proof of lemma moreau norm bound}

For each index $t$, define 
$$\zeta_t:= \frac{\partial \hat{\phi}(b_t, w_t)}{\partial b},$$ 
where $w_t$ is the maximizer of $\hat{\phi}(b_t, \cdot)$ such that $w_t \in \argmax_{w\in \Delta^{|\Ec|}} \hat{\phi}(b_t, w)$. Since we define $\widehat{\Phi}(b) = \max_{w\in \Delta^{|\Ec|}}\hat{\phi}(b, w)$, by Danskin's theorem, we have $\zeta_t \in \partial \widehat{\Phi}(b_t).$ For any $\hat{\rho}> \rho$, we set $\hat{b}_t := {\rm prox}_{\widehat{\Phi}/\hat{\rho}}(b_t).$

By definition of $\widehat{\Phi}_{1/\hat{\rho}}(\cdot)$, we have
\begin{equation}
    \widehat{\Phi}_{1/\hat{\rho}}(b_{t+1}) = \min_{x}\left\{\widehat{\Phi}(x)+ \frac{\hat{\rho}}{2}\|x - b_{t+1}\|^2_2\right\} \leq \widehat{\Phi}(\hat{b}_t) + \frac{\hat{\rho}}{2}\|\hat{b}_t - b_{t+1}\|^2_2.
    \label{eq: proof grad 1}
\end{equation}
In Algorithm \ref{alg: GDmax subgrad}, we update $b_{t+1} = b_t - \alpha_t \zeta_t$, thus
\begin{equation*}
\begin{aligned}
    \|\hat{b}_t - b_{t+1}\|^2_2 &= \|(b_t - \hat{b}_t) - \alpha_t \zeta_t\|^2_2 \\
    &= \|b_t - \hat{b}_t\|^2_2 + 2\alpha_t\langle \hat{b}_t - {b}_t, \zeta_t\rangle + \alpha_t^2 \|\zeta_t\|^2_2 \\
    &\leq \|b_t - \hat{b}_t\|^2_2 + 2\alpha_t\langle \hat{b}_t - {b}_t, \zeta_t\rangle + \alpha_t^2 L^2,
\end{aligned}
\end{equation*}
where the last inequality is due to the Lemma \ref{Lemma: bounded subdiff} that $\sup_{x \in \partial \widehat{\Phi}(b_t)} \|x\|^2_2 \leq L^2$ and $\zeta_t \in \partial \widehat{\Phi}(b_t).$
Putting the above inequality back to \eqref{eq: proof grad 1}, we have
\begin{equation}
    \begin{aligned}
        \widehat{\Phi}_{1/\hat{\rho}}(b_{t+1}) &\leq \widehat{\Phi}(\hat{b}_t) + \frac{\hat{\rho}}{2}\|b_t - \hat{b}_t\|^2_2 + \hat{\rho}\alpha_t  \langle\hat{b}_t - {b}_t, \zeta_t\rangle  + \frac{\alpha_t^2\hat{\rho}}{2} L^2.
    \end{aligned}
    \label{eq: proof grad 2}
\end{equation}

By definition of $\hat{b}_t$ and $\widehat{\Phi}_{1/\hat{\rho}}(b_t)$ as follows:
\[
\hat{b}_t:= {\rm prox}_{\widehat{\Phi}/\hat{\rho}}(b_t) = \argmin_{b}\left\{\widehat{\Phi}(b) + \frac{\hat{\rho}}{2}\|b - b_t\|^2_2\right\},
\]
and
\[
\widehat{\Phi}_{1/\hat{\rho}}(b_t) := \min_{b} \left\{\widehat{\Phi}(b) + \frac{\hat{\rho}}{2}\|b - b_t\|^2_2\right\},
\]
we have
\[
\widehat{\Phi}(\hat{b}_t) + \frac{\hat{\rho}}{2}\|\hat{b}_t - b_t\|^2_2 = \widehat{\Phi}_{1/\hat{\rho}}(b_t).
\]
Combining the above equality with \eqref{eq: proof grad 2}, we have
\begin{equation}
    \widehat{\Phi}_{1/\hat{\rho}}(b_{t+1})\leq \widehat{\Phi}_{1/\hat{\rho}}(b_t) + \hat{\rho}\alpha_t \langle \hat{b}_t - {u}_t, \zeta_t\rangle + \frac{\alpha_t^2\hat{\rho}}{2} L^2.
\end{equation}
Recall that $\widehat{\Phi}(\cdot)$ is $\rho$-weakly convex, it holds that:
\[
\langle \hat{b}_t - b_t, \zeta_t\rangle \leq \widehat{\Phi}(\hat{b}_t) - \widehat{\Phi}(b_t) + \frac{\rho}{2}\|\hat{b}_t - b_t\|^2_2.
\]
Then it holds that
\begin{equation}
    \widehat{\Phi}_{1/\hat{\rho}}(b_{t+1})\leq \widehat{\Phi}_{1/\hat{\rho}}(b_t) + \hat{\rho}\alpha_t \left(\widehat{\Phi}(\hat{b}_t) - \widehat{\Phi}(b_t) + \frac{\rho}{2}\|\hat{b}_t - b_t\|^2_2\right) + \frac{\alpha_t^2\hat{\rho}}{2} L^2.
    \label{eq: proof grad 3}
\end{equation}

With recursion from index $t=0$ to $t=T$, it follows from \eqref{eq: proof grad 3} that
\begin{equation}
    \widehat{\Phi}_{1/\hat{\rho}}(b_{T+1})\leq \widehat{\Phi}_{1/\hat{\rho}}(b_0) + \frac{\hat{\rho}L^2}{2}\sum_{t=0}^T \alpha_t^2 - \hat{\rho}\sum_{t=0}^T \alpha_t \left(\widehat{\Phi}(b_t) - \widehat{\Phi}(\hat{b}_t) - \frac{\rho}{2}\|b_t -\hat{b}_t\|^2_2\right).
    \label{eq: proof grad 3.5}
\end{equation}

Next, we show that the left-hand side is lower bounded by $0$. Notice that
\[
\begin{aligned}
    \widehat{\Phi}_{1/\hat{\rho}}(b_{T+1}) = \min_{u}\left\{\widehat{\Phi}(b) + \frac{\hat{\rho}}{2}\|b - b_{T+1}\|^2_2\right\} \geq \min_b \widehat{\Phi}(b),
\end{aligned}
\]
and for any $b\in \Rb^p$
\[
\begin{aligned}
    \widehat{\Phi}(b) &= \max_{w\in \Delta^{|\Ec|}}(1+\gamma|\Ec|)\left(w_e -\frac{\gamma}{1+\gamma |\Ec|}\right) \widehat{\Eb}[\Y{e}-b^\intercal\X{e}]^2 \\
    &\geq 0.
\end{aligned}
\]
Therefore, $\widehat{\Phi}_{1/\hat{\rho}}(b_{T+1})\geq 0$, and then \eqref{eq: proof grad 3.5} implies that: 
\begin{equation*}
   0\leq \widehat{\Phi}_{1/\hat{\rho}}(b_0) + \frac{\hat{\rho}L^2}{2}\sum_{t=0}^T \alpha_t^2 - \hat{\rho}\sum_{t=0}^T \alpha_t \left(\widehat{\Phi}(b_t) - \widehat{\Phi}(\hat{b}_t) - \frac{\rho}{2}\|b_t -\hat{b}_t\|^2_2\right).
    \label{eq: proof grad 3.9}
\end{equation*}

By setting step sizes $\alpha_t = \alpha$ for all indices $t$ and rearranging the above inequality, we obtain the bound:
\begin{equation}
    \frac{1}{T+1}\sum_{t=0}^T \left(\widehat{\Phi}(b_t) - \widehat{\Phi}(\hat{b}_t) - \frac{\rho}{2}\|b_t -\hat{b}_t\|^2_2\right)\leq \frac{\widehat{\Phi}_{1/\hat{\rho}}(b_0) + \frac{\hat{\rho} L^2}{2} (T+1) \alpha^2}{\hat{\rho}(T+1)\alpha}.
    \label{eq: proof grad 4}
\end{equation}

We observe that
\begin{equation}
    \begin{aligned}
    \widehat{\Phi}(b_t) - \widehat{\Phi}(\hat{b}_t) - \frac{\rho}{2}\|b_t-\hat{b}_t\|^2_2 &= 
    \left(\widehat{\Phi}(b_t) + \frac{\hat{\rho}}{2}\|b_t - {b}_t\|^2_2\right) - \left(\widehat{\Phi}(\hat{b}_t) + \frac{\hat{\rho}}{2}\|\hat{b}_t - b_t\|^2_2\right) + \frac{\hat{\rho}-\rho}{2}\|b_t - \hat{b}_t\|^2_2
\end{aligned}
\label{eq: proof grad 5}
\end{equation}
The function $b\mapsto \widehat{\Phi}(b) + \frac{\hat{\rho}}{2}\|b - b_t\|^2_2$ is strongly convex with parameter $\hat{\rho} - \rho$, therefore by the property of strongly convexity
\[
\begin{aligned}
    &\left(\widehat{\Phi}(b_t) + \frac{\hat{\rho}}{2}\|b_t - {b}_t\|^2_2\right) - \left(\widehat{\Phi}(\hat{b}_t) + \frac{\hat{\rho}}{2}\|\hat{b}_t - b_t\|^2_2\right) \\
    &\geq \left[\nabla_{b} \left(\widehat{\Phi}(b) + \frac{\hat{\rho}}{2}\|b_t - b\|^2_2\right)\right]_{{b = \hat{b}_t}} \|b_t - \hat{b}_t\| + \frac{\hat{\rho}-\rho}{2}\|b_t - \hat{b}_t\|^2_2 \\
    &= \frac{\hat{\rho}-\rho}{2}\|b_t - \hat{b}_t\|^2_2,
\end{aligned}
\]
where the last equality is due to the definition of $\hat{b}_t$ which minimizes the strongly convex function $b\mapsto \widehat{\Phi}(b) + \frac{\hat{\rho}}{2}\|b - b_t\|^2_2$
and thus its gradient is zero. Putting the above result back to \eqref{eq: proof grad 5}, we establish that
\[
\widehat{\Phi}(b_t) - \widehat{\Phi}(\hat{b}_t) - \frac{\rho}{2}\|b_t-\hat{b}_t\|^2_2 \geq (\hat{\rho} - \rho)\|b_t -\hat{b}_t\|^2_2 = \frac{\hat{\rho} - \rho}{\hat{\rho}^2} \|\nabla \widehat{\Phi}_{1/\hat{\rho}}(b_t)\|^2_2,
\]
where the last equality follows from \eqref{eq: b_prox - b dist upsilon} in Lemma \ref{Lemma: subdiff dist bound}. 
Using the result in \eqref{eq: proof grad 4}, we obtain that
\begin{equation*}
    \frac{\hat{\rho}-\rho}{\hat{\rho}^2(T+1)}\sum_{t=0}^T \|\nabla_{1/\hat{\rho}} (b_t)\|^2_2 \leq \frac{\widehat{\Phi}_{1/\hat{\rho}}(b_0)  + \frac{\hat{\rho} L^2}{2}(T+1)\alpha^2}{\hat{\rho}(T+1)\alpha}.
\end{equation*}
By setting $\hat{\rho} = 2\rho$, and we assume that there exists some constant $R$ satisfying that $\widehat{\Phi}_{1/(2\rho)}(b_0)\leq R$, then the above inequality implies that
\begin{equation}
    \frac{1}{2(T+1)}\sum_{t=0}^T \|\nabla_{1/(2\rho)} (b_t)\|^2_2 \leq \frac{R + \rho L^2(T+1)\alpha^2}{(T+1)\alpha}.
    \label{eq: proof grad 6}
\end{equation}
With the step-size $\alpha$ being set as:
\[
\alpha = \left(\frac{R}{(T+1)\rho L^2}\right)^{1/2},
\]
it follows from \eqref{eq: proof grad 6} that:
\[
\frac{1}{T+1}\sum_{t=0}^T \|\nabla_{1/(2\rho)} (b_t)\|^2_2 \leq 4 L\sqrt{\frac{R\rho}{T+1}}.
\]

\section{Additional Illustration and Simulations}
\label{appendix: secE}

\subsection{Illustration of SEMs}
\label{appendix: SEMs}

The SEMs in Definition \ref{def: SEM} naturally induce a directed causal graph $G = (V,E)$, within the framework of causal graphical models \citep{pearl2009causality, spirtes2001causation}. Here $V = \{1,..., p+1\}$ represents the set of nodes (vertices), and $E$ represents the set of directed edges, where $(i,j)\in E$ if and only if node $i$ is the parent of $j$. A node $j$ is said to be a (direct) \emph{child} of node $i$ if and only if $i$ is a parent of $j$.
\begin{Definition}[Causal Graph]
    We say there is a \emph{directed path} from node $i$ to $j$ if there exists a sequence of nodes $(v_1,...,v_k)$ with $k\geq 2$ such that $v_1 = i$, $v_k = j$, and $(v_l, v_{l+1})\in E$ for any $1\leq l\leq k-1$. We call a directed graph $G$ is a \emph{directed acyclic graph} (DAG) if there does not exist a direct path from node $j$ to itself for any node $j\in V$. Any node connected by a directed path to $i$ is an \emph{ancestor} of $i$, and any node connected by a directed path from $i$ is a \emph{descendant} of $i$.
    \label{def: DAG}
\end{Definition}

In the example of Section \ref{subsec: illu}, the SEMs in \eqref{eq: illus-1} indicate that $\X{e}_1$ is the sole cause of $\Y{e}$, resulting in $\beta^* = (1,0)^\intercal$; $\Y{e}$ directly causes $\X{e}_2$, leading to $\Bbf_{Y X} = (0,1)^\intercal$. There are no direct causal relationships among covariates, making $\Bbf_{XX} = 0_{2\times 2}$.
    We consider two more cases of $\Bbf$ as follows, whose causal graphs are depicted in Figure \ref{fig:more_egs_B}.
    \begin{equation}
        \begin{aligned}
        &\textbf{\textrm{Case (a):}}\quad \beta^* = \begin{pmatrix}
            1 \\ 0 \\ 0
        \end{pmatrix}; \Bbf_{YX} = \begin{pmatrix}
            0 \\ -1 \\ -2
        \end{pmatrix}; \Bbf_{XX} = \begin{pmatrix}
            0 & 0 & 0 \\
            0 & 0 & 0 \\
            2 & -3 & 0
        \end{pmatrix} \\
        &\textbf{\textrm{Case (b):}}\quad \beta^* = \begin{pmatrix}
            0 \\ 1 \\ 2 \\ 0
        \end{pmatrix}; \Bbf_{YX} = \begin{pmatrix}
            0 \\ 0 \\ 0 \\ 3
        \end{pmatrix}; \Bbf_{XX} = \begin{pmatrix}
            0 & 0 & 0 & 0 \\
            0 & 0 & 0 & 0\\
            -1 & -2 & 0 & 0\\
            0 & -3 & 0 & 0
        \end{pmatrix}
    \end{aligned} 
    \label{eq: add sems}
    \end{equation}
    \begin{figure}[H]
        \centering
        \includegraphics[width=0.6\linewidth]{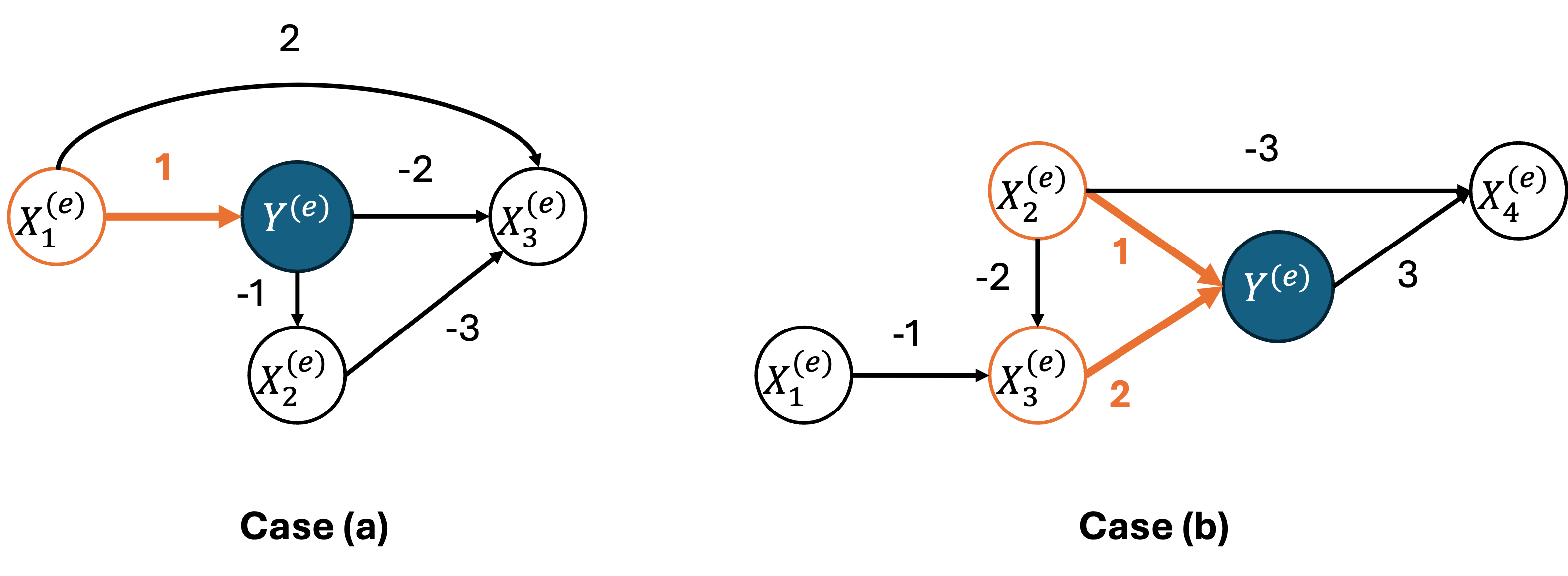}
        \caption{Causal Graphs of \eqref{eq: add sems}.}
        \label{fig:more_egs_B}
    \end{figure}

\subsection{Additional experiments with hidden confounders}
\label{appendix: add simus}
While the experiments in Section~\ref{sec: numerical} exclude hidden confounders between the covariates and the outcome, we now examine the performance of NegDRO and competing approaches when hidden confounding is present.

The data-generation mechanism follows exactly the same structural equations as in~\eqref{eq: simu - SEMs}; the only difference lies in the noise specification.
We introduce a hidden confounder $H\sim \Nc(0,1)$ that simultaneously influences both the outcome and certain covariates. Specifically, the outcome noise is generated as:
\[
\eps{e}_Y = \mathcal{N}(0,1) + 0.5 H
\]
and the covariate noise $\eps{e}_{1:p}$ are defined as follows for each environment $e\in \Ec=\{1,2,3,4\}$,
\[
\eps{e}_{1} = \mathcal{N}(0,1) + (0.5+0.2 e)H + \delta^e_1,\quad \eps{e}_{2:p} = \mathcal{N}(0,1) + \delta^e_{2:p}.
\]
Here, the hidden confounder $H$ affects both $Y$ and $X_1$. The intervention components $\delta^e$ are generated as:
\[
\delta^1_{1:5} =0_5,\; \delta^2_{1:5}\stackrel{i.i.d.}{\sim}\mathcal{N}(0,9),\; \delta^3_{1:5}=(1,2,-1,-2,1)^\intercal,\; \delta^4_{1:5} \stackrel{i.i.d.}{\sim}{\rm Unif}(-1,1),
\]
and for the irrelevant covariates,
\[
\delta^e_{6:p}\sim \mathcal{N}(0_{p-5}, 0.25 e^2{\bf I}_{p-5}).
\]

Following the setup in Section~\ref{sec: numerical}, we compare the proposed \texttt{NegDRO} with existing causal invariance learning methods in terms of both estimation accuracy and runtime, varying the covariate dimension $[5,100]$. The competing methods include \texttt{ICP}~\citep{peters2016causal}, \texttt{Anchor}~\citep{rothenhausler2021anchor}, and \texttt{EILLS}~\citep{fan2023environment}.
For reference, we also include the \texttt{ERM} baseline, which pools all environments and fits a least-squares model.
All other experimental configurations are identical to those in Section~\ref{sec: numerical}.

\begin{figure}[ht!]
    \centering
    \includegraphics[width=0.85\linewidth]{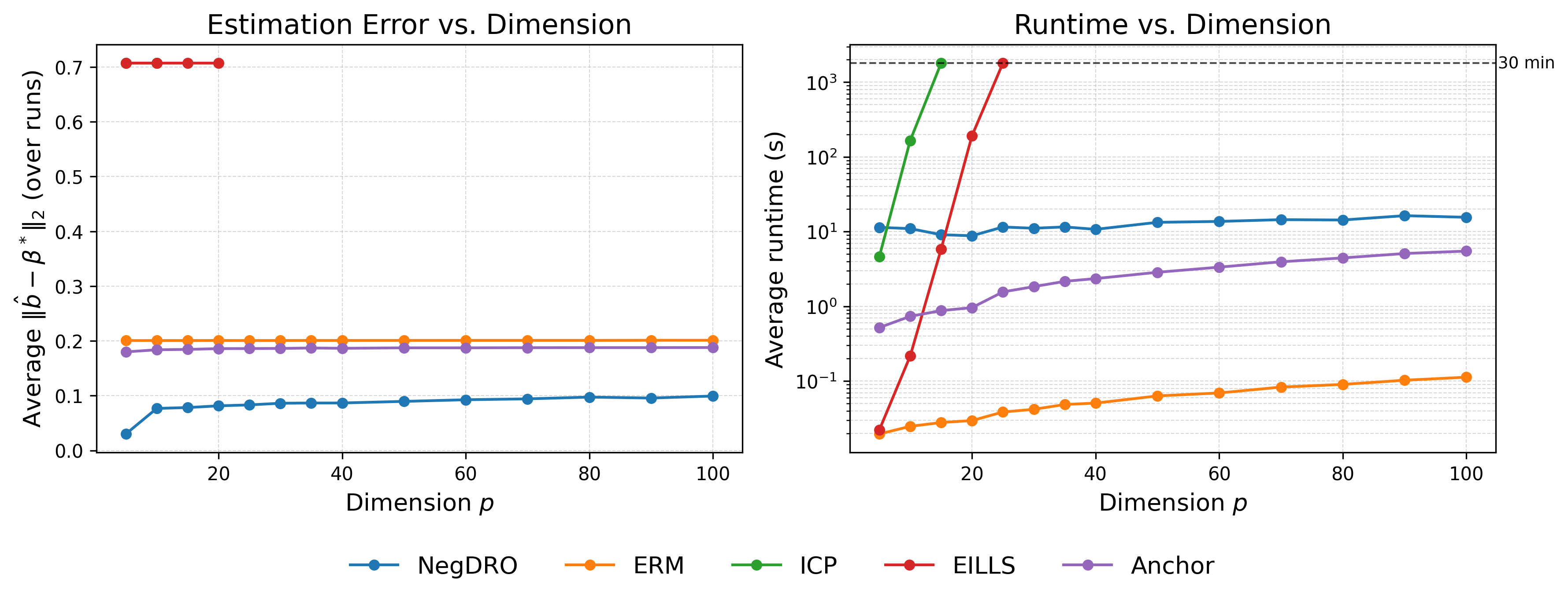}
    \caption{Comparison of different causal invariance learning methods under hidden confounding. The covariate dimension $p$ varies within the range of $[5, 100]$, while the sample size is fixed at $n = 20,000$. The left panel displays the $\ell_2$ distance between each method’s estimator and the causal outcome model $\beta^*$.  The right panel depicts the average runtime for each method, with a black dashed horizontal line marking the 30-minute time limit. Methods exceeding this limit are terminated. Results are averaged over 200 simulations.}
    \label{fig: comparison with methods - 2}
\end{figure}

Figure~\ref{fig: comparison with methods - 2} presents the results.
In the left panel, \texttt{NegDRO} consistently approximates the true causal model $\beta^*$ across all dimensions, demonstrating robustness to hidden confounding.
By contrast, \texttt{ICP}, \texttt{Anchor}, and \texttt{EILLS} exhibit substantial estimation errors, and notably, \texttt{ICP} coincides with \texttt{EILLS} in performance.
Unlike in the confounder-free case shown in Figure~\ref{fig: comparison with methods}, \texttt{EILLS} fails to recover $\beta^*$ here, highlighting its sensitivity to latent confounding.
Overall, only \texttt{NegDRO} successfully identifies the invariant causal relationship when hidden confounders are present.

In the right panel, we report computational time.
Both \texttt{ICP} and \texttt{EILLS} rely on exhaustive subset enumeration and quickly exceed the 30-minute limit when the dimension reaches $p=15$ and $p=25$, respectively.
In contrast, \texttt{NegDRO}, \texttt{Anchor}, and \texttt{ERM} maintain polynomial-time scalability, with runtime growing smoothly as $p$ increases.

\subsection{Numerical Implementation of the Proposed Algorithm}
\label{appendix: subsec adapated algorithm}
We start with the practical issue of the current Algorithm \ref{alg: GDmax subgrad}, which is designed to locate a stationary point of the NegDRO objective $\widehat{\Phi}(b)$, as defined in \eqref{eq: obj unpenal empirical}. We reformulate $\widehat{\Phi}(b)$ as follows:
\[
\widehat{\Phi}(b) = \frac{1}{1+\gamma|\Ec|} \max_{e\in \Ec} \Eb[\ell(\X{e},\Y{e};b)] + \frac{\gamma|\Ec|}{1+\gamma|\Ec|}\cdot\left(\max_{e\in \Ec}\Eb[\ell(\X{e},\Y{e};b)] - \frac{1}{|\Ec|} \sum_{e\in \Ec}\Eb[\ell(\X{f},\Y{f};b)]\right).
\]
The equivalence arises because the maximization over the linear weight $w$ in \eqref{eq: obj unpenal empirical} is achieved by assigning weight $1+\gamma(|\Ec|-1)$ to the highest-risk environment and $-\gamma$ to all other environments. As $\gamma$ becomes large, the first term $\frac{1}{1+\gamma|\Ec|}\max_{e\in \Ec}\Eb[\ell(\X{e},\Y{e};b)]$ diminishes, resulting in
\[
\widehat{\Phi}(b)\approx \max_{e\in \Ec}\Eb[\ell(\X{e},\Y{e};b)] - \frac{1}{|\Ec|} \sum_{e\in \Ec}\Eb[\ell(\X{f},\Y{f};b)].
\]
Therefore, minimizing $\widehat{\Phi}(b)$ when $\gamma$ is set large becomes solely about ensuring risk parity across environments, i.e., the risk of all environments is nearly identical.

While risk parity is critical for causal discovery, it is also important to minimize risk, particularly in limited intervention scenarios, as discussed in Section \ref{sec: minimization}. To address this, we introduce a constant $\tau>0$ and identify a stationary point of the augmented objective
\begin{equation}
    \widehat{\Phi}(b) + \tau \max_{e\in \Ec} \Eb[\ell(\X{e},\Y{e};b)].
    \label{eq: obj augmented}
\end{equation}
Notably, the preceding objective can be expressed in another form of NegDRO by replacing $\Uc(\gamma)$, defined in \eqref{eq: obj original}, with the following set.
For the given $\gamma,\tau\geq 0$, we define the set $\Uc(\tau, \gamma)$:
\[
\Uc(\gamma,\tau) = \left\{w\in \Rb^{|\Ec|}: ~~\sum_{e\in \Ec}w_e = 1+\tau\gamma|\Ec|, ~~\min_{e\in \Ec}w_e\geq -\gamma\right\},
\]
Then we have
\[
\widehat{\Phi}(b) + \tau \max_{e\in \Ec} \Eb[\ell(\X{e},\Y{e};b)] = \frac{1}{1+\gamma|\Ec|} \max_{w\in \Uc(\gamma,\tau)}\sum_{e\in \Ec} w_e \Eb[\ell(\X{e},\Y{e};b)].
\]

Given a specified $\tau>0$, we search for a stationary point of the augmented objective function \eqref{eq: obj augmented}.
We then apply a hard thresholding step to refine its support:
\[
S^\tau = \{j\in [p]: |\hat{b}^{\gamma,\tau}_j| \geq c_0\},
\]
where $c_0>0$ is a small, pre-specified threshold. After updating $S^\tau$, we reduce $\tau$ by half and find the stationary point of the augmented objective in \eqref{eq: obj augmented} again, but using only the covariates within $S^\tau$. This process is repeated until the support $S^\tau$ stabilizes, i.e., when $S^\tau$ no longer changes. Finally, restricted to the stabilized support $S^\tau$, we search for a stationary point of $\widehat{\Phi}(b)$ without adjustment for the final solution. The adapted algorithm is illustrated in Figure \ref{fig: illus adapt}.
\begin{figure}[H]
    \centering
    \includegraphics[width=0.8\linewidth]{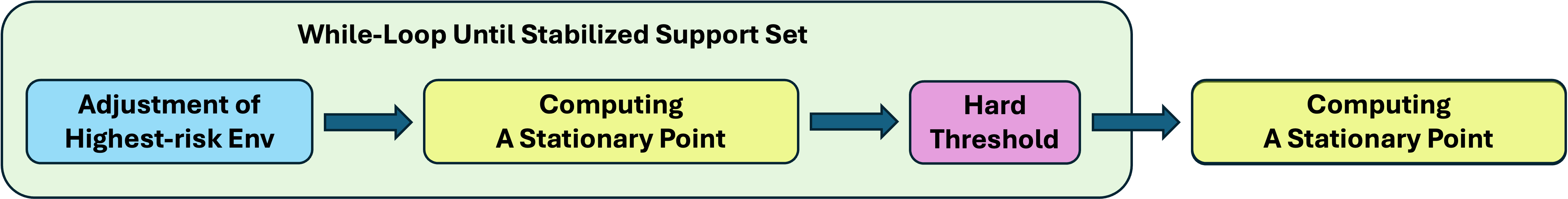}
    \caption{Illustration of adapted Algorithm \ref{alg: GDmax unpenal adapt} for numerical implementation. Compared to Algorithm \ref{alg: GDmax subgrad}, it includes an adjustment to account for the highest-risk environment and application of hard thresholding.}
    \label{fig: illus adapt}
\end{figure}

We summarize the adapted algorithm in the following Algorithm \ref{alg: GDmax unpenal adapt}.

\begin{algorithm}[H]
    \DontPrintSemicolon
    \SetAlgoLined
    \SetNoFillComment
    \LinesNotNumbered 
    \caption{Implementation Algorithm for NegDRO in \eqref{eq: obj unpenal empirical}}
    \SetKwInOut{Input}{Input}
    \SetKwInOut{Output}{Output}
    \Input{Iteration time $T$, step sizes $\{\alpha^{t}\}_{t=0}^T$, initial point $b^0$, regularization parameter $\gamma>0$, adjustment parameter $\tau>0$ ,threshold $c_0>0$}
    \Output{$\hat{b}^\gamma$}

    Initialize the current support set $S^\tau = [p]$ and the previous support set $S^{\tau}_{\rm prev} = \emptyset$
    
    \While{$S^\tau\neq S^{\tau}_{\rm prev}$}{

    Set $S^\tau_{\rm prev} = S^\tau$;

    Compute $\hat{b}^{\gamma,\tau}$, a stationary point of
    \[
    \widehat{\Phi}_\mu(b) + \tau \max_{e\in \Ec} \Eb[\ell(\X{e},\Y{e};b)], ~~{\rm s.t.}~~ b_j = 0~~\textrm{for all $j\notin S^\tau$.}
    \]

    Update the support set with hard thresholding: $S^\tau = \{j\in [p]: |\hat{b}^{\gamma,\tau}_j|\geq c_0\}.$

    Half the adjustment parameter $\tau = \tau/2$.
    }

    Compute $\hat{b}^\gamma$, a stationary point of 
    $$\widehat{\Phi}_\mu(b), ~~ ~~{\rm s.t.}~~ b_j = 0~~\textrm{for all $j\notin S^\tau$.}$$
    
    \label{alg: GDmax unpenal adapt}
\end{algorithm}

\subsection{Implementation Details for Other Causal Invariance Learning Methods}
\label{appendix: subsec implementation}
For ELLIS, we use its brute-force search implementation, with code available at \url{https://github.com/wmyw96/EILLS/tree/main}. For CausalDantzig, we apply its regularized version, using the implementation provided in the R package ``InvariantCausalPrediction'' (version 0.7-1), where the regularization parameter is internally tuned by its algorithm. Through experimentation, we find that the regularized version offers greater stability compared to the unregularized implementation, particularly in scenarios where the difference of Gram matrices $\Eb[\X{e}X^{e\intercal}]-\Eb[\X{f}X^{f\intercal}]$ is not invertible, for some pair $e, f\in \Ec$.
ICP includes a Type-I error hyperparameter $\alpha$, while DRIG and Anchor regression require hyperparameters to balance the least-squares and invariance regularizers. For ICP, DRIG, and Anchor regression, we select hyperparameters in an oracle manner by enumerating a range of possible values and choosing the one that minimizes the $\ell_2$ distance to $\beta^*$.

\end{document}